\documentclass[aps,prb,reprint,showpacs,groupedaddress]{revtex4-1}
\pdfoutput=1
\usepackage{graphicx,color}
\usepackage{amsmath, amssymb}
\usepackage{longtable}
\usepackage{natbib}

\begin{document}
\title{
Dynamical Jahn--Teller instability in metallic fullerides
}
\author{Naoya Iwahara}
\author{Liviu F. Chibotaru}
%\email[]{Liviu.Chibotaru@chem.kuleuven.be}
\affiliation{Theory of Nanomaterials Group, 
Katholieke Universiteit Leuven, 
Celestijnenlaan 200F, B-3001 Leuven, Belgium}
\date{\today}

\begin{abstract}
Dynamical Jahn-Teller effect has escaped so far direct observation in metallic systems. 
It is particularly believed to be quenched also in correlated conductors with 
orbitally degenerate sites such as cubic fullerides.
Here the Gutzwiller approach is extended to treat electron correlation 
over metals with Jahn-Teller active sites 
and applied to the investigation of the ground
state of K$_3$C$_{60}$. 
It is shown that dynamical Jahn-Teller instability fully develops in this material
when the interelectron repulsion $U$ on C$_{60}$ sites exceeds some critical value.
The latter is found to be lower than the current estimates of $U$,
meaning that dynamical Jahn-Teller effect takes place in all cubic fullerides.
This leads to strong splitting of LUMO orbitals on C$_{60}$ sites and calls for 
reconsideration of the role of orbital degeneracy in the Mott-Hubbard transition in fullerides.
\end{abstract}

% 71.70.Ej : Jahn--Teller effect in condensed matter
% 71.20.Tx : Fullerenes - electronic structure 
% 71.27.+a : Strongly correlated electron systems 

\maketitle

%%%%%%%%%%%%%%%%%%%%%%%%%%%%%%%%%%%%%%%%%%%%%%%%%%%%%%%%%%%%%
%
% Introduction
%
%%%%%%%%%%%%%%%%%%%%%%%%%%%%%%%%%%%%%%%%%%%%%%%%%%%%%%%%%%%%% 

\section{Introduction}
\label{Sec:Introduction}
Dynamical Jahn-Teller effect (JTE) is an ubiquitous phenomenon in molecules and isolated
impurity centers with orbital degeneracy. \cite{Bersuker1989a, Bersuker2006a}
Its presence in Jahn-Teller crystals is encountered less often, where cooperative
ordering of static Jahn-Teller distortion is the most probable scenario. \cite{Kaplan1995a}
Dynamical JTE has been advocated as a reason for the lack of orbital ordering in some 
insulating materials such as LiNiO$_2$, \cite{Barra1999a}
Ba$_3$CuSb$_2$O$_9$, \cite{Nakatsuji2012a} and FeSc$_2$S$_4$. \cite{Krimmel2005a}
It was also assumed to take place in insulating fullerides $A_4$C$_{60}$ with 
$A =$ K, Rb, Cs, \cite{Fabrizio1997a} and Li$_3$(NH$_3$)$_6$C$_{60}$. \cite{Chibotaru2005a, Chibotaru2007a}
Recently, {\it ab initio} calculations have shown that dynamical JTE is the reason for 
the lack of orbital ordering in Cs$_3$C$_{60}$ fullerides, which explains their conventional
antiferromagnetic ordering. \cite{Iwahara2013a}
As for metallic systems, no direct evidence for %developing 
development of dynamical JT instability in their ground state has been obtained so far. 
Materials where such instability is likely to be realized are metallic cubic fullerides $A_3$C$_{60}$,
\cite{Gunnarsson1997a} for the %a %simple 
reason that these correlated conductors are 
close to Mott-Hubbard insulators \cite{Ganin2008a, Takabayashi2009a, Ganin2010a} 
for which the existence of dynamical JTE was already proved.
\cite{Chibotaru2005a, Chibotaru2007a, Iwahara2013a}

In $A_3$C$_{60}$, the conduction band originating from the triply degenerate 
$t_{1u}$ lowest unoccupied molecular orbitals
(LUMO) on fullerene sites strongly couples to the intramolecular JT active
fivefold degenerate $h_g$ modes. \cite{Gunnarsson1995a, Iwahara2010a}
Despite the JT coupling, the symmetry lowering has not been observed in x-ray diffraction data
\cite{Stephens1991a, Stephens1992a, Ganin2008a, Takabayashi2009a, Ganin2010a} implying
that the JT effect is either quenched by the formation of the band or dynamical.
An adequate description of the ground state in metallic $A_3$C$_{60}$ requires a concomitant 
treatment of the JT effect and the electron correlation.
One of the simplest methods to treat the electron correlation is 
by variational approach with the Gutzwiller's wave function. \cite{Gutzwiller1963a, Vollhardt1984a}
Despite the simplicity, Gutzwiller's approach allows to take into account the main contribution to the correlation energy.
Concerning the ground state of metallic phase, the description by this method is comparable 
in accuracy to dynamical mean-field theory. \cite{Lanata2012a}
Moreover, it has been extended to treat various situations, for instance, the multiband systems. \cite{Bunemann1998a}
However an adequate approach suitable for degenerate conductors with JT effect on sites
is still lacking.

In this work, we propose a method to treat electron correlation in metallic JT systems based
on a self-consistent multiband Gutzwiller ansatz and apply it to metallic fullerides.
We find that a dynamical JT instability takes place in $A_3$C$_{60}$ already at intermediate
strength of electron correlation, leading to large amplitudes of JT distortions on the fullerene 
sites and to the removal of the degeneracy of three LUMO levels. % by ca 300 meV.
This means, in particular, that the electron correlation in $A_3$C$_{60}$ develops not in a
threefold degenerate LUMO band as thought before \cite{Gunnarsson1997a} but in three split
subbands. The immediate implication is that the degeneracy of the LUMO band as a reason for 
the high critical value $U/w$ for the Mott-Hubbard transition 
\cite{Gunnarsson1996a, Iwasa2003a} ($w$ is the bandwidth of the degenerate LUMO band)
%is not relevant 
should be reconsidered for fullerides.

%%%%%%%%%%%%%%%%%%%%%%%%%%%%%%%%%%%%%%%%%%%%%%%%%%%%%%%%%%%%%
% 
% Self-consistent Gutzwiller wavefunction
%
%%%%%%%%%%%%%%%%%%%%%%%%%%%%%%%%%%%%%%%%%%%%%%%%%%%%%%%%%%%%%

%\section{Jahn-Teller effect and electron correlation in metallic fulleride}
%\label{Sec:H}
\section{Electronic and vibronic model for the LUMO band in metallic fullerides}
\begin{figure}[bt]
\begin{center}
 \includegraphics[bb=0 0 360 361, width=3.5cm]{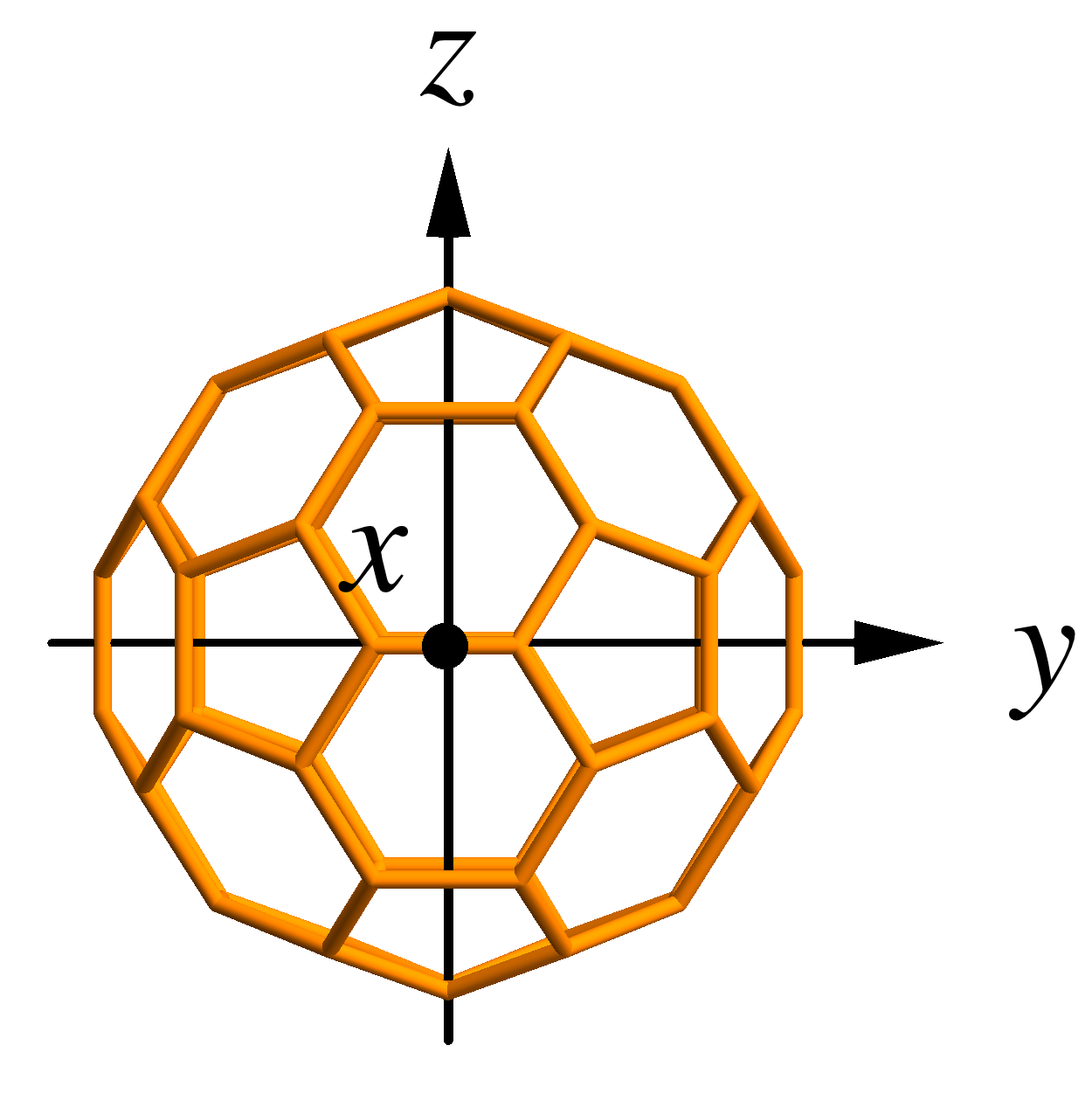}
\end{center}
\caption{(color online) Orientation of C$_{60}$ with respect to tetragonal axes of fcc lattice.}
\label{Fig:A1}
\end{figure}

The model Hamiltonian of $A_3$C$_{60}$ consists of 
transfer $\hat{H}_{\rm t}$, Jahn-Teller $\hat{H}_{\rm JT}$, 
and on-site bielectronic $\hat{H}_{\rm bi}$ parts \cite{Chibotaru1996a, Ceulemans1997a}:
\begin{eqnarray}
 \hat{H} &=& \hat{H}_{\rm t} + \hat{H}_{\rm JT} + \hat{H}_{\rm bi},
\label{Eq:H}
\\
 \hat{H}_{\rm t} &=& \sum_{\mathbf{m},\Delta \mathbf{m}}\sum_{\lambda \lambda' \sigma} 
                     t_{\lambda \lambda'}^{\Delta \mathbf{m}} 
                     \hat{c}_{\mathbf{m}+\Delta \mathbf{m} \lambda\sigma}^\dagger \hat{c}_{\mathbf{m} \lambda'\sigma}, 
\label{Eq:Ht}
\\ 
 \hat{H}_{\rm JT} &=& \sum_{\mathbf{m}} 
                    \hslash \omega
                    \left[
                     \sum_\gamma \frac{1}{2} \left(p_{\mathbf{m}\gamma}^2 + q_{\mathbf{m}\gamma}^2\right) 
                    \right.
\nonumber\\
                  &+&
                    \left.
                     g \sum_{\lambda \lambda'\sigma} \sum_\gamma 
                    G_{\lambda \lambda'}^\gamma
                    \hat{c}_{\mathbf{m}\lambda \sigma}^\dagger 
                    \hat{c}_{\mathbf{m}\lambda' \sigma}
                    q_{\mathbf{m}\gamma} 
                   \right],
%+
% \hat{H}_{\rm JT} &=& 
% \sum_{\mathbf{m}} \sum_\gamma 
% \frac{\hslash \omega}{2} \left(p_{\mathbf{m}\gamma}^2 + q_{\mathbf{m}\gamma}^2\right) 
%\nonumber\\
% &+& 
% \sum_\mathbf{m} \sum_\sigma g \left(
% \hat{c}_{\mathbf{m}x\sigma}^\dagger,
% \hat{c}_{\mathbf{m}y\sigma}^\dagger,
% \hat{c}_{\mathbf{m}z\sigma}^\dagger
% \right)
%\nonumber\\
% &\times&
% \begin{pmatrix}
%  \frac{1}{2}q_{\mathbf{m}\theta}-\frac{\sqrt{3}}{2}q_{\mathbf{m}\epsilon} & 
% -\frac{\sqrt{3}}{2} q_{\mathbf{m}\zeta} &
% -\frac{\sqrt{3}}{2} q_{\mathbf{m}\eta}\\
% -\frac{\sqrt{3}}{2} q_{\mathbf{m}\zeta} &
%  \frac{1}{2}q_{\mathbf{m}\theta}+\frac{\sqrt{3}}{2}q_{\mathbf{m}\epsilon} & 
% -\frac{\sqrt{3}}{2} q_{\mathbf{m}\xi} \\
% -\frac{\sqrt{3}}{2} q_{\mathbf{m}\xi} &
% -\frac{\sqrt{3}}{2} q_{\mathbf{m}\eta} &
% -q_{\mathbf{m}\theta}
% \end{pmatrix}
% \begin{pmatrix}
% \hat{c}_{\mathbf{m}x\sigma}\\
% \hat{c}_{\mathbf{m}y\sigma}\\
% \hat{c}_{\mathbf{m}z\sigma}
% \end{pmatrix},
%\nonumber\\
\label{Eq:HJT}
\\
 \hat{H}_{\rm bi} &=& \frac{1}{2}\sum_\mathbf{m}\sum_{\lambda \sigma} \left[ 
                      U_\parallel \hat{n}_{\mathbf{m} \lambda \sigma} \hat{n}_{\mathbf{m} \lambda -\sigma}  
                      \right.
\nonumber\\
                  &+& U_\perp \sum_{\lambda' (\ne \lambda)\sigma'} 
                      \hat{n}_{\mathbf{m} \lambda \sigma} \hat{n}_{\mathbf{m} \lambda' \sigma'}  
                  - J_{\rm H} \sum_{\lambda' (\ne \lambda)} \left(
                      \hat{n}_{\mathbf{m} \lambda \sigma} \hat{n}_{\mathbf{m} \lambda' \sigma}  
                      \right.
\nonumber\\
                  &-& \hat{c}_{\mathbf{m} \lambda \sigma}^\dagger \hat{c}_{\mathbf{m} \lambda' \sigma}
                      \hat{c}_{\mathbf{m} \lambda -\sigma}^\dagger \hat{c}_{\mathbf{m} \lambda' -\sigma}  
\nonumber\\
                  &-& \left.\left.
                      \hat{c}_{\mathbf{m} \lambda \sigma}^\dagger \hat{c}_{\mathbf{m} \lambda' \sigma}
                      \hat{c}_{\mathbf{m} \lambda' -\sigma}^\dagger \hat{c}_{\mathbf{m} \lambda -\sigma}  
                 \right)
                 \right],
\label{Eq:Hbi}
\end{eqnarray}
%\end{widetext}
where $\mathbf{m}$ is a site, $\Delta \mathbf{m}$ is a position relative to $\mathbf{m}$,
$\lambda, \lambda' = x,y,z$ are the components of the $t_{1u}$ LUMO (Fig. \ref{Fig:A1}),
$\sigma$ is the spin projection, 
$\gamma = \theta, \epsilon, \xi, \eta, \zeta$ is the component of the $h_g$ vibrational mode
($\gamma = 1,4,5,2,3$ in Ref. \onlinecite{OBrien1996a}, respectively),
$\hat{c}_{\mathbf{m}\lambda \sigma}^\dagger (\hat{c}_{\mathbf{m}\lambda \sigma})$
is the creation (annihilation) operator of an electron in orbital $\lambda \sigma$ at site $\mathbf{m}$, 
$\hat{n}_{\mathbf{m} \lambda \sigma} = \hat{c}_{\mathbf{m}\lambda \sigma}^\dagger \hat{c}_{\mathbf{m}\lambda \sigma}$, and 
$q_{\mathbf{m}\gamma}$ and $p_{\mathbf{m}\gamma}$ are the 
dimensionless normal coordinate and its conjugate momentum, \cite{Auerbach1994a} respectively. 
$G_{\lambda \lambda'}^{\gamma}$ is the Clebsch-Gordan coefficient, \cite{OBrien1996a}
$\omega$ and $g$ are the frequency and the dimensionless vibronic coupling constant for the effective $h_g$ mode, 
$t_{\lambda \lambda'}^{\Delta \mathbf{m}}$ is the transfer parameter, 
$U_\parallel$ and $U_\perp = U_\parallel - 2J_{\rm H}$ are the intra and interorbital Coulomb repulsion on the fullerene site, respectively, 
and $J_{\rm H}$ is the Hund's rule coupling.
%The JT part $\hat{H}_{\rm JT}$ is expressed by the linear JT Hamiltonian.
%First, we treat the ground state within the static JTE, and then include the effect of the JT dynamics.

%The transfer parameters are derived from the fit of the band structure of K$_3$C$_{60}$ 
The tight-binding Hamiltonian (\ref{Eq:Ht}) has been parametrized on the basis of 
density functional theory (DFT) band structure 
calculation of K$_3$C$_{60}$ and includes nearest neighbor and next-nearest-neighbor electron transfer
(see Appendix \ref{Sec:Ht} for details).
Although nearest neighbor tight-binding models were intensively used in the past to 
describe the LUMO bands of fullerides, \cite{Gelfand1992a, Chibotaru2000}
the inclusion of next-nearest-neighbor electron transfer is necessary for realistic
description of the band dispersion. \cite{nomura2012, Iwahara2013a}
The JT effect in fullerene anions involves eight vibrational $h_g$ modes, i.e.,
40 vibrational coordinates. \cite{Gunnarsson1997a} %, Ceulemans1997a}
The corresponding vibronic coupling parameters for C$_{60}^{3-}$ have been recently 
extracted from DFT calculation, \cite{Iwahara2013a} while the reliability of this approach was proven by a 
satisfactory reproduction of photoemission spectrum for C$_{60}^-$. \cite{Iwahara2010a}
Nevertheless, in the present calculations the use of a full multimode description of JTE on fullerene
sites seems to be impractical. 
For this reason, the eight-mode JT interaction on fullerenes has been replaced with an effective
single-mode one (\ref{Eq:HJT}).
Thus the two parameters, 
$\hslash\omega =$ 87.7 meV and $g =$ 1.07 were obtained via the reproduction of the 
JT stabilization energy and the energies of the 
lowest vibronic excitation of C$_{60}^{3-}$ ion. \cite{Iwahara2013a} 
%and Hund's rule coupling constant $J_{\rm H} =$ 44 meV are taken from Ref. \onlinecite{Iwahara2013a}.
In the model JT Hamiltonian (\ref{Eq:HJT}), the quadratic vibronic couplings 
are not included because, as we discussed in Ref. \onlinecite{Iwahara2013a},
they are weak in C$_{60}$ anions and do not give significant effect 
on the JT dynamics of C$_{60}^{n-}$ in cubic fullerides.
Finally, the Hund's rule coupling parameter, $J_{\rm H} = $ 44 meV, was also taken from the DFT
calculations. \cite{Iwahara2013a}
This is not the case of interelectron repulsion parameters of fullerene site, which are 
strongly renormalized by screening in fullerides. \cite{Gunnarsson1997a}
In the present work, the Coulomb repulsion $U$ is treated as a free parameter.
$U$ is defined here as the average repulsion of two electrons in C$_{60}^{3-}$ for a cubic (undistorted) 
LUMO band: 
\begin{eqnarray}
%U=(U_\parallel +4U_\perp )/5 = U_\perp +2J_{\rm H}/5. 
U=\frac{1}{5}\left(U_\parallel +4U_\perp\right) = U_\perp +\frac{2}{5}J_{\rm H}. 
\label{Eq:U}
\end{eqnarray}
%, where $U_\parallel$ and $U_\perp$ are intra- and interorbital Coulomb repulsions"
%The main parameter in Eq. (\ref{Eq:Hbi}), defining the strength of electron correlation, is the average
%Coulomb repulsion of two electrons at a fullerene site, $U$. \cite{U_av}

\begin{figure}[tb]
\begin{center}
\includegraphics[height=6cm, bb = 0 0 414 376]{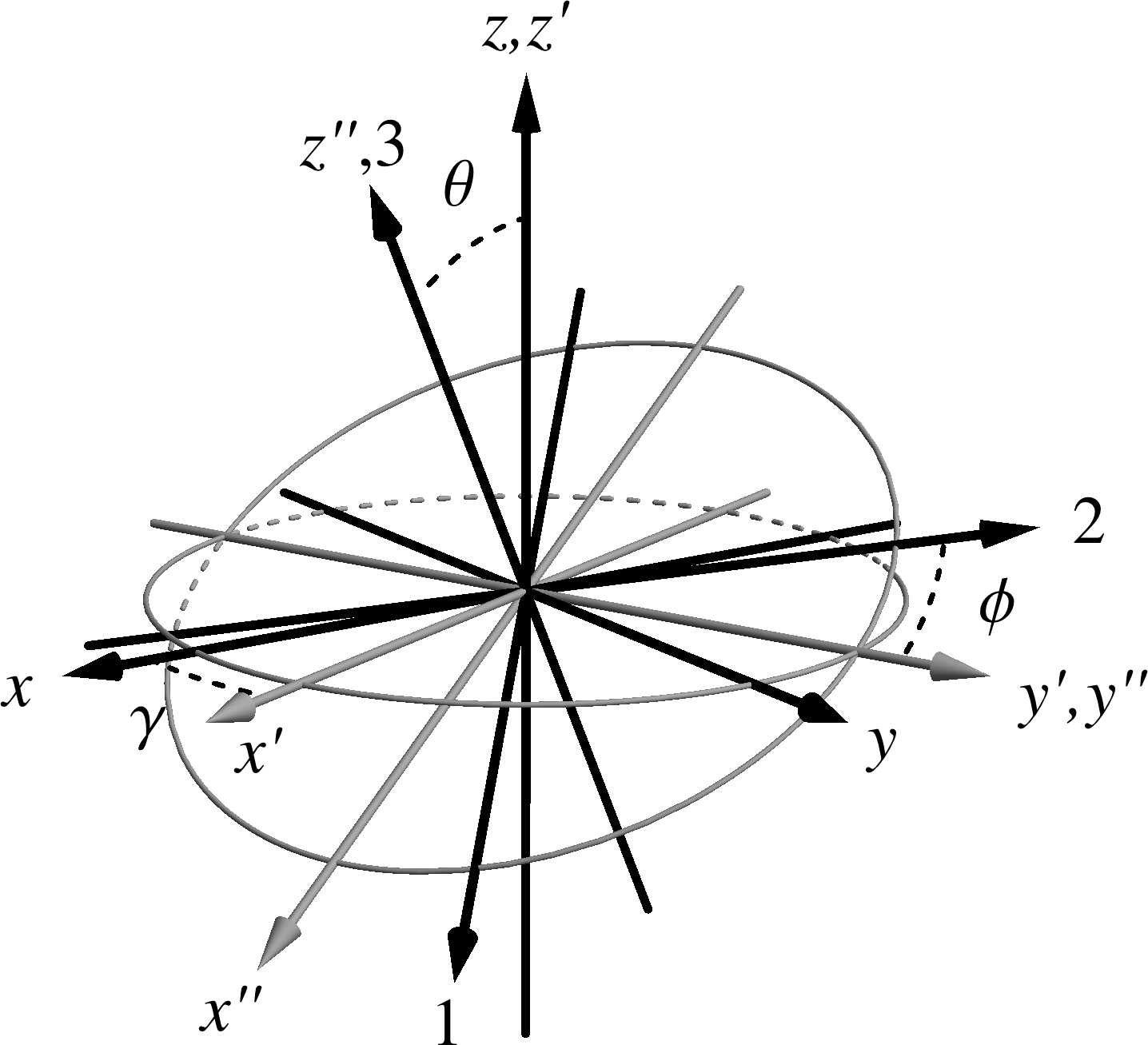}
\end{center}
\caption{(color online) The coordinate systems used to describe JT effect on a fullerene site. 
$x,y,z$ correspond to the orthorhombic LUMO orbitals (Fig. \ref{Fig:A1})
and $1,2,3$ to the adiabatic orbitals.
%$x',y',z'$ are obtained by the rotation of $x,y,z$ axes by $\gamma$ around the $z$ axis, 
%$x'',y'',z''$ are by rotation of $x',y',z'$ axes by $\theta$ around $y'$ axis, and 
%$1,2,3$ are rotation of $x'',y'',z''$ by $\phi$ around $z''$ axis.
$x',y',z'$ and $x'',y'',z''$ are intermediate coordinate systems \cite{Landau1977a}
appearing during the Euler rotation of the orbitals from $x,y,z$ to $1,2,3$, Eq. (\ref{Eq:Sllambda}).
}
\label{Fig:xyz}
\end{figure}

\subsection{Adiabatic orbitals}
\label{Sec:adiabatic}
The $h_g$ normal coordinates on each site $\mathbf{m}$, $q_{\mathbf{m}\gamma}$,
are expressed by polar coordinates, 
$(q_\mathbf{m}, \alpha_\mathbf{m}, \gamma_\mathbf{m}, \theta_\mathbf{m}, \phi_\mathbf{m})$.
\cite{OBrien1996a}
Introducing a unitary matrix,
\begin{eqnarray}
 S_{l\lambda}(\Omega_\mathbf{m}) &=& 
 \left[B_P(\gamma_\mathbf{m}) C_P(\theta_\mathbf{m}) D_P(\phi_\mathbf{m})\right]_{l\lambda},
\label{Eq:Sllambda}
\end{eqnarray}
we transform the electronic basis $(\lambda = x,y,z)$ into adiabatic basis $(l = 1,2,3)$ on each C$_{60}$ site
(Fig. \ref{Fig:xyz}),
\begin{eqnarray}
 \hat{c}_{\mathbf{m}l\sigma}^\dagger &=& \sum_{\lambda = x,y,z} 
 S_{l\lambda}(\Omega_{\mathbf{m}}) \hat{c}_{\mathbf{m}\lambda \sigma}^\dagger.
\label{Eq:cl+}
\end{eqnarray}
Here, $\Omega_\mathbf{m} = (\gamma_\mathbf{m}, \theta_\mathbf{m}, \phi_\mathbf{m})$, and 
$B_P$, $C_P$, $D_P$ are the Euler rotational matrices defined in Ref. \onlinecite{OBrien1996a}.
%Using the unitary operator, $\hat{S}_{\mathbf{m}}$, 
%whose matrix element is given by Eq. (\ref{Eq:Sllambda})
%the linear vibronic term $\hat{U}_{\rm LJT}$ of the JT Hamiltonian (\ref{Eq:HJT}) 
%is transformed into diagonal form:
By the transformation of the electronic basis, Eq. (\ref{Eq:cl+}), 
the linear vibronic term $\hat{U}_{\rm LJT}$ of the JT Hamiltonian (\ref{Eq:HJT}) 
becomes diagonal:
\begin{eqnarray}
 \hat{\tilde{U}}_{\rm LJT}
 &=&
 \hat{S}^\dagger \hat{U}_{\rm LJT} \hat{S}
\nonumber\\
 &=&
 \sum_\mathbf{m} \sum_\sigma \hslash \omega g q_\mathbf{m}
     \left[
      \cos\left(\alpha_\mathbf{m}+\frac{\pi}{3}\right)
      \hat{n}_{\mathbf{m}1\sigma}
     \right.
\nonumber\\
      &+&
      \left.
        \cos\left(\alpha_\mathbf{m}-\frac{\pi}{3}\right)
        \hat{n}_{\mathbf{m}2\sigma}
      - \cos \alpha_\mathbf{m} \hat{n}_{\mathbf{m}3\sigma}
     \right],
\label{Eq:ULJT}
\end{eqnarray}
where $\hat{S} = \prod_\mathbf{m}\hat{S}_\mathbf{m}$,
and $\hat{S}_{\mathbf{m}}$ is the unitary operator whose
matrix element is given by Eq. (\ref{Eq:Sllambda}).
Eq. (\ref{Eq:ULJT}) shows that the amplitude of the JT distortion is determined by 
radial coordinates $q_\mathbf{m}$ and $\alpha_\mathbf{m}$, and the direction of the JT distortion 
in the space of the five dimensional $h_g$ normal coordinates
is defined by Euler angular coordinates $\Omega_\mathbf{m}$ (Fig. \ref{Fig:xyz}).
In the described coordinate system, the elastic energy term in Eq. (\ref{Eq:HJT}) is written as
\begin{eqnarray}
 U_{\rm el} &=& \sum_\mathbf{m} \frac{\hslash \omega}{2}q_\mathbf{m}^2,
\label{Eq:Uel}
\end{eqnarray}
i.e., is invariant under the unitary transformation (\ref{Eq:Sllambda}).
%and does not change by the unitary transformation.
%The unitary transformation of the kinetic energy term is discussed in Sec. \ref{Sec:DJT}.
On the other hand, the kinetic energy term changes, which is discussed in Sec. \ref{Sec:vibronic}.

Under the transformation of the electronic basis (\ref{Eq:cl+}),
the transfer Hamiltonian (\ref{Eq:Ht}) becomes
\begin{eqnarray}
 \hat{\tilde{H}}_{\rm t} = 
 \hat{S}^\dagger \hat{H}_{\rm t} \hat{S} 
 = \sum_{\mathbf{m}, \mathbf{m}'} \sum_{ll'\sigma} t^{\mathbf{m}\mathbf{m}'}_{ll'}
     \hat{c}_{\mathbf{m}l\sigma}^\dagger \hat{c}_{\mathbf{m}'l'\sigma},
\label{Eq:Ht_adiabatic}
\end{eqnarray}
where $t_{ll'}^{\mathbf{m}\mathbf{m}'}$ is 
\begin{eqnarray}
 t^{\mathbf{m}\mathbf{m}'}_{ll'}(\Omega_{\mathbf{m}}, \Omega_{\mathbf{m}'}) &=&
 \sum_{\lambda \lambda'=x,y,z} S_{\lambda l}(\Omega_\mathbf{m}) 
 t_{\lambda \lambda'}^{\mathbf{m}-\mathbf{m}'} 
 S_{\lambda'l'}(\Omega_{\mathbf{m}'}).
\nonumber\\
\label{Eq:t_adiabatic}
\end{eqnarray}
We note also that $\hat{H}_{\rm bi}$ is invariant under the unitary transformation (\ref{Eq:Sllambda})
due to the isomorphism of $t_{1u}^n$ LUMO shell of C$_{60}^{n-}$ to the atomic $p^n$ shell.

For any Euler angles, $\Omega_\mathbf{m}$, the JT potential term, 
Eqs. (\ref{Eq:ULJT}) and (\ref{Eq:Uel}), and the bielectronic term has the same form. 
Therefore, the adiabatic potential energy surface of an isolated C$_{60}^{n-}$ 
has continuous minima (trough) \cite{Auerbach1994a, OBrien1996a}
even in the presence of the term splitting.
%as long as the term splitting is weak. 
In the case of C$_{60}^{3-}$, the potential surface has three dimensional (3D) trough at
\begin{eqnarray}
q = \sqrt{3}g \sqrt{1-\left(\frac{J_{\rm H}/\hslash \omega}{3g^2}\right)^2}
\label{Eq:qmin}
\end{eqnarray} 
and $\alpha=\pi/2$.
Substituting $g$, $\omega$, and $J_{\rm H}$ above into Eq. (\ref{Eq:qmin}), 
the amplitude of the JT distortion is $q = 0.989 \times \sqrt{3}g$,
indicating that the effect of the Hund's rule coupling on the JT potential surface of C$_{60}^{3-}$ is small. 
%of C$_{60}^{3-}$ ion is 

\begin{figure}[bt]
\begin{center}
 \includegraphics[bb=0 0 360 319, width=6.4cm]{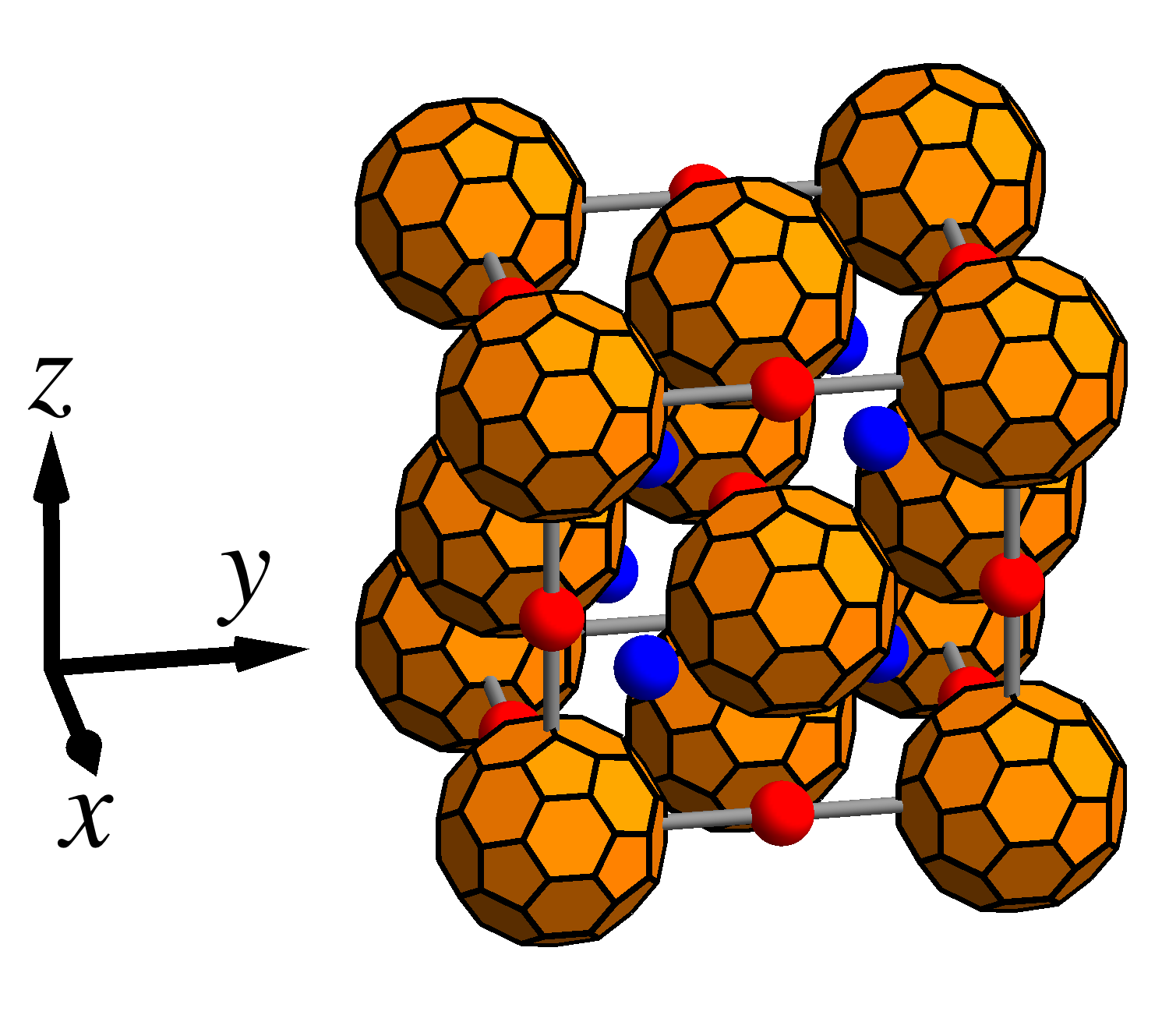}
\end{center}
\caption{
(color online) 
Ordered fcc K$_3$C$_{60}$. 
The orange ball is C$_{60}$, 
and the red and blue spheres are K atoms in octahedral and tetrahedral interstices.
}
\label{Fig:fccK3C60}
\end{figure}

\section{Gutzwiller approach to Static Jahn-Teller systems}
\label{Sec:G+SJT}
\subsection{Self-consistent Gutzwiller approach for the LUMO bands}
The merohedral disorder in the K$_3$C$_{60}$ lattice and the orientation of the JT distortions on the fullerene sites do not have important effect on the band energy. \cite{Ceulemans1997a,Chibotaru2000} 
The change in Hartree-Fock energy per C$_{60}$ site due to the disorders is only 14 meV,
\cite{Ceulemans1997a} which is smaller than the JT energy of C$_{60}^{3-}$ by one order of magnitude.
The variation will be further reduced by the electron correlation
as is discussed in Sec. \ref{Sec:vibronic}.
Therefore, for the sake of simplicity, we further consider a K$_3$C$_{60}$ in an ordered fcc lattice
(Fig. \ref{Fig:fccK3C60}).
%In the case of static JT distortions we consider equal JT distortions 
As a possible scenario of static JT effect we consider equal JT distortions 
on fullerene sites of the following form:
\begin{eqnarray}
(q_\mathbf{m}, \alpha_\mathbf{m}, \gamma_\mathbf{m}, \theta_\mathbf{m}, \phi_\mathbf{m}) 
 = (q, \pi/2, 0, 0, 0),
% = (-\sqrt{3}q/2,q/2,0,0,0),
\label{Eq:q}
\end{eqnarray}
or in conventional coordinates:
\begin{eqnarray}
(q_{\mathbf{m}\theta}, q_{\mathbf{m}\epsilon}, q_{\mathbf{m}\xi}, q_{\mathbf{m}\eta}, q_{\mathbf{m}\zeta}) 
= (0,\sqrt{3}q,0,0,0),
\end{eqnarray}
which do not remove translational symmetry of the lattice. 
Here, $q$ is a variable. The direction of 
the JT distortion (\ref{Eq:q}) corresponds to the one which gives 
the maximal static JT stabilization in the case of isolated C$_{60}^{3-}$ ion. 
\cite{Auerbach1994a,OBrien1996a}
Under the distortion (\ref{Eq:q}), 
the adiabatic orbitals $l=1,2,3$ correspond to $x,y,z$, respectively (Fig. \ref{Fig:xyz}), and 
the linear JT term (\ref{Eq:ULJT}) reduces to
\begin{eqnarray}
 \hat{\tilde{U}}_{\rm LJT} &=&
 \sum_\mathbf{m} \sum_\sigma -\frac{\sqrt{3}}{2} \hslash \omega g q
     \left(
      \hat{n}_{\mathbf{m}x\sigma} - \hat{n}_{\mathbf{m}y\sigma}
     \right).
\end{eqnarray}
This expression shows that the $z$ orbital level remains unchanged while the $x$ and the $y$ levels are 
stabilized and destabilized, % by $-\sqrt{3}\hslash \omega gqn_x/2$ and $\sqrt{3}\hslash \omega gqn_y/2$, 
respectively. %due to JT distortion (\ref{Eq:q}).
\cite{Auerbach1994a,OBrien1996a}

The Gutzwiller wave function, $|\Psi_{\rm G}\rangle$, is expressed as
\begin{eqnarray}
 |\Psi_{\rm G}\rangle &=& \hat{P}_{\rm G} |\Phi_{\rm S}\rangle,
\label{Eq:gwf}
\end{eqnarray}
where $|\Phi_{\rm S}\rangle$ is a Slater determinant, and $\hat{P}_{\rm G}$ is a Gutzwiller projector.
The Slater determinant is written as follows:
\begin{eqnarray}
 |\Phi_{\rm S}\rangle &=& \prod_{p\mathbf{k}\sigma}^{\rm occ} \hat{a}_{p \mathbf{k} \sigma}^\dagger |0\rangle, 
\label{Eq:slater}
\\
 \hat{a}_{p \mathbf{k} \sigma}^\dagger &=& 
 \sum_{\mathbf{m} \lambda} \frac{e^{i\mathbf{k}\cdot \mathbf{m}}}{\sqrt{N}} u_{\lambda, p \mathbf{k}}
 \hat{c}_{\mathbf{m} \lambda \sigma}^\dagger,
\label{Eq:a+ordered}
\end{eqnarray}
where $p$ is a band index, %$\mathbf{k}$ is a $k$-point, 
$N$ is the number of sites in the system, 
%$\lambda (= x,y,z)$ is a component of the $t_{1u}$ LUMO orbitals of C$_{60}$ sites, 
and $u_{\lambda, p \mathbf{k}}$ is a variational orbital coefficient.
We note that the band described by $|\Phi_{\rm S}\rangle$ is not constrained to obey the cubic symmetry.
%The orbital coefficients $\{u_{\lambda, p\mathbf{k}}\}$ are treated as variational parameters.
In order to include properly the effect of JT distortions and of electron correlation, 
%the Gutzwiller projector $\hat{P}_{\rm G}$ has to have different variational parameters for different orbitals:
the variational parameters ($A$) in $\hat{P}_{\rm G}$ have to be orbital-specific:
\begin{eqnarray}
 \hat{P}_{\rm G} &=& \prod_\mathbf{m} \exp\left(-\frac{1}{2}\sum_{\lambda \sigma \ne \lambda' \sigma'} 
                     A_{\lambda \lambda'} \hat{n}_{\mathbf{m} \lambda \sigma} \hat{n}_{\mathbf{m} \lambda' \sigma'} \right),
\label{Eq:pg}
\end{eqnarray}
where $A_{\lambda \lambda'}$ are real and symmetric with respect to interchange of indices.
%$\lambda, \lambda'$ denote natural orbitals on site $\mathbf{m}$.
Therefore, the projector (\ref{Eq:pg}) is described by six independent Gutzwiller parameters 
($A_{\lambda \lambda'}$) instead of a single parameter used in conventional Gutzwiller wave function.
\cite{Gutzwiller1963a, Vollhardt1984a}
In a general case, $\lambda, \lambda'$ denote natural orbitals on the site $\mathbf{m}$.
For the chosen JT distortions (\ref{Eq:q}), preserving the orthorhombic site-symmetry, 
these natural orbitals coincide with the orthorhombic $x,y,z$ $t_{1u}$ LUMO orbitals.
Due to equal distortions (\ref{Eq:q}) on all fullerene sites, 
$A_{\lambda \lambda'}$ are independent from the index $\mathbf{m}$.
%The relatively simple form of Gutzwiller projector (\ref{Eq:pg}), involving only diagonal elements of the density matrix operators on sites, is due to the chosen JT distortions (\ref{Eq:q}) preserving the orthorhombic symmetry of each fullerene site.

The calculations of expectation values with $|\Psi_{\rm G}\rangle$ 
have been done within the Gutzwiller's approximation. \cite{Vollhardt1984a, Ogawa1975a}
Within this approximation, the energy per site,
\begin{eqnarray}
E_{\rm g} &=& \frac{1}{N}
 \frac{\langle \Psi_{\rm G}| \hat{H} |\Psi_{\rm G} \rangle}{\langle \Psi_{\rm G}|\Psi_{\rm G}\rangle},
\label{Eq:Eg}
%\\
% &=& E_{\rm t} + U_{\rm el} + U_{\rm LJT} + E_{\rm bi},
\end{eqnarray}
consists of the band energy,
\begin{eqnarray}
 E_{\rm t} &=& \sum_{\lambda \lambda'\sigma} q_{\lambda \lambda'} \tau_{\lambda \lambda'},
\label{Eq:Et}
\end{eqnarray}
the elastic energy (\ref{Eq:Uel}), the linear vibronic energy, 
\begin{eqnarray}
 U_{\rm LJT} &=& \sum_{\sigma} -\frac{\sqrt{3}}{2}\hslash \omega g q 
 \left(n_x - n_y\right), 
\label{Eq:ULJT_G}
\end{eqnarray}
and the bielectronic energy $E_{\rm bi}$.
Here, $q_{\lambda \lambda'}$ is the Gutzwiller's reduction factor, 
$\tau_{\lambda \lambda'}$ is 
\begin{eqnarray}
 \tau_{\lambda \lambda'}
 &=&
 \frac{1}{N} \sum_\mathbf{k} t_{\lambda \lambda'}^\mathbf{k} \rho_{\lambda \lambda'}^\mathbf{k},
\end{eqnarray}
where $t_{\lambda \lambda'}^\mathbf{k}$ is the Fourier transform of 
$t_{\lambda \lambda'}^{\Delta \mathbf{m}}$,
\begin{eqnarray}
 t_{\lambda\lambda'}^\mathbf{k} &=& 
 \sum_{\Delta \mathbf{m}} e^{-i\mathbf{k}\cdot \Delta \mathbf{m}} t_{\lambda \lambda'}^{\Delta \mathbf{m}},
\end{eqnarray}
$\rho_{\lambda \lambda'}^\mathbf{k}$ is the density matrix at a $\mathbf{k}$ point,
\begin{eqnarray}
 \rho_{\lambda \lambda'}^\mathbf{k} &=& 
 \sum_p^{\rm occ} u_{\lambda, p\mathbf{k}}^* u_{\lambda', p\mathbf{k}},
\end{eqnarray}
and $n_\lambda$ $(\lambda = x,y,z)$ is the occupation number, 
\begin{eqnarray}
 n_\lambda &=& 
 \frac{\langle \Psi_{\rm G}| \hat{n}_{\mathbf{m}\lambda \sigma}|\Psi_{\rm G} \rangle}
 {\langle \Psi_{\rm G}|\Psi_{\rm G}\rangle}.
% \frac{1}{N}\sum_\mathbf{m}
% \frac{\langle \Psi_{\rm G}| \hat{n}_{\mathbf{m}\lambda \sigma}|\Psi_{\rm G} \rangle}
% {\langle \Psi_{\rm G}|\Psi_{\rm G}\rangle}.
\label{Eq:n1}
\end{eqnarray}
The explicit forms of the occupation number $n_\lambda$, the Gutzwiller's reduction factor
$q_{\lambda \lambda'}$, and the bielectronic energy $E_{\rm bi}$ are given in Appendix \ref{Sec:q}.
The Gutzwiller projector does not influence the on-site density matrix, 
hence, Eq. (\ref{Eq:n1}) corresponds to 
\begin{eqnarray}
 n_\lambda 
 &=& 
 \langle \Phi_{\rm S}| \hat{n}_{\mathbf{m}\lambda \sigma}|\Phi_{\rm S} \rangle
 = 
 \frac{1}{N}\sum_\mathbf{k} \rho_{\lambda \lambda}^\mathbf{k}.
\label{Eq:n2}
\end{eqnarray}
Hereafter, we use the form (\ref{Eq:n2}) for $n_\lambda$.

The ground state for different amplitudes of JT distortion
$q$ is obtained by minimizing the energy per site (\ref{Eq:Eg})
with respect to $\{u_{\lambda, p\mathbf{k}}\}$ and $\{A_{\lambda \lambda'}\}$, 
which is performed in two steps. 
The first one is the variational calculation of $\tilde{E}_{\rm g}$ with respect to 
$\{u_{\lambda, p\mathbf{k}}\}$ for fixed $\{A_{\lambda \lambda'}\}$.
%and occupation numbers of fullerene orbitals $\{n_{\lambda}\}$. 
The resulting self-consistent equations in the case of static JT effect are obtained in the form:
\begin{eqnarray}
\sum_{\lambda'} h_{\lambda \lambda'}^\mathbf{k} u_{\lambda', p \mathbf{k}}
 = \epsilon_{p \mathbf{k}} u_{\lambda, p \mathbf{k}},
\label{Eq:step1}
\end{eqnarray}
where the one-particle Hamiltonian is 
\begin{eqnarray}
h_{\lambda \lambda'}^\mathbf{k} 
&=& q_{\lambda \lambda'} t_{\lambda \lambda'}^\mathbf{k} 
+ \delta_{\lambda \lambda'} 
  \left[
  \sum_{\kappa \kappa'}
  \frac{\partial q_{\kappa \kappa'}}{\partial n_{\lambda}}
  \tau_{\kappa \kappa'}
 + \frac{1}{2} \frac{\partial E_{\rm bi}}{\partial {n}_{\lambda}}
\right.
\nonumber\\
&-& 
\left.
\frac{\sqrt{3}}{2}\hslash \omega g q\left(\delta_{\lambda x} - \delta_{\lambda y}\right)\right],
\label{Eq:hk}
\end{eqnarray}
and $\epsilon_{p \mathbf{k}}$ is the Gutzwiller's orbital energy. 
Using the solutions of Eq. (\ref{Eq:step1}), $\{u_{\lambda, p \mathbf{k}}\}$, 
the occupation numbers $\{n_{\lambda}\}$ are recalculated via Eq. (\ref{Eq:n2}).
The chemical potential is found by consecutive population of Gutzwiller's orbitals
following the aufbau principle. 
%For the self-consistency, the occupation numbers obtained by Eqs. (\ref{Eq:n1}) and 
%(\ref{Eq:n2}) should correspond to each other. 
%In the derivation of Eq. (\ref{Eq:hk}), Eq. (\ref{Eq:n2}) is used.
% in the order of increase of their energy $\epsilon_{p \mathbf{k}}$. 
The second step is the minimization of $\tilde{E}_{\rm g}$ 
with respect to $\{A_{\lambda \lambda'}\}$ for fixed 
$\{u_{\lambda,p\mathbf{k}}\}$ and $\{n_\lambda\}$, 
%
%\begin{eqnarray}
% E_{\rm g} &=& \min_{\{A_{\lambda \lambda'}\}} \left[\tilde{E}_{\rm g}\right].
%\label{Eq:step2}
%\end{eqnarray} 
%
\begin{eqnarray}
 \frac{\partial \tilde{E}_{\rm g}}{\partial A_{\lambda \lambda'}} 
 &=&
 \sum_{\sigma \kappa \kappa'} \frac{\partial q_{\kappa \kappa'}}{\partial A_{\lambda \lambda'}} 
 \tau_{\kappa \kappa'}
 + \frac{\partial E_{\rm bi}}{\partial A_{\lambda \lambda'}}
 = 0,
\label{Eq:step2}
\end{eqnarray} 
using the numerical algorithm proposed in Ref. \onlinecite{Bunemann2012a}.
%The minimization (\ref{Eq:step2}) is done under the constraint that Eq. (\ref{Eq:n1}) 
%corresponds to $\{n_\lambda\}$.
The two minimizations, (\ref{Eq:step1}) and (\ref{Eq:step2}), are repeated iteratively until 
variations in the occupation numbers and the ground state energy become smaller than thresholds.
%

%%%%%%%%%%%%%%%%%%%%%%%%%%%%%%%%%%%%%%%%%%%%%%%%%%%%%%%%%%%%%
%
% Static JT
%
%%%%%%%%%%%%%%%%%%%%%%%%%%%%%%%%%%%%%%%%%%%%%%%%%%%%%%%%%%%%%
\begin{figure}
\begin{center}
\begin{tabular}{l}
(a)
\\
 \includegraphics[bb= 0 0 3968 2176, width=8.6cm]{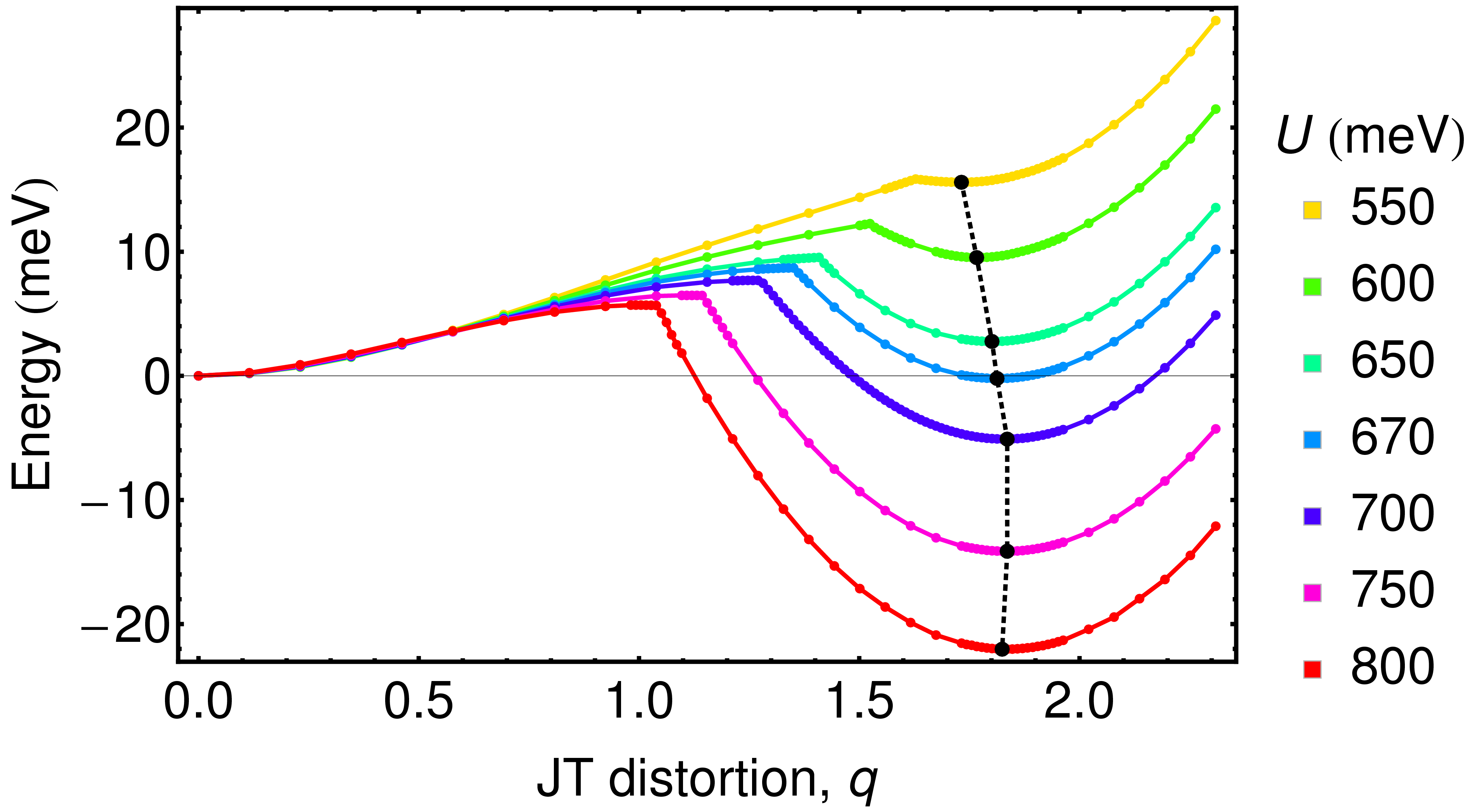}
\\
(b)
\\
 \includegraphics[bb = 0 0 3968 3176, width=8.6cm]{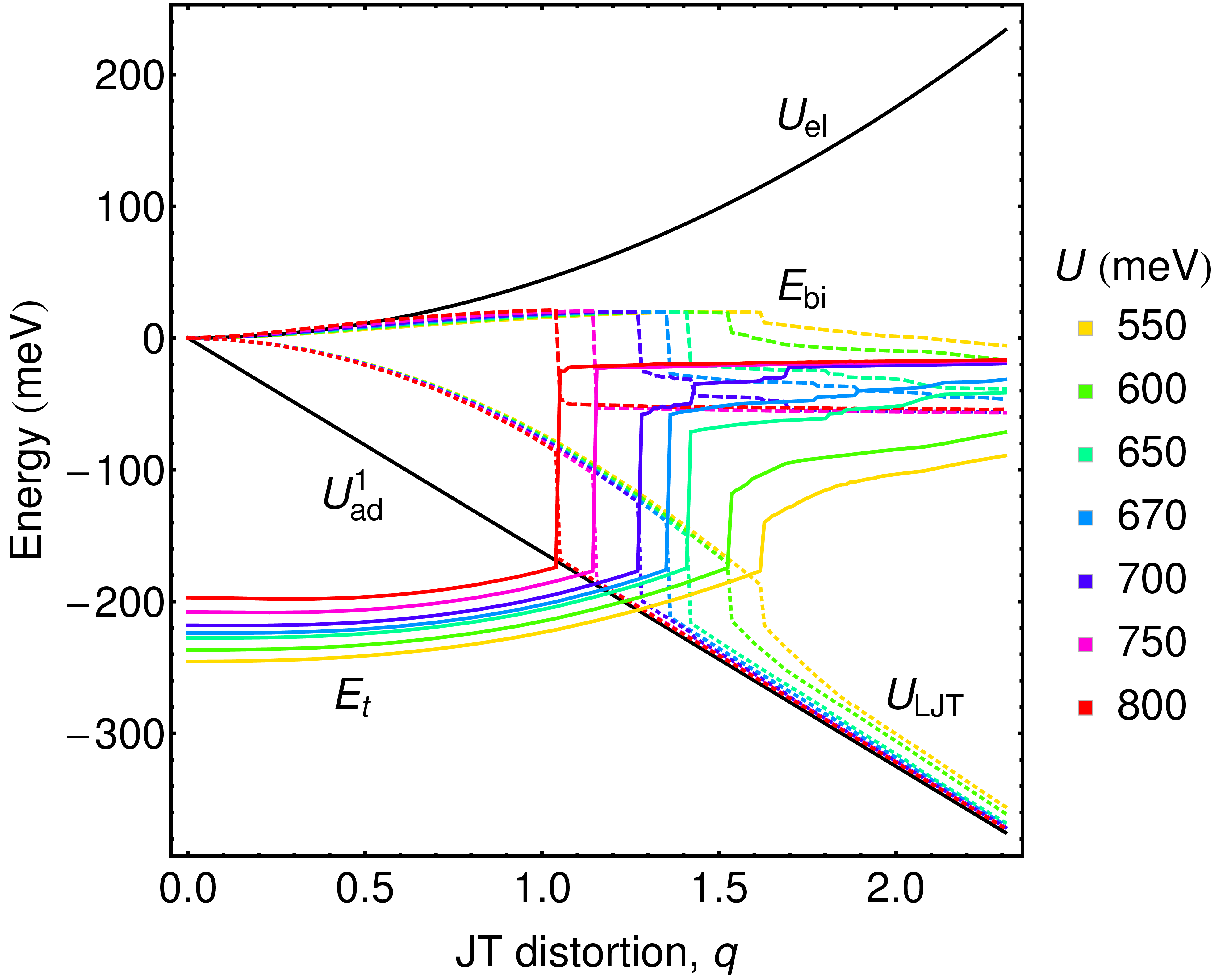}
\\
(c)
\\
 \includegraphics[bb = 0 0 3968 2208, width=8.6cm]{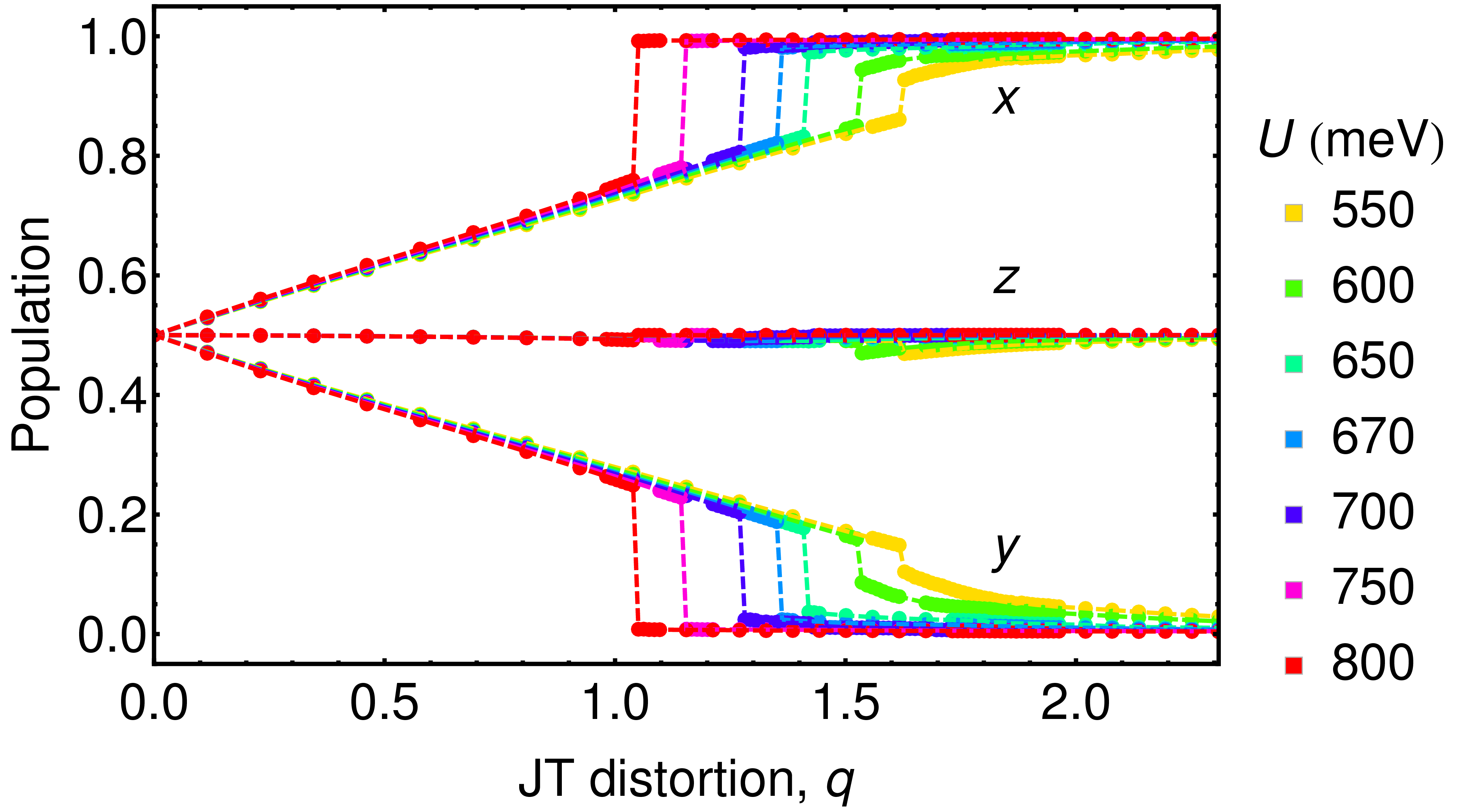}
\end{tabular}
\end{center}
\caption{
(color online) (a) Total energy $E_{\rm g}(q)$ of K$_3$C$_{60}$ as function of amplitude %($q$) 
of static JT distortion (\ref{Eq:q}) for several values of $U$ (\ref{Eq:U}). %$U_\perp$. 
The dashed line indicates the minimum of $E_{\rm g}$. % with JTE. 
(b) Energy components contributing to the total energy.
$U_{\rm LJT}$ is the linear JT energy (\ref{Eq:ULJT}) 
and $U_{\rm el}$ is the elastic energy (\ref{Eq:Uel}) in $\hat{H}_{\rm JT}$;
$E_{\rm t}$ is the band energy (\ref{Eq:Et}), % ($\hat{H}_{\rm t}$), 
and $E_{\rm bi}$ is the bielectronic energy. % ($\hat{H}_{\rm bi}$).
%The same components for one isolated C$_{60}^{3-}$ are shown by lines.
$U_{\rm ad}^{1}$ is the ground state adiabatic potential of an isolated C$_{60}^{3-}$. 
%For convenience all energy components are set to zero at $q$=0.
$E_{\rm g}$ and $E_{\rm bi}$ are set to zero at $q$=0.
The unit of $q$ is the amplitude of zero-vibration of $h_g$ mode. \cite{Bersuker1989a, Bersuker2006a}
(c) Occupation numbers of LUMO orbitals $n_\lambda$ as function of $q$.
$x$, $y$, and $z$ are orbital components.
}
\label{Fig1}
\end{figure}

%\section{Ground state with static Jahn-Teller effect}
\subsection{Static Jahn-Teller instability in K$_3$C$_{60}$}
\label{Sec:SJT}
%\subsection{Adiabatic potential energy surface}
The ground state energy $E_{\rm g}$ (\ref{Eq:Eg}) 
as a function of the JT distortion $q$ is plotted in Fig. \ref{Fig1}a.
Quite unexpectedly, the energy curve $E_{\rm g}(q)$ has two minima,
one at the undistorted configuration $q = 0$ %, 
and the other at $q \approx \sqrt{3}g$ $(= 1.85)$ which corresponds approximately to the 
amplitude of JT distortion in an isolated C$_{60}^{3-}$ ion.
For $U$ smaller than the critical value $U_c =$ 670 meV %, corresponding to average electron repulsion parameter $U$= 668 meV \cite{U_av}, 
the static JT distortion is quenched ($q = 0$). %\cite{U_av}
The minimum corresponding to JT-distorted sites lowers with the increase of $U$,
at $U = U_{c}$ the values of the two minima %in $E_{\rm g}$ correspond to each other, 
equalize and for $U > U_c$ %$U_\perp > U_{c\perp}$
the JT distortion achieves its equilibrium value matching approximately 
the distortion in an isolated C$_{60}^{3-}$.
%ion 
%and while $n_x$ and $n_z$ take saturated values. 

%%% comparison
The static JT effect has been investigated in $A_4$C$_{60}$ within the local density approximation (LDA) of 
DFT, for which completely quenched JT distortions have been found. \cite{Capone2000a}
Since the static JT effect in C$_{60}^{4-}$ is stronger than in C$_{60}^{3-}$ anion, \cite{Auerbach1994a}
it was concluded that the JT distortions in $A_3$C$_{60}$ are also quenched.
However it was recently revealed that the LDA calculations underestimate the JT stabilization energy of %fullerene monoanion 
C$_{60}^-$ by ca 30 \%. \cite{Iwahara2010a}
On the other hand the broken-symmetry Hartree-Fock (HF) calculations %, although do
predict smaller $U_c$ for static JT instability than the present calculations
and orbital disproportionation of the intrasite charge density in fullerides. 
\cite{Chibotaru1996a, Ceulemans1997a} 
%predict static JT instability and orbital disproportionation of the intrasite charge density in fullerides \cite{Chibotaru1996a, Ceulemans1997a} at smaller $U$ than the present calculations. 
%suffer from artifactual errors.
%obviously 
The reason of this discrepancy is that the broken-symmetry HF calculations 
exaggerate the %band character of electronic structure in $A_3$C$_{60}$.
tendency towards the stabilization of low-symmetry electronic phases.
Thus the splitting of the LUMO band is overestimated and is mainly contributed by the interelectron repulsion, 
\cite{Ceulemans1997a}
suggesting that an approach based on a single Slater determinant is not flexible enough to include 
properly the effects of electron correlation in orbitally degenerate bands 
(see Sec. \ref{Sec:disproportionation} for detailed discussion).
%Hence an appropriate description of these effects should be based on a multideterminantal approach.
%Compared to broken-symmetry HF calculations \cite{Ceulemans1997a}, 
%the JT distortions arise in the present calculations at larger $U$.

The structure of $E_{\rm g}(q)$ is mainly determined by the $q$-dependent contributions, the band energy 
$E_{\rm t}$ (\ref{Eq:Et}) and the JT potential $U_{\rm LJT}$ (\ref{Eq:ULJT_G}) 
which are found in competition (Fig. \ref{Fig1}b).
The contribution of the band energy term is the largest
when the orbitals are hybridized and equally populated, which takes place in the 
weak correlation limit ($U\rightarrow 0$).
On the other hand, 
%the linear JT term lifts the degeneracy of the $t_{1u}$ LUMO on each fullerene site.
the contribution from the JT term is the largest %when 
%in the case of full orbital disproportionation, $(n_x, n_y, n_z) = (1,1/2,0)$.
when the disproportionation of the electronic charge among the LUMO subbands,
accompanying the JT distortions, is full, $(n_x, n_y, n_z) = (1,0,1/2)$.
At the same time, the splitting of the orbital levels prevents the hybridization and vice versa.
As Fig. \ref{Fig1}a suggests, 
the ground state energy $E_{\rm g}(q)$ consists of 
two potential energy surfaces which cross at $q = q_c$.
%Figure \ref{Fig1}b shows that for small distortion, the 
For small distortions $q < q_c$, the band energy exceeds the JT energy 
and the latter is quenched compared to the case of an isolated C$_{60}^{3-}$ ($U_{\rm ad}^1$),
while for $q > q_c$ the JT energy takes over.
Because of the hybridization, the occupation numbers $n_\lambda$ are fractional (Fig. \ref{Fig1}c)
and the linear JT energy $U_{\rm LJT}$ %= -\sqrt{3}\hslash \omega gq(n_x-n_y)$ 
is not proportional to $q$ unlike the isolated C$_{60}^{3-}$ molecule. % ($U_{\rm ad}^1$).
The band energy is reduced by the intrasite Coulomb repulsion  
by quenching charge fluctuations on C$_{60}$'s.
Since the band energy in Rb$_3$C$_{60}$ and Cs$_3$C$_{60}$ is smaller than in K$_3$C$_{60}$, 
while $U$ is larger, \cite{nomura2012} 
the static JT instability is favored even more %than in the latter.
in these fullerides.

%\section{Ground state with dynamical Jahn-Teller effect}
\section{Gutzwiller approach to dynamical Jahn-Teller systems}
\label{Sec:DJT}
\subsection{Dynamical Jahn-Teller contribution}
\label{Sec:EDJT}
Another important ingredient is the energy gain arising from the dynamical delocalization of JT distortions 
%in the trough of the lowest potential energy surface of 
at each C$_{60}^{3-}$ anion. 
%Recently, {\it ab initio} calculations have shown that 
The energy gain %due to JT dynamics 
in isolated C$_{60}^{3-}$ amounts to ca 90 meV which is more than a half of the static JT stabilization 
(ca 150 meV) in this anion. \cite{Iwahara2013a}
%Therefore, the dynamical contribution to the JT stabilization is not negligible and should be taken into account along with other interactions in order to establish the true ground state of metallic fullerides.
%The dynamical JT effect, when present, leads to additional energy gain due to the delocalization of JT deformations in the 3D trough of the lowest adiabatic energy surface. 
To assess this energy gain in fullerides, one should take into account that 
the JT effect on C$_{60}^{3-}$ sites in fullerides is different from the case of isolated fullerene anions.
The main difference is that the LUMO orbitals 
on the fullerene sites do not have the same populations as in an isolated C$_{60}^{3-}$ (Fig. \ref{Fig1}c),
which leads, in particular, to lower values of the amplitude of dynamical JT deformation in fullerides. 
Only in the case of full orbital disproportionation 
(Sec. \ref{Sec:SJT})
the deformation achieves the equilibrium value in a free ion ($q=1.85$) 
and the corresponding energy gain owing to dynamical JT effect is maximal. 
One should stress that in the case of dynamic JT effect the adiabatic orbitals $l = 1,2,3$ on fullerene 
sites are not fixed electronic orbitals $\lambda = x, y, z$, considered in the 
previous section but are their linear combinations with $\Omega_{\mathbf{m}}$-dependent coefficients,
Eq. (\ref{Eq:cl+}). \cite{OBrien1996a}
%(Sec. \ref{Sec:adiabatic}). \cite{OBrien1996a}
To simulate the dependence of dynamical JT effect on the extent of orbital disproportionation, 
we introduce the effective vibronic coupling constant, 
\begin{eqnarray}
g_{\rm eff} = g(n_1 - n_2),
\label{Eq:geff}
\end{eqnarray}
which varies from 0 to $g$ when the orbital disproportionation $n_1-n_2$ varies from the minimal value
(0) to the maximal value (1). 
Note that $n_3$ has an unchanged value $1/2$.

\begin{figure}[tb]
\begin{center}
\includegraphics[bb = 0 0 3320 2160, width=8.6cm]{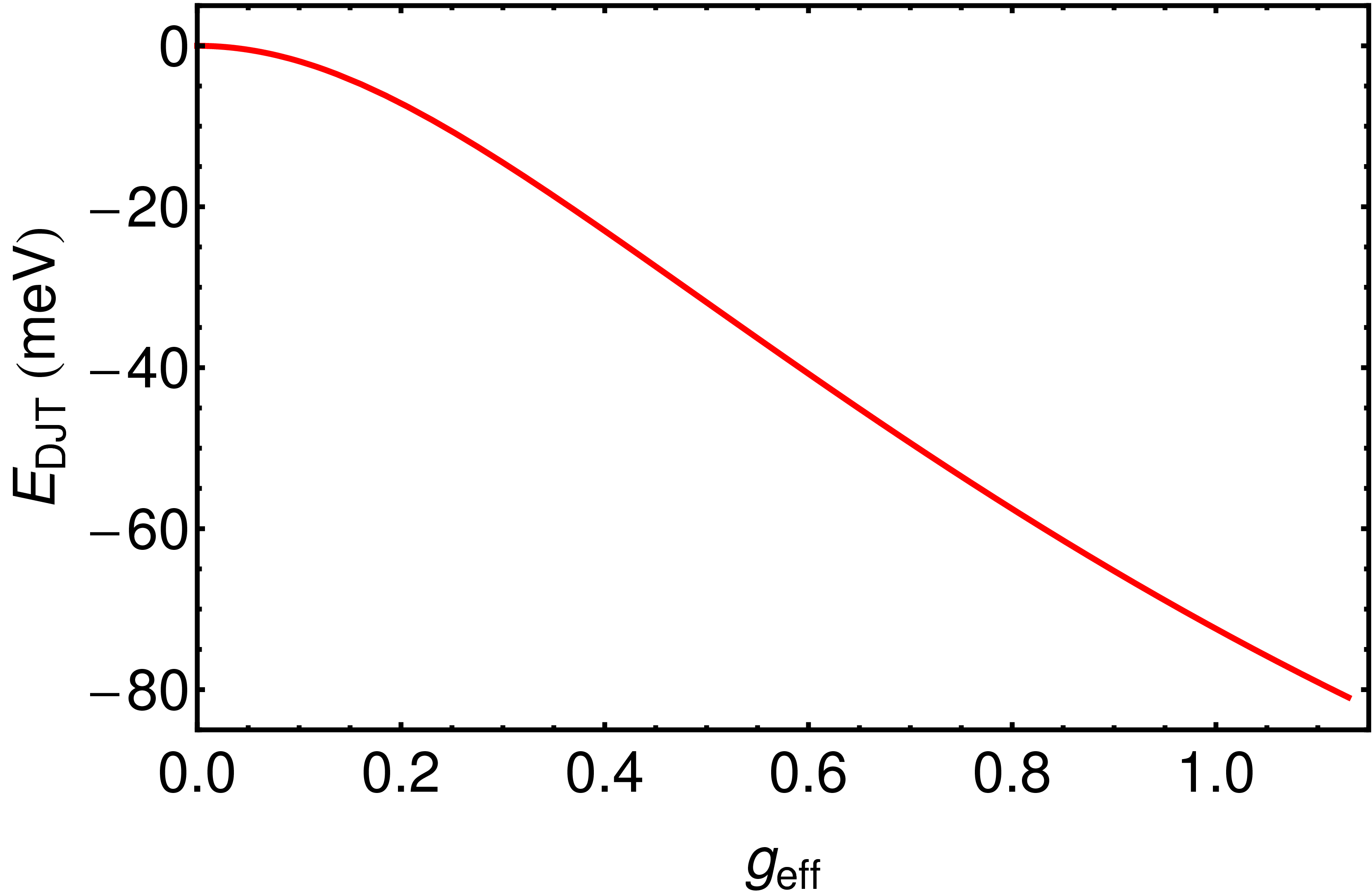}
\end{center}
\caption{
(color online)
The gain of JT stabilization energy (meV)
due to dynamical JTE in function of $g_{\rm eff}$.
%$E_{\rm DJT}(g_{\rm eff})$.
%The ground vibronic level is set to zero at $g_{\rm eff} = 0$.
}
\label{Fig:DJT}
\end{figure}

Diagonalizing the JT Hamiltonian (\ref{Eq:HJT}) for different values of $g_{\rm eff}$,
and extracting the ground state energy at corresponding static JT distortion, 
$3\hslash \omega g_{\rm eff}^2/2$, 
together with the energy of zero-vibrations at distorted point, $5\hslash \omega/2$,
we obtain the dynamical contribution, $E_{\rm DJT}$, to the %energy of the 
ground vibronic level. 
Figure \ref{Fig:DJT} shows the dependence of this contribution on $g_{\rm eff}$ for the 
case of effective single-mode JT Hamiltonian of C$_{60}^{3-}$. \cite{Iwahara2013a}
Note that the existence of the energy gain due to dynamical delocalization of JT deformations 
does not guarantee by itself the development of dynamical JT effect on fullerene sites.
For the latter to take place, an additional condition are the small variations of the band energy 
under arbitrary JT distortion on C$_{60}^{3-}$ sites, which is investigated below.

\subsection{Form of the ground vibronic state}
\label{Sec:vibronic}
Previous {\it ab initio} investigations have shown that the low-lying vibronic states in 
an isolated C$_{60}^{3-}$ can be described satisfactorily within the adiabatic approximation. 
\cite{Iwahara2013a}
This approximation can be extended over the $A_3$C$_{60}$ crystal.
Following the molecular approach, \cite{Bersuker1989a}
first we perform the unitary transformation (\ref{Eq:cl+}) 
to diagonalize the linear vibronic term in Eq. (\ref{Eq:HJT})
\cite{Auerbach1994a, OBrien1996a}:
\begin{eqnarray}
\hat{\tilde{H}} &=& \hat{S}^\dagger \hat{H}\hat{S}
 = \hat{H}_{\rm rad} + \hat{H}_{\rm rot} + \hat{H}^{(1)}_{\rm el} + \hat{H}_{\rm bi},
\label{Eq:Htilde}
\\
 \hat{H}_{\rm rad} 
 &=&
 \sum_{\mathbf{m}}
  -\frac{\hslash \omega}{2}
 \left[
  q_\mathbf{m}^{-4}\frac{\partial}{\partial q_\mathbf{m}}\left(q_\mathbf{m}^4 \frac{\partial}{\partial q_\mathbf{m}}\right) 
 \right.
\nonumber\\
 &+& 
 \left.
  \frac{1}{q_\mathbf{m}^2 \sin 3\alpha_\mathbf{m}}\frac{\partial}{\partial \alpha_\mathbf{m}} 
   \left(\sin 3\alpha_\mathbf{m} \frac{\partial}{\partial \alpha_\mathbf{m}}\right)
 \right]
 + \frac{\hslash \omega}{2} q_\mathbf{m}^2,
\nonumber\\
 \hat{H}_{\rm rot} 
 &=&
 \sum_{\mathbf{m}}
 \frac{\hslash \omega}{8q_\mathbf{m}^2} 
 \left[
   \frac{(\hat{L}^{\rm nuc}_{\mathbf{m}1} + \hat{L}^{\rm el}_{\mathbf{m}1})^2}
   {\sin^2(\alpha_{\mathbf{m}}-2\pi/3)}
  + \frac{(\hat{L}^{\rm nuc}_{\mathbf{m}2} + \hat{L}^{\rm el}_{\mathbf{m}2})^2}
   {\sin^2(\alpha_{\mathbf{m}}+2\pi/3)}
 \right.
\nonumber\\
 &+& 
 \left.
   \frac{(\hat{L}^{\rm nuc}_{\mathbf{m}3} + \hat{L}^{\rm el}_{\mathbf{m}3})^2}
   {\sin^2\alpha_{\mathbf{m}}}
 \right].
%\\
% \hat{H}_{\rm el}^{(1)} 
% &=& 
% \hat{\tilde{U}}_{\rm LJT} + \hat{\tilde{H}}_{\rm t}. 
%\sum_\mathbf{m} \sum_\sigma \hslash \omega g q_\mathbf{m}
%     \left[
%      \cos\left(\alpha_\mathbf{m}+\frac{\pi}{3}\right)
%      \hat{n}_{\mathbf{m}1\sigma}
%     \right.
%\nonumber\\
%      &+&
%      \left.
%        \cos\left(\alpha_\mathbf{m}-\frac{\pi}{3}\right)
%        \hat{n}_{\mathbf{m}2\sigma}
%      - \cos \alpha_\mathbf{m} \hat{n}_{\mathbf{m}3\sigma}
%     \right]
%\nonumber\\
% &+& \sum_{\mathbf{m}, \mathbf{m}'} \sum_{ll'\sigma} t^{\mathbf{m}\mathbf{m}'}_{ll'}
%     \hat{c}_{\mathbf{m}l\sigma}^\dagger \hat{c}_{\mathbf{m}'l'\sigma}.
\end{eqnarray}
%Here, $(q_\mathbf{m}, \alpha_\mathbf{m}, \gamma_\mathbf{m}, \theta_\mathbf{m}, \phi_\mathbf{m})$
%are the polar coordinates of the $h_g$ normal coordinates, \cite{OBrien1996a}
%the unitary operator $\hat{S} = \prod_\mathbf{m} \hat{S}_\mathbf{m}$ transforms the 
%electronic basis $(\lambda = x, y, z)$ into adiabatic basis $(l=1,2,3)$ on each C$_{60}$ site,
%$t_{ll'}^{\mathbf{m}\mathbf{m}'}$ is 
%\begin{eqnarray}
% t^{\mathbf{m}\mathbf{m}'}_{ll'}(\Omega_{\mathbf{m}}, \Omega_{\mathbf{m}'}) &=&
% \sum_{\lambda \lambda'=x,y,z} S^{\mathbf{m}}_{\lambda l} t_{\lambda \lambda'}^{\mathbf{m}-\mathbf{m}'} 
% S^{\mathbf{m}'}_{\lambda'l'},
%\end{eqnarray}
%$\Omega_\mathbf{m} = (\gamma_\mathbf{m}, \theta_\mathbf{m}, \phi_\mathbf{m})$
%denotes the angular (trough) Euler coordinates of C$_{60}^{3-}$ site, \cite{OBrien1996a}
%$S_{\lambda l}^\mathbf{m} = S_{\lambda l}(\{\Omega_\mathbf{m}\})$ 
%is a real matrix element of unitary operator $\hat{S}_\mathbf{m}$,
Here, 
$\hat{H}^{(1)}_{\rm el}$ is the sum of the linear vibronic term (\ref{Eq:ULJT}) and 
the transfer part (\ref{Eq:Ht_adiabatic}),
$\hat{L}^{\rm nuc}_{\mathbf{m}1}$,  
$\hat{L}^{\rm nuc}_{\mathbf{m}2}$,  
$\hat{L}^{\rm nuc}_{\mathbf{m}3}$ %correspond to 
are nuclear angular momenta in the initial orbital basis
($\lambda_x, \lambda_y, \lambda_z$ in Ref. \onlinecite{OBrien1996a}, respectively), 
and $\hat{L}^{\rm el}_{\mathbf{m}j} (j = 1, 2, 3)$ %is defined by
are electronic angular momenta:
\begin{eqnarray}
 \hat{L}^{\rm el}_{\mathbf{m}1} &=& \sum_\sigma i 
 \left(\hat{c}_{\mathbf{m}2\sigma}^\dagger \hat{c}_{\mathbf{m}3\sigma} - 
       \hat{c}_{\mathbf{m}3\sigma}^\dagger \hat{c}_{\mathbf{m}2\sigma} \right),
\\
 \hat{L}^{\rm el}_{\mathbf{m}2} &=& \sum_\sigma i 
 \left(\hat{c}_{\mathbf{m}3\sigma}^\dagger \hat{c}_{\mathbf{m}1\sigma} - 
       \hat{c}_{\mathbf{m}1\sigma}^\dagger \hat{c}_{\mathbf{m}3\sigma} \right),
\\
 \hat{L}^{\rm el}_{\mathbf{m}3} &=& \sum_\sigma i 
 \left(\hat{c}_{\mathbf{m}1\sigma}^\dagger \hat{c}_{\mathbf{m}2\sigma} - 
       \hat{c}_{\mathbf{m}2\sigma}^\dagger \hat{c}_{\mathbf{m}1\sigma} \right).
\end{eqnarray}
%We note also that $\hat{H}_{\rm bi}$ is invariant under the above unitary transformation
%due to the isomorphism of $t_{1u}^n$ LUMO shell of C$_{60}^{n-}$ to the atomic $p^n$ shell.
For arbitrary JT deformations on sites, 
the system does not possess translational symmetry anymore. 
In the case of intermediate to strong vibronic coupling, the amplitude of dynamical JT deformation
$q_0$ is not small.
Since the LUMO orbitals of each fullerene are, on average, occupied by three electrons,
the vibronic term has a minimum at $\alpha = \pi/2$. \cite{OBrien1996a}
%where $\alpha$ is one of the polar coordinates of the $h_g$ normal coordinates,
%$(q, \alpha, \gamma, \theta, \phi)$. \cite{OBrien1996a}
Substituting
\begin{eqnarray}
 q_\mathbf{m} = q_0 + q'_\mathbf{m}, \qquad
 \alpha_\mathbf{m} = \frac{\pi}{2} + \alpha'_\mathbf{m}, 
\label{Eq:deviation}
\end{eqnarray}
into Eq. (\ref{Eq:Htilde}), we obtain 
\begin{eqnarray}
 \hat{\tilde{H}} 
 &=& 
 \hat{H}_{\rm rad} + 
 \hat{H}_{\rm rot} + 
 \hat{H}_{\rm el}^{(1)} + 
 \hat{H}_{\rm bi}, 
\label{Eq:HstrongJT}
\\
 \hat{H}_{\rm rad} &=&
 \sum_{\mathbf{m}}
  -\frac{\hslash \omega}{2}
 \left[
  q_0^{-4}\frac{\partial}{\partial q'_\mathbf{m}}
  \left(q_\mathbf{m}^4 \frac{\partial}{\partial q'_\mathbf{m}}\right) 
 + \frac{1}{q_0^2}\frac{\partial^2}{\partial \alpha'^2_\mathbf{m}} 
 \right]_{q_\mathbf{m}=q_0}
%, \alpha_\mathbf{m}=\frac{\pi}{2}}
\nonumber\\
 &+& N\frac{\hslash \omega}{2} q_0^2
 + \sum_\mathbf{m} \frac{\hslash \omega}{2} {q'^2_\mathbf{m}},
\label{Eq:Hrad}
\\
 \hat{H}_{\rm rot}
 &=& 
 \sum_{\mathbf{m}}
 \frac{\hslash \omega}{8q_0^2} 
 \left[
   4\left(\hat{L}^{\rm nuc}_{\mathbf{m} 1} + \hat{L}^{\rm el}_{\mathbf{m} 1}\right)^2
 + 4\left(\hat{L}^{\rm nuc}_{\mathbf{m} 2} + \hat{L}^{\rm el}_{\mathbf{m} 2}\right)^2
 \right.
\nonumber\\
 &+& 
 \left.
   \left(\hat{L}^{\rm nuc}_{\mathbf{m} 3} + \hat{L}^{\rm el}_{\mathbf{m} 3}\right)^2 
 \right],
\label{Eq:Hrot}
\\
 \hat{H}_{\rm el}^{(1)}
 &=& \sum_\mathbf{m} \sum_\sigma -\frac{\sqrt{3}}{2}\hslash \omega g q_0 
     \left(\hat{n}_{\mathbf{m}1\sigma} - \hat{n}_{\mathbf{m}2\sigma}\right)
\nonumber\\
 &+& \sum_{\mathbf{m}, \mathbf{m}'} \sum_{ll'\sigma} t^{\mathbf{m}\mathbf{m}'}_{ll'}
     \hat{c}_{\mathbf{m}l\sigma}^\dagger \hat{c}_{\mathbf{m}'l'\sigma},
\label{Eq:Hel}
\end{eqnarray}
where $q'_\mathbf{m}$ and $\alpha'_\mathbf{m}$ are the deviations from the equilibrium point.
%Eq. (\ref{Eq:deviation}).
The above derivation is based on the assumption that the radial JT coordinates (\ref{Eq:deviation})
remain unchanged under the electron transfer.
The justification for that, i.e., for the neglect of JT polaronic effect will be given in %the last section 
Sec. \ref{Sec:polaron}.
Following the adiabatic approximation, \cite{Bersuker1989a}
in Eq. (\ref{Eq:HstrongJT}) the terms smaller than $1/q_0^2$ are neglected and, consequently,
the radial degrees of freedom ($q',\alpha'$) are decoupled from the other degrees of freedoms
corresponding to the rotation of JT deformation in the 3D trough \cite{Auerbach1994a,OBrien1996a}
(see Sec. \ref{Sec:adiabatic} for the trough).
The rotational Hamiltonian (\ref{Eq:Hrot}) has nonadiabatic terms,
$\hat{V} = \hslash \omega \hat{L}^{\rm nuc}_{\mathbf{m} j} \hat{L}^{\rm el}_{\mathbf{m} j}/q_0^2$.
%Applying the adiabatic approximation \cite{adiabatic}
Neglecting these terms, \cite{Bersuker1989a} we obtain 
%we obtain the adiabatic Hamiltonian for the crystal:
\begin{eqnarray}
 \hat{H}_{\rm ad} &=& \hat{H}_{\rm rad} + \hat{H}_{\rm rot}^{\rm nuc} + \hat{H}_{\rm rot}^{\rm el} 
 + \hat{H}_{\rm el}^{(1)} + \hat{H}_{\rm bi},
\label{Eq:Had}
\\ 
 \hat{H}_{\rm rot}^{\rm nuc} &=& 
 \sum_{\mathbf{m}}
 \frac{\hslash^2}{8q_0^2} 
 \left[
   4\left(\hat{L}^{\rm nuc}_{\mathbf{m}}\right)^2 - 3\left(\hat{L}^{\rm nuc}_{\mathbf{m}3}\right)^2 
 \right],
\label{Eq:Hrotnuc}
\\ 
 \hat{H}_{\rm rot}^{\rm el} &=& 
 \sum_{\mathbf{m}}
 \frac{\hslash^2}{8q_0^2} 
 \left[
   4\left(\hat{L}_{\mathbf{m}}^{\rm el}\right)^2 - 3\left(\hat{L}_{\mathbf{m}3}^{\rm el}\right)^2 
 \right],
\label{Eq:Hrotel}
\end{eqnarray}
where $\hat{L}_\mathbf{m}^2 = \hat{L}_{\mathbf{m}1}^2+\hat{L}_{\mathbf{m}2}^2+\hat{L}_{\mathbf{m}3}^2$.
The adiabatic approximation is valid when the energy gap between the ground and the first excited
energies $\Delta E$ is large compared with the matrix element of the nonadiabatic term 
$|\hat{V}|$.
The ratio of $|\hat{V}| \approx \hslash \omega/(3g^2) $ and 
$\Delta E \approx 3\hslash \omega g^2/2$ for C$_{60}^{3-}$ is $|\hat{V}|/\Delta E \approx 1/5$,
which justifies the application of adiabatic approximation in the present case.
In this estimation, a value $|\hat{L}| \approx 1$ was taken.

Diagonalizing $\hat{H}_{\rm el}^{(1)}$ (\ref{Eq:Hel}), %we obtain the Hamiltonian in the basis of adiabatic band orbitals:
the Hamiltonian is written in the basis of adiabatic band orbitals:
\begin{eqnarray}
 \hat{H}_{\rm el}^{(1)}
 &=& 
 \sum_{i\sigma} \epsilon_i(\Omega) \hat{a}_{i\sigma}^\dagger(\Omega) \hat{a}_{i\sigma}(\Omega),
\label{Eq:Helband}
\end{eqnarray}
where $\Omega=\{\Omega_\mathbf{m}\}$ is the set of all Euler angles on all C$_{60}$ sites in the system
(Fig. \ref{Fig:xyz}),
$i$ indicates adiabatic band orbital, $\epsilon_i$ denotes its energy,
and $\hat{a}_{i\sigma}^\dagger$ is given by
\begin{eqnarray}
 \hat{a}_{i\sigma}^\dagger(\Omega) &=& \sum_{\mathbf{m}} \sum_l 
 U_{\mathbf{m}l i}(\Omega) \hat{c}_{\mathbf{m}l\sigma}^\dagger.
\label{Eq:a+}
\end{eqnarray}
For the ordered system (\ref{Eq:q}),
the coefficient $U_{\mathbf{m}li}$ reduces to 
$e^{i\mathbf{k}\cdot \mathbf{m}}u_{\lambda,p\mathbf{k}}/\sqrt{N}$ appearing in Eq. (\ref{Eq:a+ordered}).

Then the solution of the Hamiltonian (\ref{Eq:Had}) in the adiabatic approximation 
for the ground and low-lying vibronic states has the form:
\begin{eqnarray}
 |\Psi(R,\Omega)\rangle &=& 
 |\Phi_{\rm S}^{\rm ad}(\Omega)\rangle %\chi_\mu(R,\Omega)
 \chi^{\rm rad}(R)
 \chi^{\rm rot}(\Omega),
\label{Eq:Psi_vibr_prod}
\end{eqnarray}
where $|\Phi_{\rm S}^{\rm ad}\rangle$ is the 
Slater determinant of occupied adiabatic band orbitals (\ref{Eq:a+}):
\begin{eqnarray}
 |\Phi_{\rm S}^{\rm ad}(\Omega)\rangle = \prod_{i\sigma}^{\rm occ} \hat{a}_{i\sigma}^\dagger(\Omega)|0\rangle,
\end{eqnarray}
and $\chi^{\rm rad}(R)$ and $\chi^{\rm rot}(\Omega)$ are nuclear wave functions
depending on radial $R = \{q'_\mathbf{m}, \alpha'_\mathbf{m}\}$ and rotational 
$\Omega$ % = \{\Omega_\mathbf{m}\}$ 
nuclear coordinates, respectively.
The factorization of nuclear wave function became possible due to the separation of radial 
and rotational degrees of freedom in the adiabatic Hamiltonian (\ref{Eq:Had}).
Furthermore, the radial coordinates of different sites are independent from each other 
(see Eq. (\ref{Eq:Hrad})), hence,
the radial part $\chi^{\rm rad}$ is the product of the ground vibrational wave functions of all sites:
\begin{eqnarray}
 \chi^{\rm rad}(R) = \prod_\mathbf{m} \chi^{\rm rad}_{\mathbf{m}}(q'_\mathbf{m},\alpha'_\mathbf{m}).
\label{Eq:chirad}
\end{eqnarray}

\begin{figure}[tb]
\begin{center}
\includegraphics[bb = 0 0 2080 2144, height=5cm]{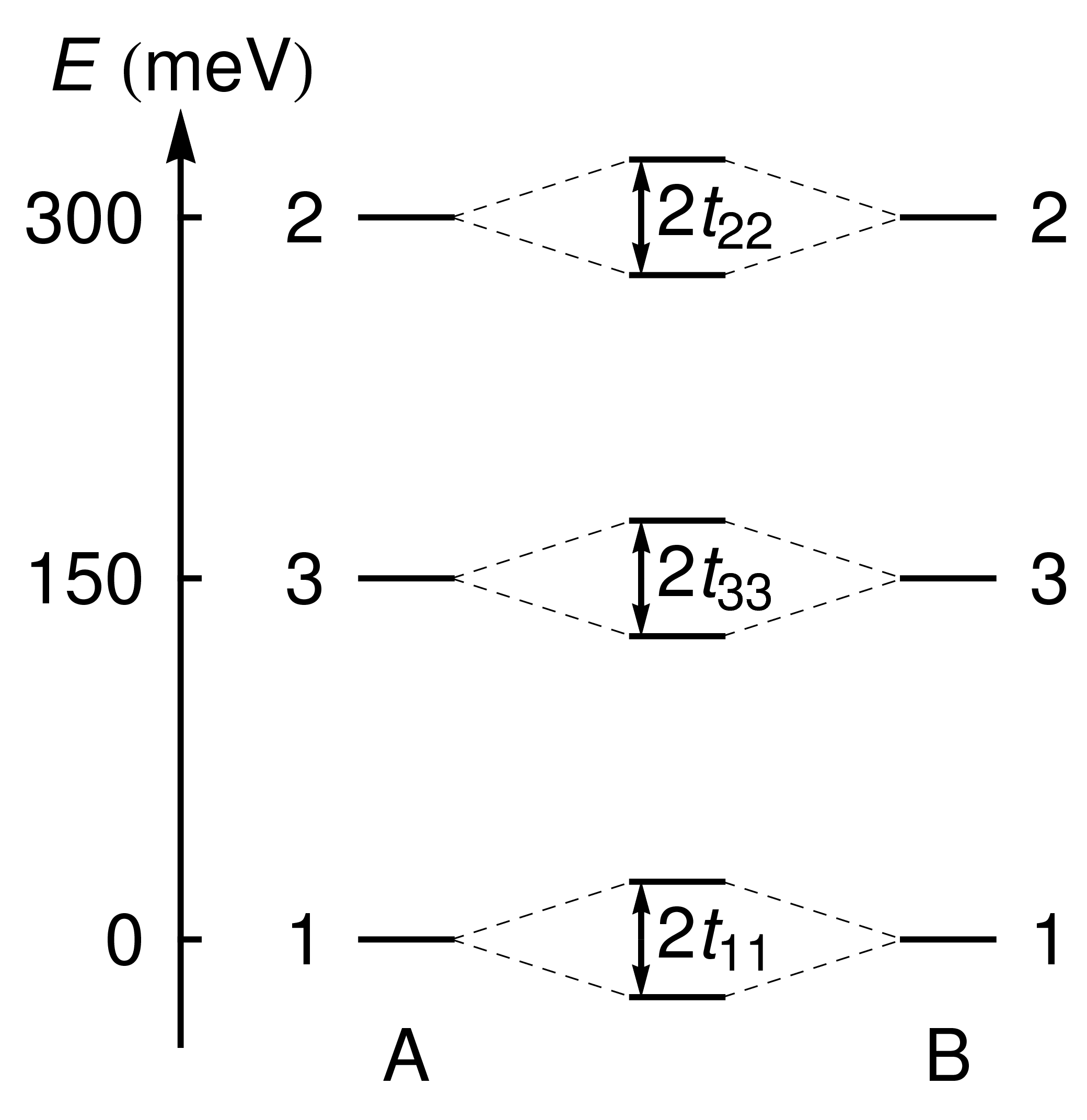}
\end{center}
\caption{Hybridization of adiabatic %band 
orbitals in a two-site model.}
\label{Fig:twosite}
\end{figure}

Further calculations are greatly simplified under the assumption that 
the dependence of $U_{\mathbf{m}li}$ (\ref{Eq:a+}) on Euler angles is relatively weak. 
This seems to be the case when correlation effects become important, 
leading to significant reduction of band energy $E_{\rm t}$ (Fig. \ref{Fig1}b) and strong 
separation of Gutzwiller bands ($x,y,z$ in Fig. \ref{Fig1}c for homogeneous JT distortions (\ref{Eq:q})).
Indeed, the hybridization of the adiabatic orbitals in this case mainly arises via resonant 
interactions (Fig. \ref{Fig:twosite}) because the width of individual bands is 
small compared with their Jahn-Teller splitting (150 meV),
thus resulting in a weak mixing of the off-resonant adiabatic orbitals.
%On the other hand, the hybridization arising from resonant interactions should be weakly
The hybridization arising from resonant interactions should be weakly
dependent on transfer parameters.
For example, in the case of two-site model (Fig. \ref{Fig:twosite}),
the ``band'' splitting of pairs of interacting resonant adiabatic orbitals $\Delta \epsilon$ is
strongly dependent on the Euler angles on two sites,
\begin{eqnarray}
% \Delta \epsilon \approx \left|2t^{AB}_{ll}(\gamma_A, \theta_A, \phi_A, \gamma_B, \theta_B, \phi_B)\right|,
 \Delta \epsilon \approx \left|2t^{AB}_{ll}(\Omega_A, \Omega_B)\right|,
\end{eqnarray}
while the adiabatic ``band'' orbitals,
\begin{eqnarray}
 |\psi_{i\sigma}(\Omega)\rangle \approx \frac{1}{\sqrt{2}} 
 \left(\hat{c}^\dagger_{Al\sigma} \pm \hat{c}^\dagger_{Bl\sigma}\right)|0\rangle,
\end{eqnarray}
have angle-independent mixing coefficients.

Neglecting the $\Omega$-dependence of coefficients $U_{\mathbf{m}li}$ in Eq. (\ref{Eq:a+}), 
the eigenvalue problem for the pseudorotational nuclear wave function reduces to the equation:
\begin{eqnarray}
 \left(\hat{H}_{\rm rot}^{\rm nuc} 
  + E_{\rm rot}^{\rm el}
  + E^{\rm el}_0(\Omega)
 \right)
 \chi^{\rm rot}(\Omega)
 = E^{\rm rot} \chi^{\rm rot}(\Omega).
\label{Eq:Hnuc}
\end{eqnarray}
where $E^{\rm el}_0(\Omega)$ is the %ground adiabatic potential energy:
adiabatic band energy:
\begin{eqnarray}
 E^{\rm el}_0(\Omega) = \sum_{i\sigma}^{\rm occ} \epsilon_i(\Omega),
\label{Eq:E0}
\end{eqnarray}
and $E^{\rm el}_{\rm rot}$ is the expectation value of $\hat{H}^{\rm el}_{\rm rot}$, Eq. (\ref{Eq:Hrotel}):
\begin{eqnarray}
 E^{\rm el}_{\rm rot} &=&  
 \langle \Phi_{\rm S}^{\rm ad}|\hat{H}_{\rm rot}^{\rm el}|\Phi_{\rm S}^{\rm ad}\rangle.
\label{Eq:Erotel}
\end{eqnarray}
The direct calculation of this matrix element gives:
\begin{eqnarray}
 E^{\rm el}_{\rm rot} &=& \sum_{\mathbf{m}}
 \frac{\hslash^2}{8q_0^2}
 \left(
   10n_{\mathbf{m}1} 
 + 10n_{\mathbf{m}2} 
 + 16n_{\mathbf{m}3} 
 \right.
\nonumber\\
 &-& 
 \left.
    4 n_{\mathbf{m}1}n_{\mathbf{m}2}
  - 16 n_{\mathbf{m}2}n_{\mathbf{m}3}
  - 16 n_{\mathbf{m}3}n_{\mathbf{m}1}
\right)
%\nonumber\\
% &-&
%\left.
% 4\left(\rho_{12}^\mathbf{m} - \rho_{21}^\mathbf{m}\right)^2
% -
% 4\left(\rho_{23}^\mathbf{m} - \rho_{32}^\mathbf{m}\right)^2
% -
% \left(\rho_{31}^\mathbf{m} - \rho_{13}^\mathbf{m}\right)^2
% \right]
\nonumber\\
 &+&
 \sum_\mathbf{m}
 \sum_{i}^{\rm occ}
 \frac{\hslash^2}{8q_0^2}
 \left(
  4
 \left|U_{\mathbf{m}1i}\right|^2
 \left|U_{\mathbf{m}2i}\right|^2
 \right.
\nonumber\\
 &+&
 \left.
 16
 \left|U_{\mathbf{m}2i}\right|^2
 \left|U_{\mathbf{m}3i}\right|^2
 + 16
 \left|U_{\mathbf{m}3i}\right|^2
 \left|U_{\mathbf{m}1i}\right|^2
 \right),
\end{eqnarray}
where $n_{\mathbf{m}l}$ are populations of the adiabatic orbitals $(l)$ on the site $\mathbf{m}$.
%where 
%$\rho_{ll'}^\mathbf{m} = \langle \hat{c}_{\mathbf{m}l\sigma}^\dagger \hat{c}_{\mathbf{m}l\sigma} \rangle$ 
%are elements of one-particle density matrix calculated on adiabatic band orbital (\ref{Eq:a+}).
The last term is smaller than the other terms by $1/N$ because $|U_{\mathbf{m}li}| \approx 1/\sqrt{N}$ and 
$\sum_i |U_{\mathbf{m}li}|^2|U_{\mathbf{m}l'i}|^2 \approx 1/N$, 
while the occupation number $n_{\mathbf{m}l} = \sum_i |U_{\mathbf{m}li}|^2 \approx 1$.
Neglecting the last term, we obtain
\begin{eqnarray}
 E^{\rm el}_{\rm rot} &=& \sum_{\mathbf{m}}
 \frac{\hslash^2}{8q_0^2}
 \left( 
   10n_{\mathbf{m}1} 
 + 10n_{\mathbf{m}2} 
 + 16n_{\mathbf{m}3} 
 \right.
\nonumber\\
 &-& 
 \left.
     4 n_{\mathbf{m}1}n_{\mathbf{m}2}
  - 16 n_{\mathbf{m}2}n_{\mathbf{m}3}
  - 16 n_{\mathbf{m}3}n_{\mathbf{m}1}
 \right).
\quad
\label{Eq:Erotel2}
\end{eqnarray}
%In Eq. (\ref{Eq:Erotel2}), the terms of the order of $1/N$ are neglected.
The obtained energy is additive over the sites, with one-site contributions being
equivalent with the corresponding energy of an isolated C$_{60}^{3-}$, 
$E_{\rm rot}^{{\rm el} (1)} = 5\hslash^2/(4q_0^2)$ (Eq. (37) in Ref. \onlinecite{OBrien1996a}),
in the case of full disproportionation of electron density among three orbitals, 
$(n_1, n_2, n_3) = (1, 0, 1/2)$ (Sec. \ref{Sec:SJT}).
Note the lack of $\Omega$-dependence of the energy in Eq. (\ref{Eq:Erotel2}),
which is the result of neglected $\Omega$-dependence of the coefficients
$U_{\mathbf{m}li}$ in Eq. (\ref{Eq:a+}).

\begin{figure}[tb]
\begin{center}
\begin{tabular}{l}
(a) \\
\includegraphics[bb = 0 0 4208 2104, width=8.6cm]{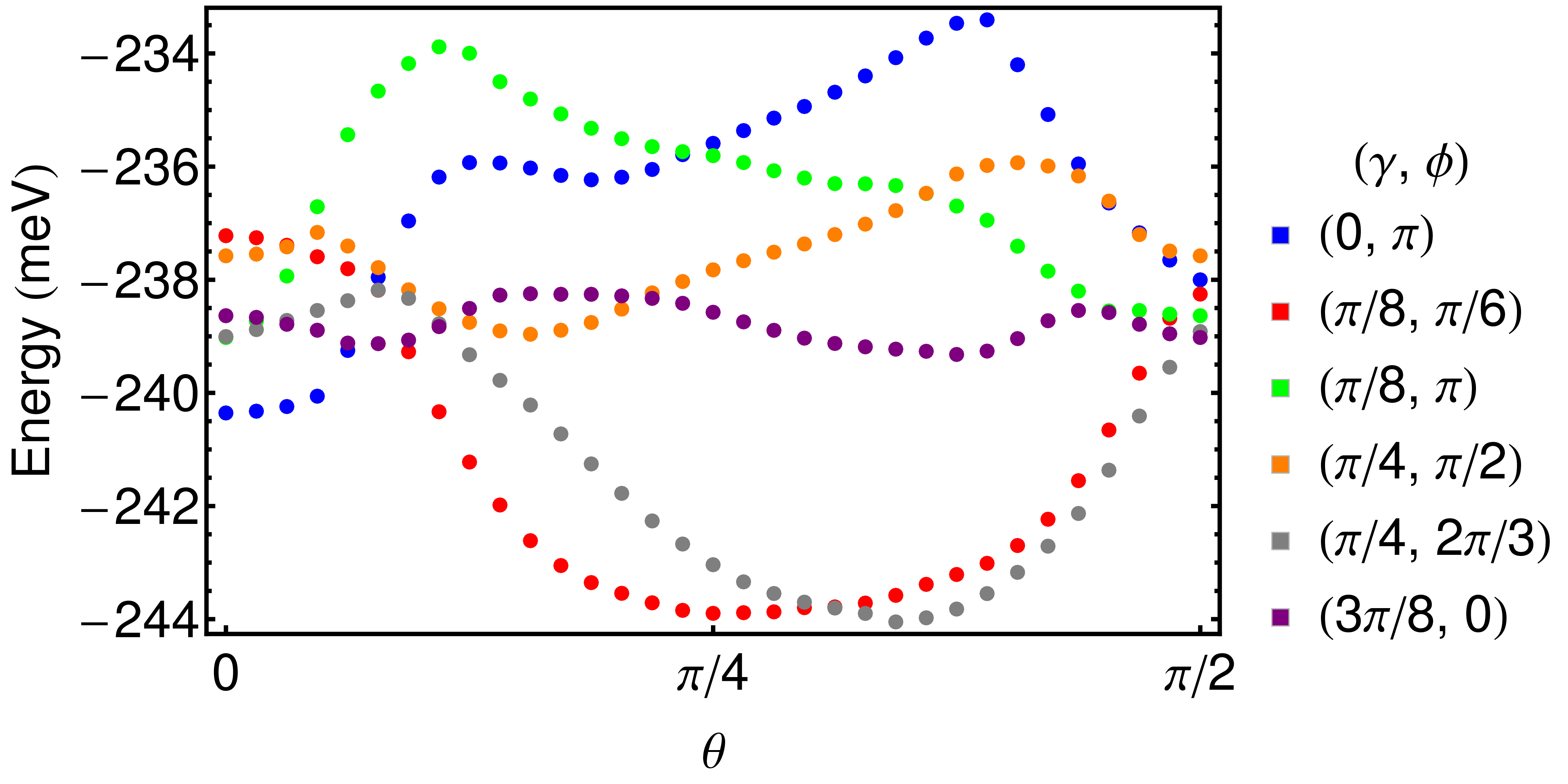}
\\
(b) \\
\includegraphics[bb = 0 0 4360 2104, width=8.6cm]{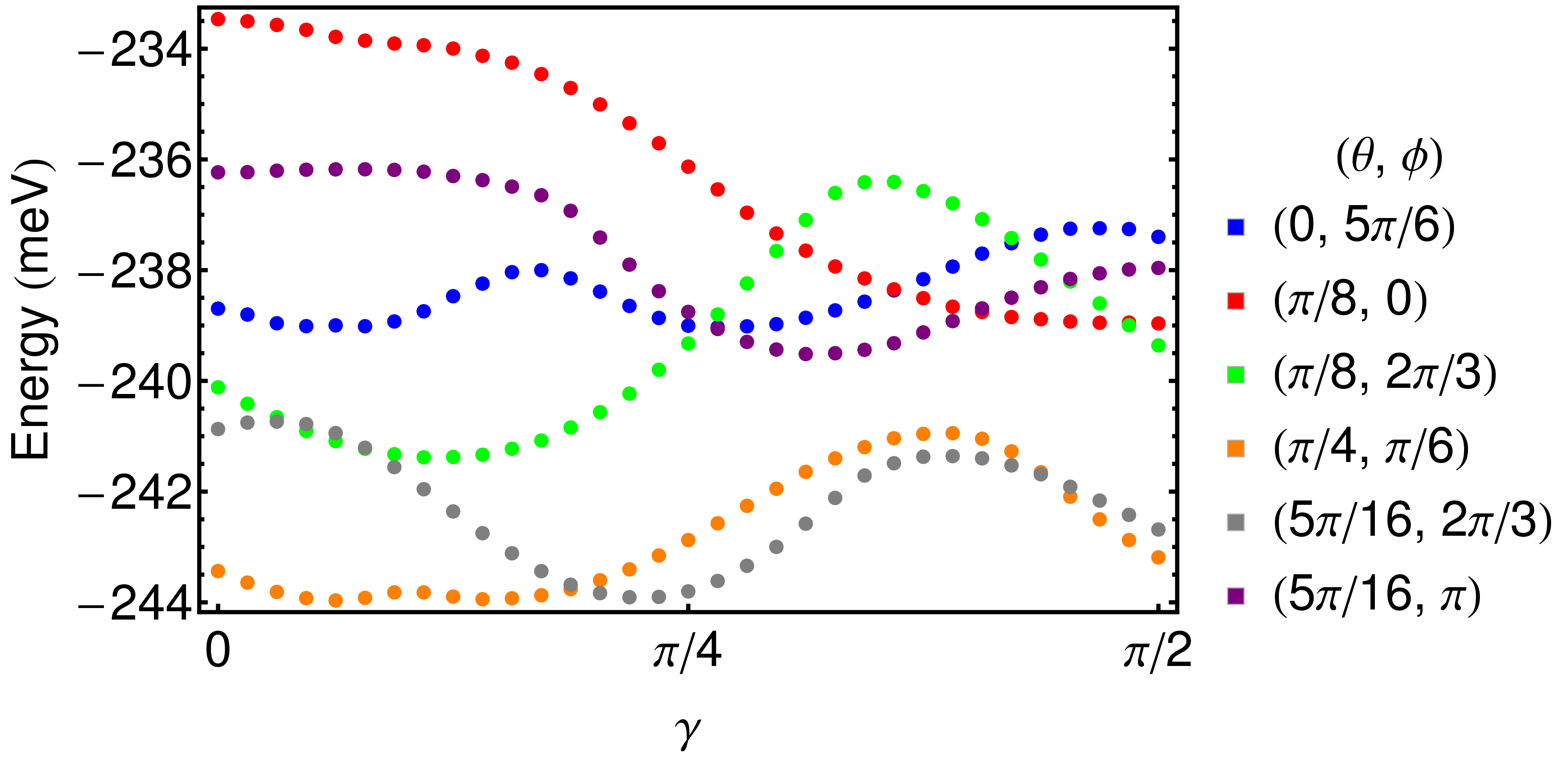}
\end{tabular}
\end{center}
\caption{
(color online)
Dependence of uncorrelated band energy $E_{\rm t}^{0}$ on 
the Euler angles of JT distortions  $\theta$ (a) and $\gamma$ (b), 
respectively, for $q = \sqrt{3}g = 1.85$.
}
\label{FigEt0}
\end{figure}

On the other hand, the adiabatic band energy $E_0^{\rm el}$ (\ref{Eq:E0}) is $\Omega$-dependent even if the 
coefficients $U_{\mathbf{m}li}$ are not, and this dependence a priori is not weak. 
This $\Omega$-dependence is estimated here by direct calculations of the uncorrelated band %kinetic 
energy $E_{\rm t}^{0}$ for different directions of ordered JT distortions, $q = \sqrt{3}g$. 
The obtained variations of $E_{\rm t}^{0}$ do not exceed 12 meV (Fig. \ref{FigEt0}). 
The variation of $E_{\rm t}^{0}$ will be even smaller for disordered system because 
the Euler angle dependence is smeared out by the disorder.
Including electron correlation effects via the Gutzwiller's ansatz described above (Sec. \ref{Sec:SJT}) 
will result in the case of $U = U_{c}$ (corresponding to $q = 1.85$, see Fig. \ref{Fig1}b) 
to a reduction of uncorrelated $E_{\rm t}^{0}$ ($\approx -240$ meV) %band energy 
by one order of magnitude (Fig. \ref{Fig1}b).
At the same extent will reduce the variations of the band energy in function of 
the direction of JT distortions, which means that they are negligible compared to 
the dynamical contribution to JT stabilization energy (Fig. \ref{Fig:DJT}). 

In $A_3$C$_{60}$ crystals the JT pseudorotations after the Euler angles $\Omega$
can be also hindered by intermolecular vibrations.
However, the energy of these vibrations ($\approx 5-10$ meV \cite{Gunnarsson1997a}) 
is much lower than the energy gain due to delocalization of JT deformations in the trough. 

Hence the vibronic dynamics is expected to be unquenched, 
like in insulating fullerides Cs$_3$C$_{60}$. \cite{Iwahara2013a}
Given the near independence of the band energy $E_0^{\rm el}$ on the pseudorotation coordinates
of C$_{60}^{3-}$ sites $(\Omega)$, and the full $\Omega$-independence of the contribution (\ref{Eq:Erotel2}),
the pseudorotational Hamiltonian (\ref{Eq:Hnuc}) becomes merely a sum of on-site contributions.
Each such contribution is an operator depending on $\Omega_\mathbf{m}$ Euler coordinates of the 
corresponding site, Eq. (\ref{Eq:Hrotnuc}), which means that the pseudorotational wave function factorizes, 
\begin{eqnarray}
 \chi^{\rm rot} &=& \prod_{\mathbf{m}} \chi_\mathbf{m}^{\rm rot}(\Omega_\mathbf{m}),
\label{Eq:chirot}
\end{eqnarray}
with $\chi_\mathbf{m}^{\rm rot}$ being eigenfunctions of one-site operators in Eq. (\ref{Eq:Hrotnuc}).
Then, taking into account the factorization of the radial part, Eq. (\ref{Eq:chirad}), 
the Gutzwiller wave function with dynamical JT effect on fullerene sites has the form:
\begin{eqnarray}
 |\Psi_{\rm G}\rangle &=& \hat{P}_{\rm G}|\Phi_{\rm S}^{\rm ad}\rangle \times
                          \prod_{\mathbf{m}} \chi_{\mathbf{m}}^{\rm rad} \chi_{\mathbf{m}}^{\rm rot},
\label{Eq:G+DJT}
\end{eqnarray}
and the Gutzwiller projector (\ref{Eq:pg}) will involve now population operators for adiabatic orbitals on the fullerene sites:
\begin{eqnarray}
 \hat{P}_{\rm G} &=& \prod_\mathbf{m} \exp\left(
 -\frac{1}{2} \sum_{l\sigma \ne l'\sigma'} A_{\mathbf{m}ll'}
 \hat{n}_{\mathbf{m}l\sigma} \hat{n}_{\mathbf{m}l'\sigma'}
 \right).
\label{Eq:PG_DJT}
\end{eqnarray}

\subsection{Self-consistent Gutzwiller approach for the ground vibronic state}
The ground state energy of the dynamical JT system is 
obtained by minimizing the total energy per site.
Although the adiabatic band orbitals (\ref{Eq:a+}) correspond to a disordered system,
this will not pose any complication if we assume that the band energy of these orbitals 
is independent on the form of adiabatic orbitals, i.e., on the three Euler angles 
characterizing the ``direction'' of JT distortions on sites. 
This seems to be indeed the case given the weak dependence of band energy on the local JT distortions established above (Fig. \ref{FigEt0}). 
Then the calculation of the electronic part of the energy can be done for a particular 
case of Euler angles equal on all sites, yielding the previous result for a translational system, 
while the nuclear part of the wave function (\ref{Eq:G+DJT}) will give the dynamical contribution.
%the band energy part in Eq. (\ref{Eq:Ead}) reduces to Eq. (\ref{Eq:Et}).

Hence, within the adiabatic approximation (\ref{Eq:Had}), 
the ground energy with the Gutzwiller's wave function (\ref{Eq:G+DJT}) is given by 
\begin{eqnarray}
 E_{\rm ad} &=& 
% \langle \chi^{\rm rad} \chi^{\rm rot}|\tilde{E}_{\rm ad}(R,\Omega)
% |\chi^{\rm rad} \chi^{\rm rot}\rangle,
%\\
% \tilde{E}_{\rm ad}(R,\Omega) &=& 
%               \sum_{ll'\sigma} q_{ll'} \tau_{ll'}
             E_{\rm t} +  E_{\rm bi}
             - \frac{3\hslash \omega g_{\rm eff}^2}{2} + E_{\rm DJT}^{\rm ad},
\label{Eq:Ead}
%\label{Eq:Ead_disorder}
\end{eqnarray}
where 
%\begin{eqnarray}
% \tau_{ll'} &=& \frac{1}{N}\sum_{\mathbf{m},\mathbf{m}'} \sum_{i}^{\rm occ}
%                t_{ll'}^{\mathbf{m}\mathbf{m'}} 
%                U_{\mathbf{m}li}^* U_{\mathbf{m}'l'i},
%\end{eqnarray}
the dynamical JT deformation $q_0$ is replaced by 
\begin{eqnarray}
 q_0 &=& \sqrt{3}g_{\rm eff},
\end{eqnarray}
and $E_{\rm DJT}^{\rm ad}$ is the dynamical JT contribution 
\begin{eqnarray}
E_{\rm DJT}^{\rm ad} &=& -  \frac{3\hslash \omega}{2}
 - \frac{3\hslash \omega}{8g_{\rm eff}^2} + E_{\rm rot}.
\label{Eq:EDJTad}
\end{eqnarray}
The zero-point energy of the five-dimensional harmonic oscillator is set to zero.
The first and the second terms in Eq. (\ref{Eq:EDJTad})
appear from the radial Hamiltonian (\ref{Eq:Hrad}) \cite{OBrien1996a}
and $E_{\rm rot}$ is the eigenvalue of the pseudorotational Hamiltonian (\ref{Eq:Hnuc}).
%$E^{\rm el}_{\rm rot}$ in Eq. (\ref{Eq:Hnuc}) is averaged by the vibrational part 
%of the vibronic wavefunction, $|\chi^{\rm rad}\chi^{\rm rot}\rangle$, and becomes constant. 
%The form of the expectation values of the on-site operators does not change 
%due to the Gutzwiller's projector (\ref{Eq:PG_DJT}).
Furthermore, the dynamical contribution (\ref{Eq:EDJTad}) is replaced by the exact 
$E_{\rm DJT}$ (Fig. \ref{Fig:DJT}), yielding 
\begin{eqnarray}
 E &=& E_{\rm t} + E_{\rm bi} - \frac{3\hslash \omega g_{\rm eff}^2}{2} + E_{\rm DJT}(g_{\rm eff}).
\label{Eq:E_DJT}
\end{eqnarray}

The ground state for dynamical JT system is obtained by self-consistent 
minimization of the energy (\ref{Eq:E_DJT}) with respect to $\{u_{\lambda, p\mathbf{k}}\}$ and 
$\{A_{\lambda \lambda'}\}$. 
We obtain similar formula as Eqs. (\ref{Eq:step1}) and (\ref{Eq:step2}),
with the only difference in the JT term of one-particle Hamiltonian:
\begin{eqnarray}
h_{\lambda \lambda'}^\mathbf{k} 
&=& q_{\lambda \lambda'} t_{\lambda \lambda'}^\mathbf{k} 
+ \delta_{\lambda \lambda'} \left[
 \sum_{\kappa \kappa'}\frac{\partial q_{\kappa \kappa'}}{\partial n_{\lambda}}
 \tau_{\kappa \kappa'}
+ \frac{1}{2} \frac{\partial E_{\rm bi}}{\partial n_{\lambda}}
\right.
\nonumber\\
&+&
 \left.
 \left(-\frac{3\hslash \omega}{2} g_{\rm eff} 
 + \frac{g}{2} \frac{\partial E_{\rm DJT}(g_{\rm eff})}{\partial g_{\rm eff}}\right) 
 \left(\delta_{\lambda x} - \delta_{\lambda y} \right)
 \right].
\nonumber\\
\label{Eq:hkDJT}
\end{eqnarray}

\subsection{Dynamical Jahn-Teller instability in K$_3$C$_{60}$}
\begin{figure}[tb]
\begin{center}
\begin{tabular}{l}
(a)\\
\includegraphics[bb = 0 0 3320 2080, width=8.6cm]{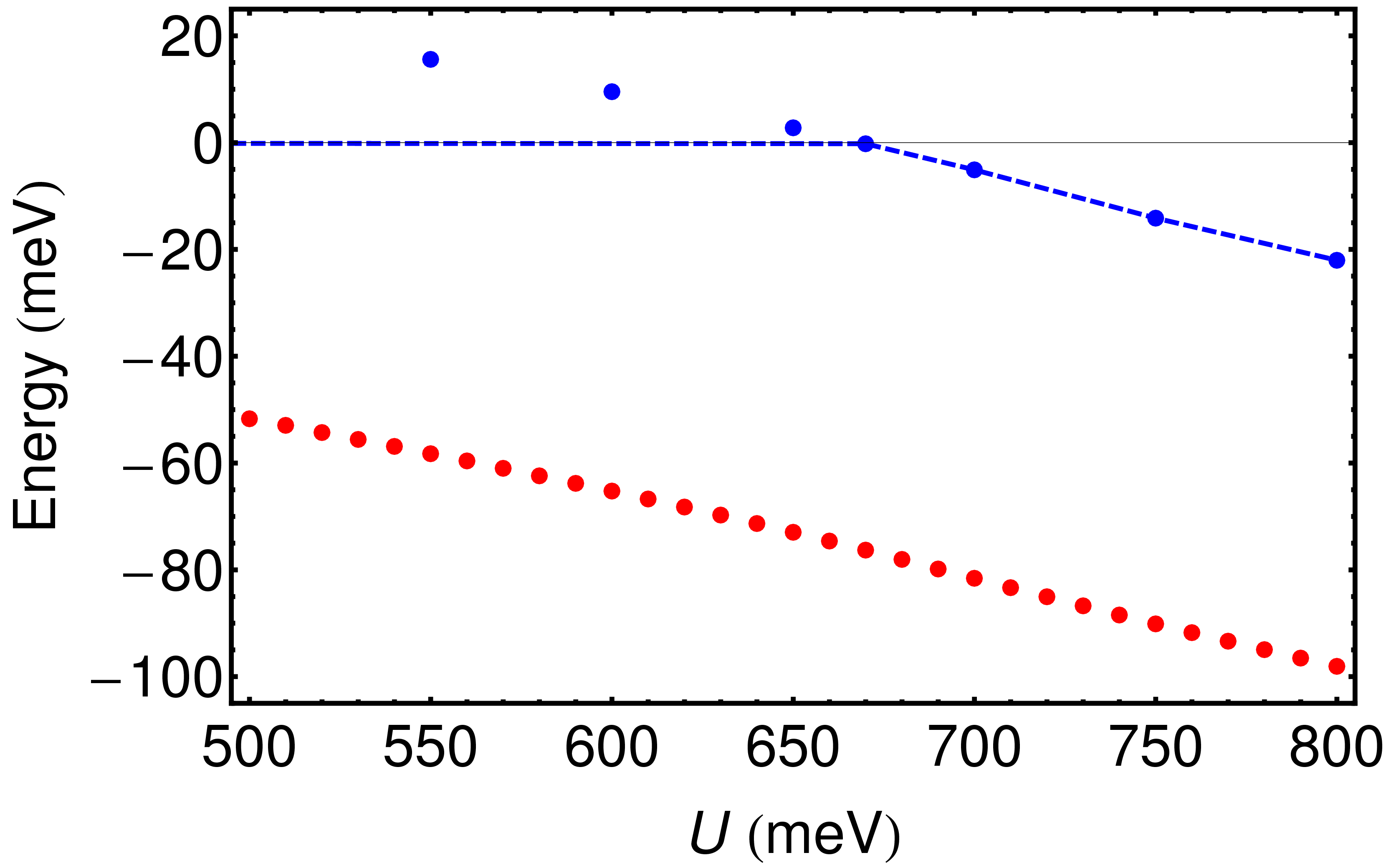}
\\
(b)\\
\includegraphics[bb = 0 0 3320 2152, width=8.6cm]{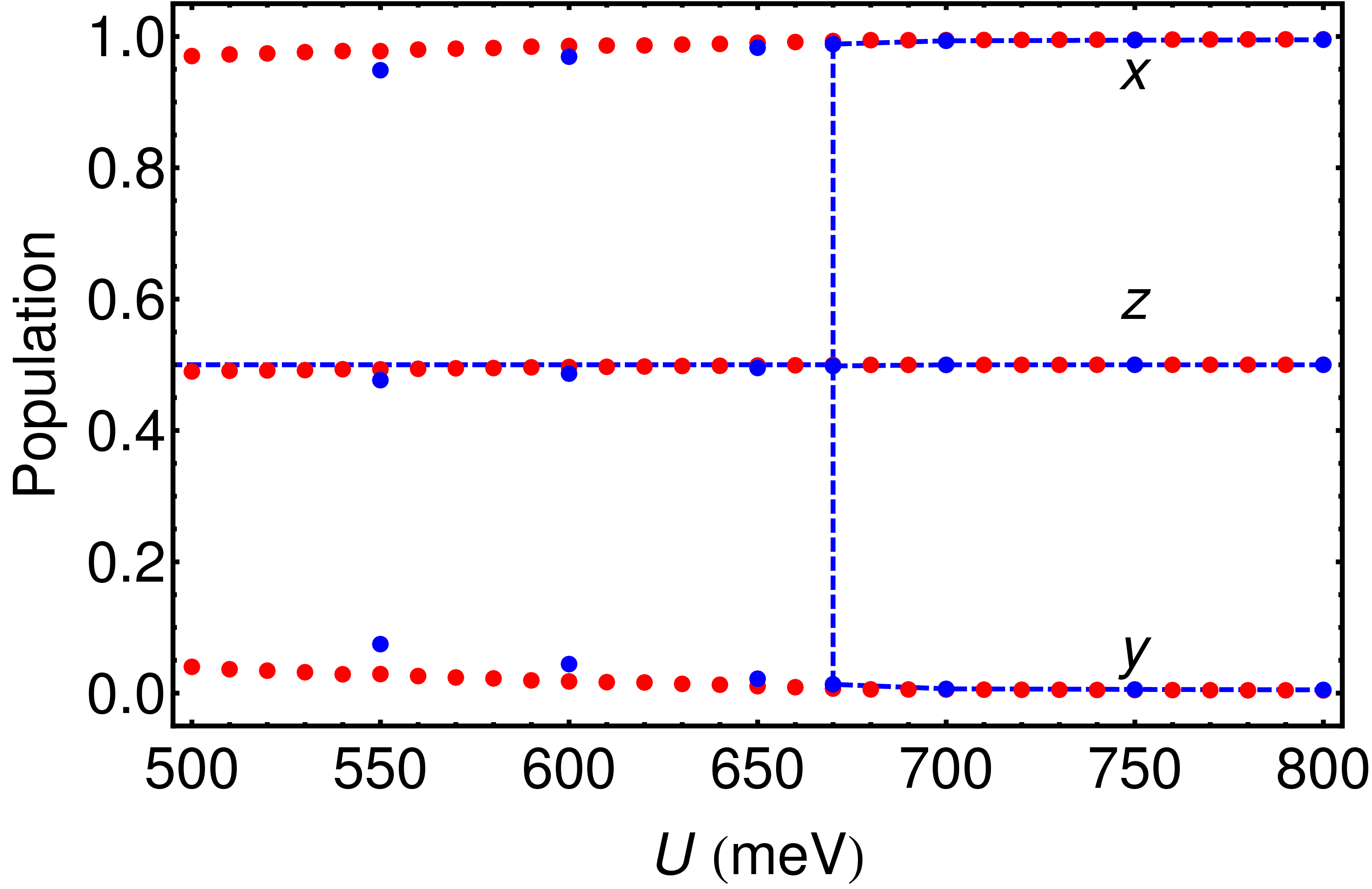}
\end{tabular}
\end{center}
\caption{
(color online) 
(a) Total energy and (b) occupation numbers of LUMO orbitals $n_\lambda$ 
with static (blue) and dynamical (red) JT effects as a function of $U$.
The blue points and the dashed line correspond to the minimum with static JT distortion
and the global minimum of $E_{\rm g}$, respectively.
Total energy at $q = 0$ is set to zero at each $U$.
$x$, $y$, and $z$ in (b) are orbital components under distortion (\ref{Eq:q}).
}  
\label{FigDJT}
\end{figure}

Minimizing the total energy (\ref{Eq:E_DJT}), we obtain the ground energy in the presence 
of the JT dynamics on sites (Fig. \ref{FigDJT}a). 
In the case of static JT effect, the JT distortion appears for $U > 670$ meV (Fig. \ref{Fig1}a).
We can see, however, that the JT dynamics enhances the dynamical JT deformation, and, consequently, 
the disproportionation of the occupation numbers in the adiabatic orbitals are also enhanced 
(Fig. \ref{FigDJT}b).
%(Fig. S3 (b) \cite{SM}). 
As a result the critical value of electron repulsion parameter for JT instability ($U_{c}$) 
is significantly reduced in the dynamical case.
%are smaller in the dynamical case.
%and for full orbital disproportionation of the  LUMO charge density ($U_{2\perp}$) are smaller in the dynamical case (Fig. \ref{Fig2}).  
%As a result the critical value of electron repulsion parameter for JT instability ($U_{1\perp}$) 
%and for full orbital disproportionation of the  LUMO charge density ($U_{2\perp}$) are smaller in the dynamical case (Fig. \ref{Fig2}).  
%Although in the case of the static JT system, the orbital disproportionation appears at $U_\perp = $ 600? meV,
%in the case of the dynamical JT system it appears at $U_\perp =$ xxx meV.
%These critical values are smaller than $U_\perp = 820$ meV for K$_3$C$_{60}$ \cite{nomura2012}, hence the metallic fullerides  
In particular,
the critical value is smaller than the estimated $U = 750$ meV for K$_3$C$_{60}$, \cite{nomura2012}
hence, the metallic fullerides always exhibit dynamical JT instability in the ground state. 
This explains the absence of staggered JT deformations in the x-ray diffraction data of $A_3$C$_{60}$. 
Furthermore, since $U >U_{c}$ the equilibrium JT distortions on sites will be close to maximal possible, 
i.e., to their values in a free C$_{60}^{3-}$ ion. % ($q=1.6$).

\section{Effect of electron correlation and Jahn-Teller instability on one-particle states}
\subsection{Orbital disproportionation}
\label{Sec:disproportionation}
The electron correlation and the JT effect induce differences in the population of the three 
LUMO orbitals on fullerene sites (orbital disproportionation).
Within the broken-symmetry Hartree-Fock approach, \cite{Chibotaru1996a} 
the JT and the bielectronic energy per site is 
\begin{eqnarray}
 E^{\rm HF} &=& -\frac{3\hslash \omega g_{\rm eff}^2}{2} 
             + U \left(\frac{5}{12}n^2-\Delta n_1^2-\Delta n_2^2-\Delta n_3^2\right),
\nonumber\\
% E^{\rm HF} &=& -\frac{3\hslash \omega g_{\rm eff}^2}{2} + 15U_\perp n^2 
% \nonumber\\ 
% &-& \left(U_\perp - 3J_{\rm H}\right)\left(\Delta n_1^2+\Delta n_2^2+\Delta n_3^2\right),
\label{Eq:EHF}
\end{eqnarray}
where $n$ is the total population of the fullerene site, 
$\Delta n_l$ is the deviation of the occupation 
of the orbital subband from the case of cubic symmetry $(1/2)$,
%thus it satisfies $\sum_{l=1}^3 \Delta n_l = 0$, 
and %$J_{\rm H}$ is neglected 
one single average electron repulsion parameter $U$ (\ref{Eq:U}) is used
for simplicity. %, and the band energy part is also neglected for simplicity.
%In the fully disproportionated case, $(n_1, n_2, n_3) = (1,0,1/2)$, 
The HF energy with full disproportionation, $(n_1, n_2, n_3) = (1,0,1/2)$, 
is lower than the energy of the degenerate system $(n_1, n_2, n_3) = (1/2,1/2,1/2)$ by
\begin{eqnarray}
% \Delta E^{\rm HF} &=& -\frac{3\hslash \omega g^2}{2} - \frac{U_\perp - 3J_{\rm H}}{2}.
 \Delta E^{\rm HF} &=& -\frac{3\hslash \omega g^2}{2} - \frac{U}{2}.
\label{Eq:DeltaEHF}
\end{eqnarray}
%However, the broken-symmetry HF approach induces the disproportionation even in the 
%absence of the JT effect, $g = 0$, which is an artifact of this approach.
%In both broken-symmetry Hartree-Fock and Gutzwiller approaches the orbital disproportionation arises, 
%while in comparison with the former the latter adequately includes the electron correlations 
%and fulfills the requirement of symmetry.
The orbital disproportionation is seen also in the present Gutzwiller's %approach with JT effect
treatment (Fig. \ref{FigDJT}).
In terms of the electron configurations, the equal population of three LUMO bands in a cubic 
band structure results in their equal probability ($1/2^6$).
The HF type symmetry breaking equally enhances the weights of four configurations, 
$\psi_1^2$, $\psi_1^2\psi_3^1$ (both spin projections) and $\psi_1^2\psi_3^2$, and quenches the others, 
leading to the gain of bielectronic energy per site $E_{\rm bi}$ by $U/2$.
%Besides the HF type disproportionation, 
The weights of configurations $\psi_1^2\psi_3^1$ among the four
are further enhanced and the rest of them are further reduced in the Gutzwiller %wave function for JT system,
%(Fig. \ref{Fig:nu})
treatment, which additionally lowers $E_{\rm bi}$ by $U/4$ in the limit of strong correlation.
The latter becomes possible because of multi determinantal character of the Gutzwiller ansatz.

Despite the larger gain of $E_{\rm bi}$ in Gutzwiller approach compared to HF one,
the latter predicts smaller $U_c$ for the static JT distortion.
This is due to the artifactual feature of the broken-symmetry HF approach mentioned above which leads, 
in particular, to orbital disproportion in K$_3$C$_{60}$ without JT effect on fullerene sites. \cite{Ceulemans1997a}
Indeed, even in the absence of the vibronic coupling, $g = 0$, the broken-symmetry HF state is more stable
than the cubic band solution by $U/2$, Eq. (\ref{Eq:DeltaEHF}).
On the other hand, the Gutzwiller's wave function is not 
disproportionated in the absence of JT effect, which is testified by equal population of three
LUMO orbitals at $q=0$ point for arbitrary $U$ (Fig. \ref{Fig1}c).
%i.e., the latter fulfills the symmetry requirement.
%Moreover, as discussed above, the Gutzwiller wave function has the degrees of freedom to include 
%more configurations with the same $E_{\rm bi}$ than the broken-symmetry HF wave function.
%i.e., the symmetry requirement is fulfilled.
This is the result of a higher flexibility of the Gutzwiller's wave function,
% fulfills the symmetry requirement 
%because it has larger degrees of freedom than the broken-symmetry HF wavefunction. 
which can include various configurations without changing the 
bielectronic energy, such as equally populated configurations of $\psi_1^1\psi_2^1\psi_3^1$ type.

%The lack of the artificial symmetry breaking in the Gutzwiller's wave function 
%is due to the larger degrees of freedom than the broken-symmetry HF wavefunction to include 
%more configurations with the same $E_{\rm bi}$ 
%such as equally populated configuration $\psi_1\psi_2\psi_3$.

%For example, in the Gutzwiller's wave function, equally populated configuration $\psi_1\psi_2\psi_3$ 
%as well as $\psi_1^2\psi_3$ is included. 
%while it is not in the broken-symmetry HF wave function.
%Since the multi determinantal approach has more degrees of freedom in comparison with HF approach, 
%the former enables us to keep the cubic symmetry of the wave function without the JTE
%and avoid the artificial splitting of the band:
%the probability of $\psi_1^2\psi_3^1$ for the Gutzwiller wave function without JTE 
%is close to equally populated case of $2 \times 1/2^6$.
%(Fig. \ref{Fig:nu}, black dotted line).
%In the Gutzwiller ansatz, the disproportionation arises only when the JTE exists.

\begin{figure}
\begin{center}
\begin{tabular}{ll}
(a) & ~~(b) \\
\includegraphics[bb = 0 0 1248 2496, height=5cm]{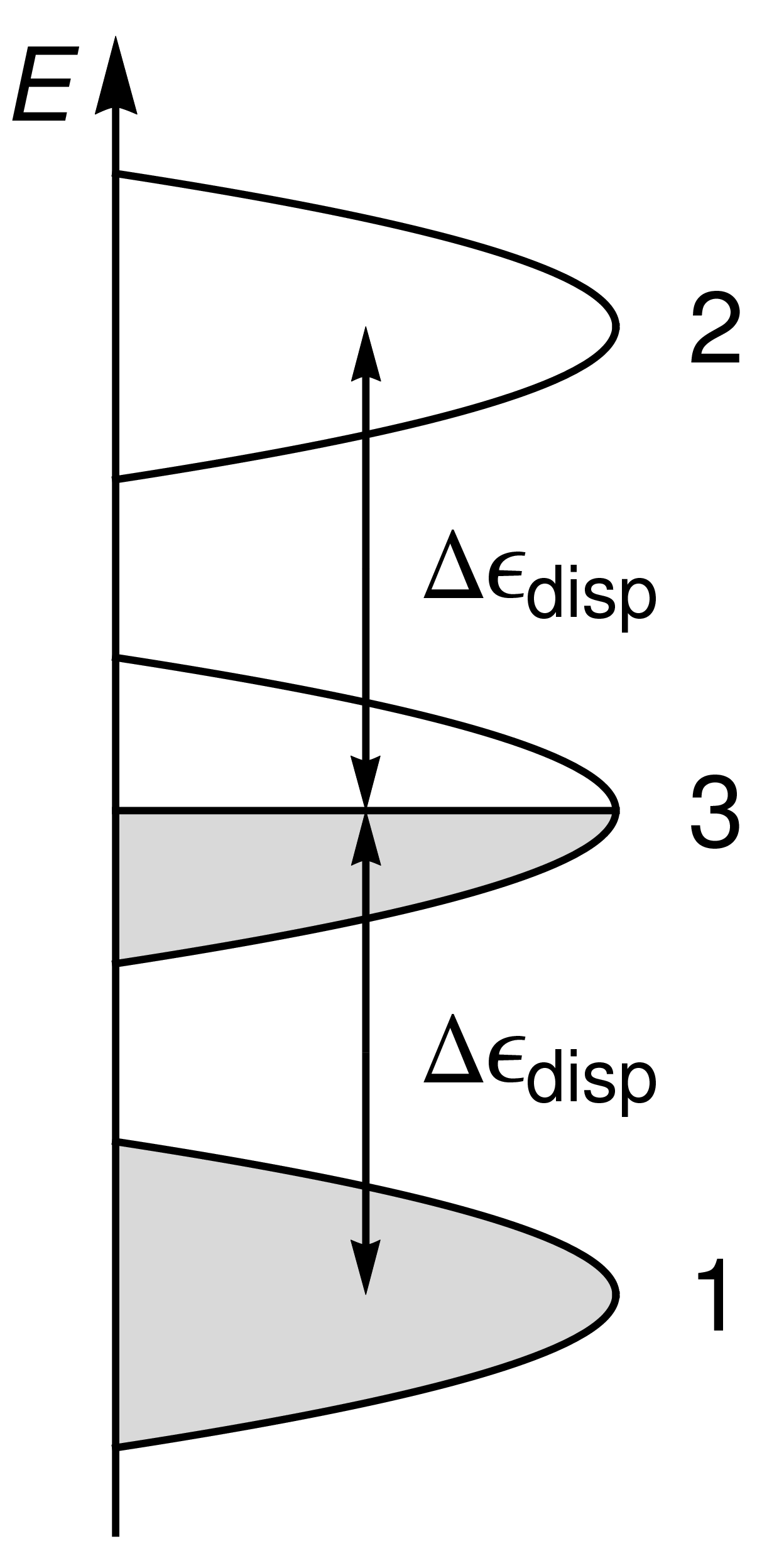}&~~
\includegraphics[bb = 0 0 3320 3096, height=5cm]{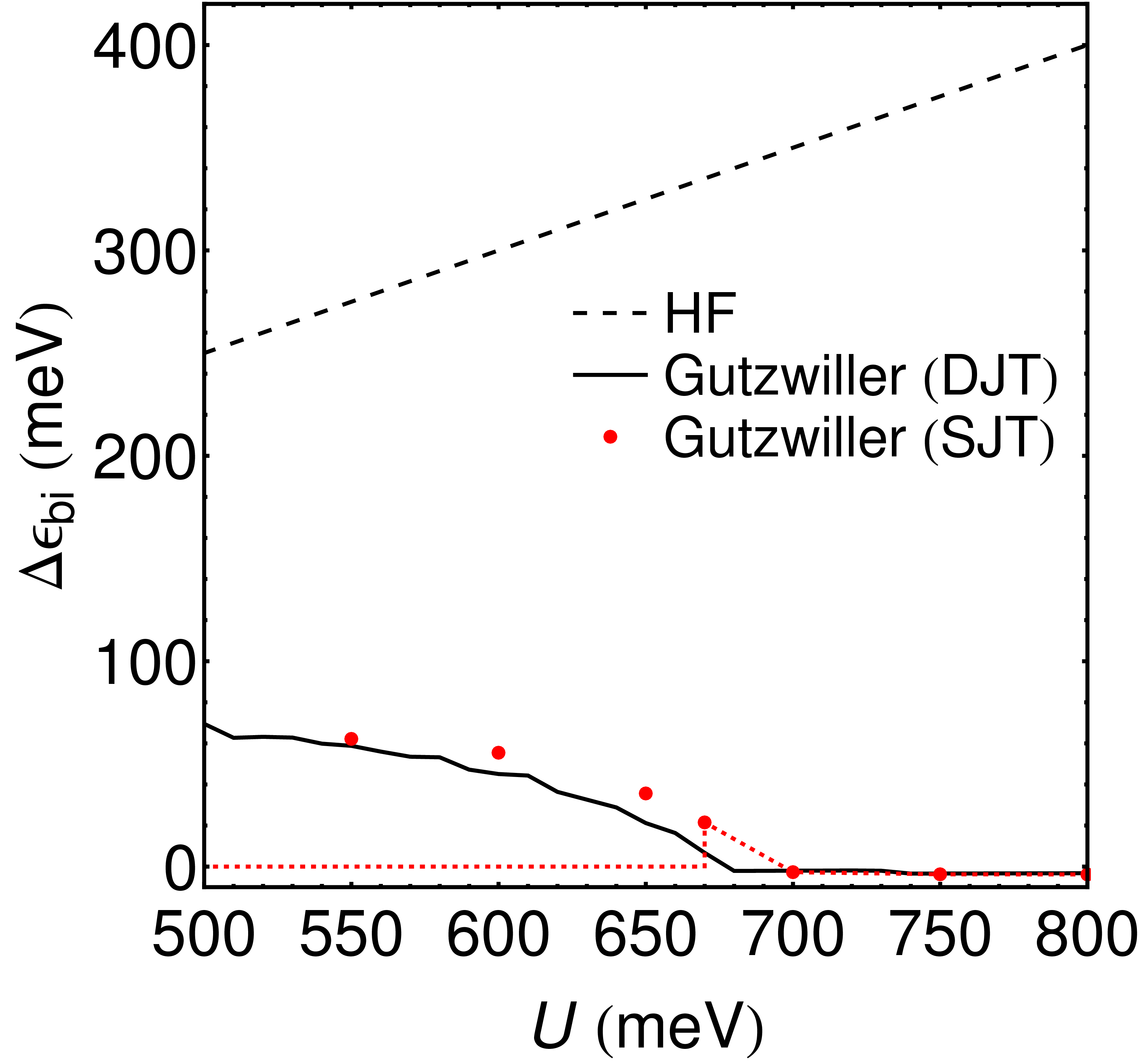}\\
\end{tabular}
\end{center}
\caption{
(color online)
(a) The energy gap between Gutzwiller subbands, $\Delta \epsilon_{\rm disp}$ (\ref{Eq:DeltaEdisp}). 
(b) The bielectronic part $\Delta \epsilon_{\rm bi}$ of Eq. (\ref{Eq:DeltaEdisp}) as function of $U$ (meV)
for HF (dashed line), Gutzwiller with dynamical JT (solid line), and 
Gutzwiller solution with static JT distortion (points).
$\Delta \epsilon_{\rm bi}$ for the global minimum with the static JT effect is shown by the red dashed line.
}
\label{Fig:Edisp}
\end{figure}

\begin{figure}
\begin{center}
\includegraphics[bb = 0 0 3320 3256, width=8.6cm]{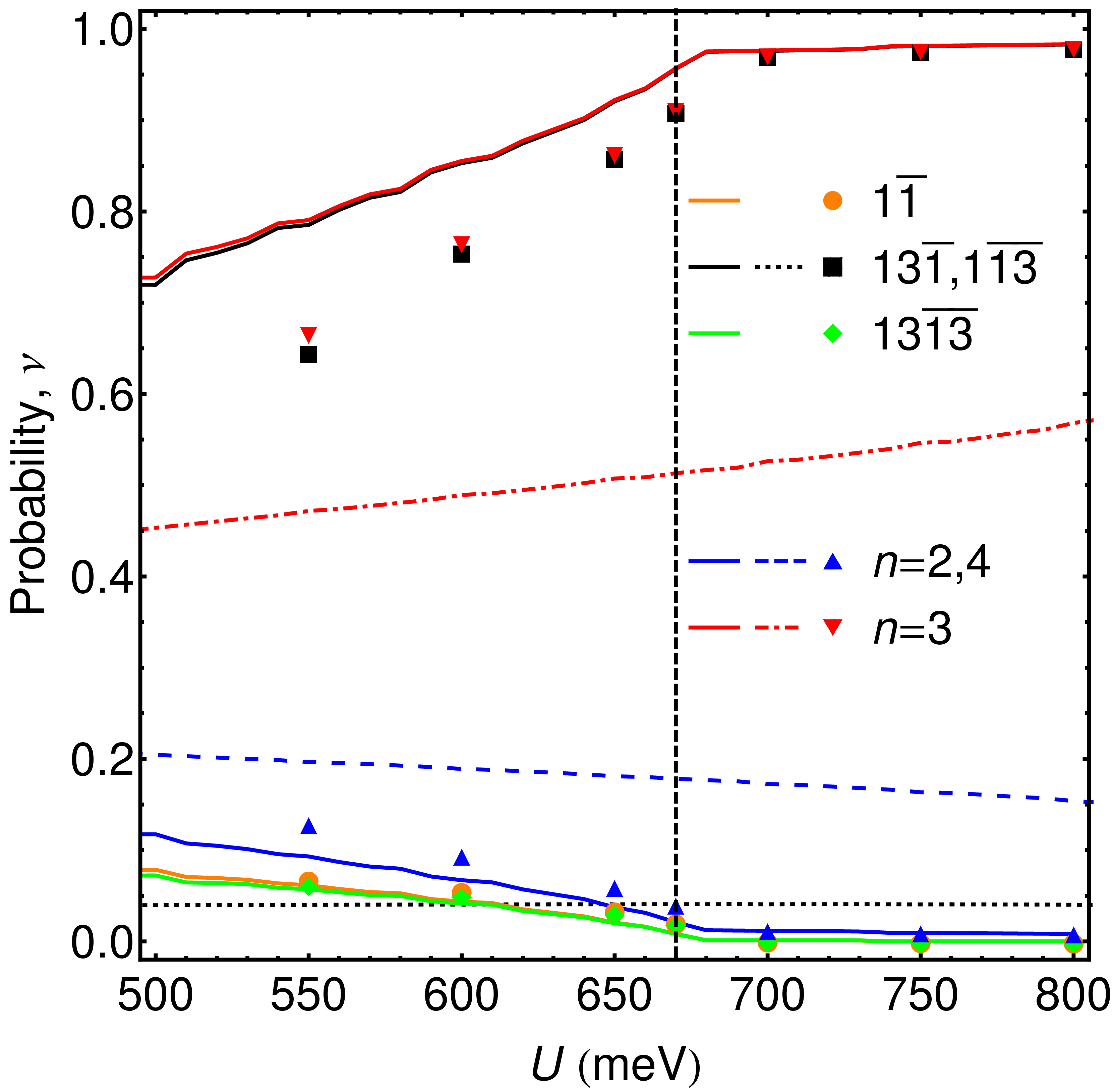}
\end{center}
\caption{(color online)
The probabilities of the electron configurations appearing in $\Psi_{\rm G}$, $\nu$,\cite{Vollhardt1984a,Ogawa1975a} as functions of $U$.
%The dotted lines, points, and solid lines correspond to the cubic, static JT, and 
%dynamical JT system, respectively.
Dotted, dashed and dot-dashed lines correspond to cubic symmetry band structure, 
solid lines correspond to dynamic JT effect, 
and symbols correspond to static JT effect. 
Black, blue, and red indicate the electron configuration $\psi_1^2\psi_2^1$, 
sum of $\nu$'s over $2$ (or $4$) electron configurations, 
and sum of $\nu$'s over $3$ electron configurations,
respectively.
The vertical dashed line indicates $U_c$ for the static JT instability.
$\nu$'s for $2$ and $4$ electrons are almost identical to each other
(only the data for $n = 2$ are shown).
$\nu$'s for $n=0,1,5,6$ are not shown here because they are close to zero.
}
\label{Fig:nu}
\end{figure}

%The manifestation of the 
Orbital disproportionation can be directly observed in spectroscopy, e.g., 
in photoemission spectra of fullerides.
%The excitation energy is %approximately evaluated from 
%approximated by 
Following the preceding discussion, the quasiparticles will belong to subbands with definite orbital index, 
$l=1,2,3$, separated by energy gaps (Fig. \ref{Fig:Edisp}a).
%On average, subband energy is expressed by 
%Neglecting the dispersion and correlation effect within subbands, their energy is
%expressed by the sum of the JT splitting and the Coulomb repulsion energy: % that each electron feels:
The centers of gravity of these subbands is expected to coincide with the centers of Gutzwiller subbands 
obtained as solutions of Eq. (\ref{Eq:step1}). 
The latter are expressed by the sum of the JT splitting and the Coulomb repulsion energy: 
\begin{eqnarray}
 \epsilon_{{\rm bi},l} &=&
 \sum_{l'\sigma'(\ne l \sigma)} U
 \frac{\langle \Psi|\hat{n}_{\mathbf{m} l\sigma} \hat{n}_{\mathbf{m} l'\sigma'}|\Psi\rangle}
 {\langle \Psi|\hat{n}_{\mathbf{m} l\sigma}|\Psi\rangle},
\label{Eq:epsilonbi}
\end{eqnarray}
where $\Psi$ is the ground state wave function.
Consequently, the energy gap between centers of weight of the subbands is expressed as:
\begin{eqnarray}
 \Delta \epsilon_{\rm disp} &=& \frac{3\hslash \omega g_{\rm eff}^2}{2} + \Delta \epsilon_{\rm bi}.
\label{Eq:DeltaEdisp}
\end{eqnarray}
The bielectronic part $\Delta \epsilon_{\rm bi}$ of Eq. (\ref{Eq:DeltaEdisp}) 
for broken-symmetry HF solution is given by $\Delta \epsilon_{\rm bi}^{\rm HF} = U/2$.
\cite{Chibotaru1996a, Ceulemans1997a}
$\Delta \epsilon_{\rm bi}$ for Gutzwiller wave function is calculated using Eq. (\ref{Eq:epsilonbix}).
%\begin{eqnarray}
% \Delta \epsilon_{\rm bi}^{\rm HF} &=& \frac{1}{2}\left(U_\perp - 3J_{\rm H}\right),
%\label{Eq:DeltaEbiHF}
%\end{eqnarray}
$\Delta \epsilon_{\rm bi}^{\rm HF}$ and $\Delta \epsilon_{\rm bi}$ 
for Gutzwiller's wave function are shown in Fig. \ref{Fig:Edisp}b.
$\Delta \epsilon_{\rm bi}^{\rm HF}$ monotonically increases with $U$, 
while $\Delta \epsilon_{\rm bi}$ for the Gutzwiller's solution approaches to zero.
The bielectronic contribution $\Delta \epsilon_{\rm bi}$ becomes zero 
because the system approaches to the isolated molecular limit:
when electrons are completely localized due to the metal-insulator transition, 
%the excitation energy corresponds to the JT splitting of isolated C$_{60}^{3-}$ ions.
the splitting of the subbands reduces to the JT splitting in isolated C$_{60}^{3-}$ ions.
%The difference between the two approaches originates from 
%the artifact of the broken-symmetry HF wave function.
%As shown above, the HF solution overestimates the stabilization of the occupied one-electron levels,
%leading to the large splitting of the LUMO band.
%%The molecular limit or metal-insulator transition occurs at $U \approx 700$ meV 
%%(Figs. \ref{Fig:nu} and \ref{Fig:Edisp}b). 
%The critical $U$ for the metal-insulator transition shown 
%in Figs. \ref{Fig:nu} and \ref{Fig:Edisp}b is discussed in Sec. \ref{Sec:Discussion}.
We can see from Fig. \ref{Fig:Edisp}b that $\Delta \epsilon_{\rm bi}$, while exaggerated in 
broken-symmetry HF approach, is not an artifactual feature but, on the contrary, gives a non-negligible
contribution to the splitting of quasiparticle subbands in the metallic phase. 
Figure \ref{Fig:nu} shows that the charge fluctuations (probabilities of configurations with $n=2,4$)
is suppressed at $U \alt$ 700 meV, signaling the arising of metal-insulator transition.

%This molecular limit is described by the additional disproportionation 
%of the multi determinantal wave function.
%$\Delta \epsilon_{\rm bi}$ for the Gutzwiller wave function with the static JTE
%is zero for $U < U_c = 670$ meV because the undistorted structure corresponds to the global minimum.
%The bielectronic contribution, the excitation energy between the subbands become 
%larger than the JT splitting.
%In the case of the Gutzwiller's wave function with static JTE, 
%the JT distortion and the orbital disproportionation is quenched for $U < U_c$ and arises for $U > U_c$.
%Therefore, the manifestation of the disproportionation is observed in the enlargement of the gap of the 
%subbaands by the bielectronic energy.

\begin{figure}
\begin{center}
\begin{tabular}{l}
(a) \\
\includegraphics[bb = 0 0 3320 2216, width=8.6cm]{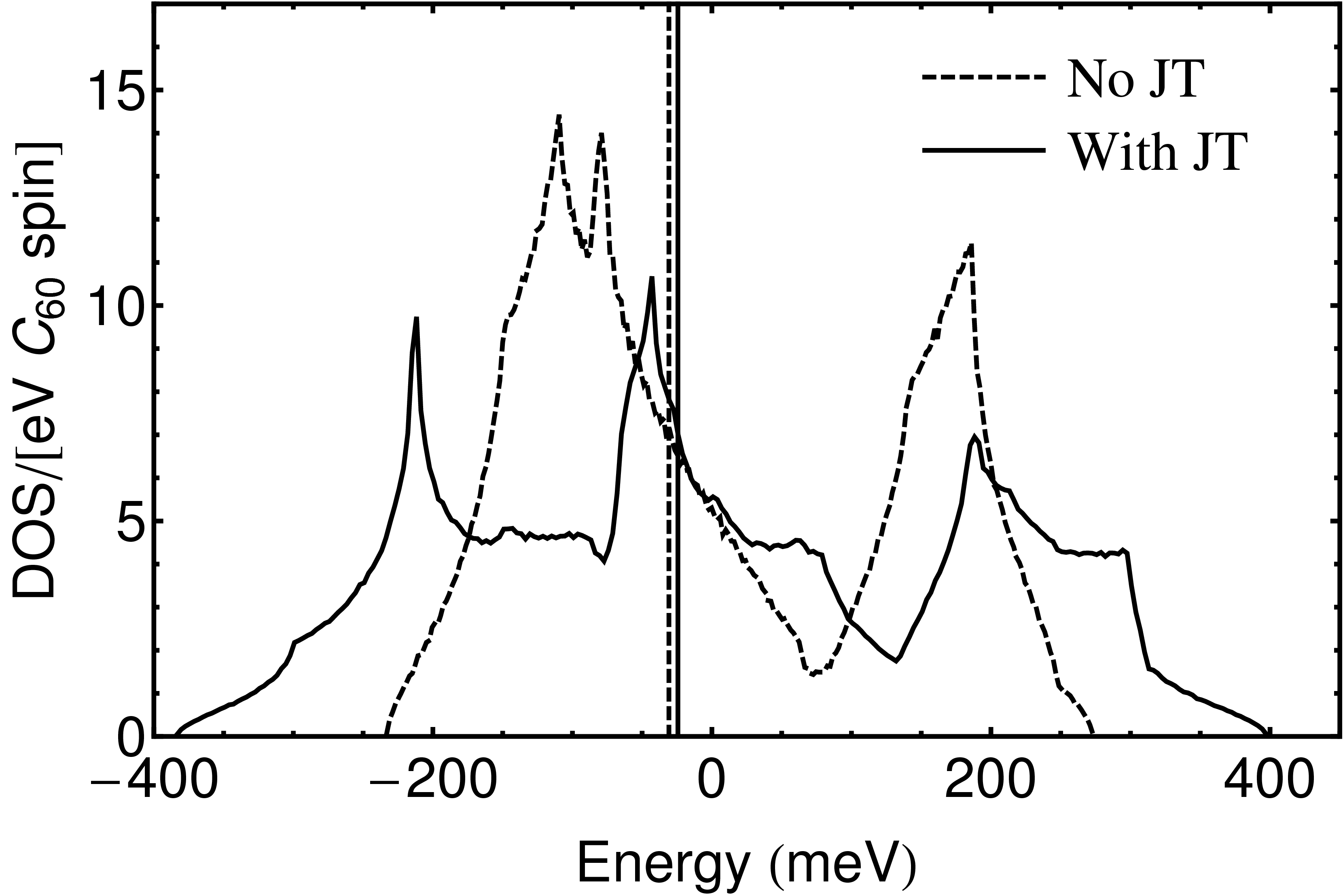}\\
(b) \\
\includegraphics[bb = 0 0 3320 2296, width=8.6cm]{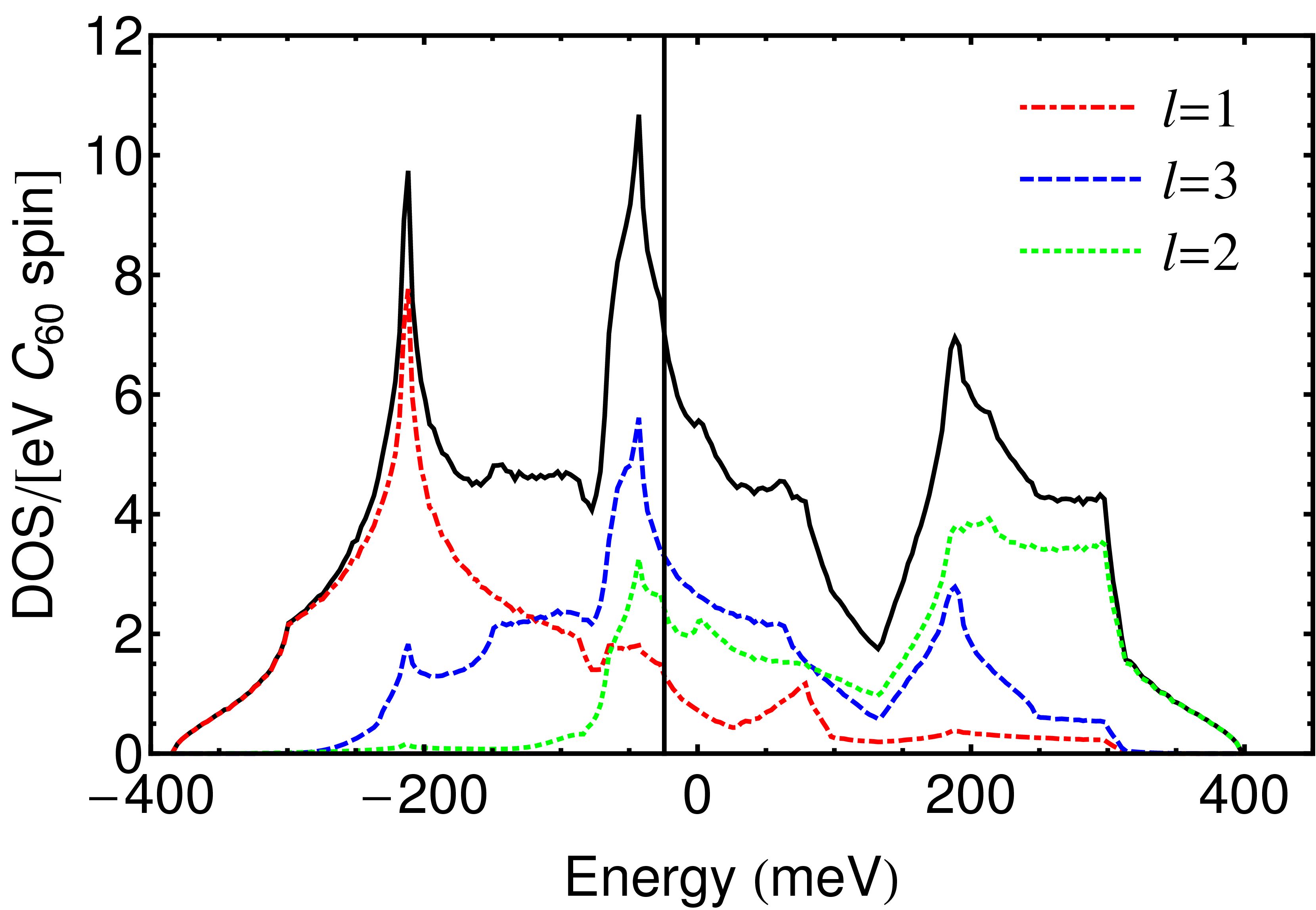}
\end{tabular}
\end{center}
\caption{
(color online)
(a) DOS's per C$_{60}$ and spin for the uncorrelated LUMO band of K$_3$C$_{60}$ in the absence (dashed) and the presence (solid) of equilibrium JT distortions on fullerene sites. 
(b) DOS with the JT distortion and partial DOS's corresponding to three adiabatic orbitals.
The vertical lines indicate Fermi levels.
}
\label{Fig:DOS}
\end{figure}

\subsection{Density of states of uncorrelated LUMO band}
\label{Sec:DOS}
It is also of interest to find out how the uncorrelated band structure is affected by JT instability. 
Figure \ref{Fig:DOS} shows the density of states (DOS) of the uncorrelated LUMO band in the presence of equilibrium JT distortion ($q=1.85$). 
Compared to cubic band structure, we see a strong enlargement of the bandwidth by ca 300 meV. 
The analysis of partial density of states shows that the degenerate LUMO band splits into three subbands (Fig. \ref{Fig:DOS}b)
mainly contributed by one of the adiabatic orbitals (these are $x$, $y$ and $z$ for the distortion (\ref{Eq:q})). 
This means that the electron correlation in fullerides does not take place in a degenerate LUMO band. 
%, as often stated. 
In particular, the Mott-Hubbard transition in cubic fullerides basically occurs 
in a split band structure, where half-filled is only the middle band.
This calls for reconsideration of the role played by orbital degeneracy in the Mott-Hubbard transition in fullerides.

\begin{figure}
\begin{center}
\begin{tabular}{l}
(a) \\
\includegraphics[bb = 0 0 3320 3320, width=8.6cm]{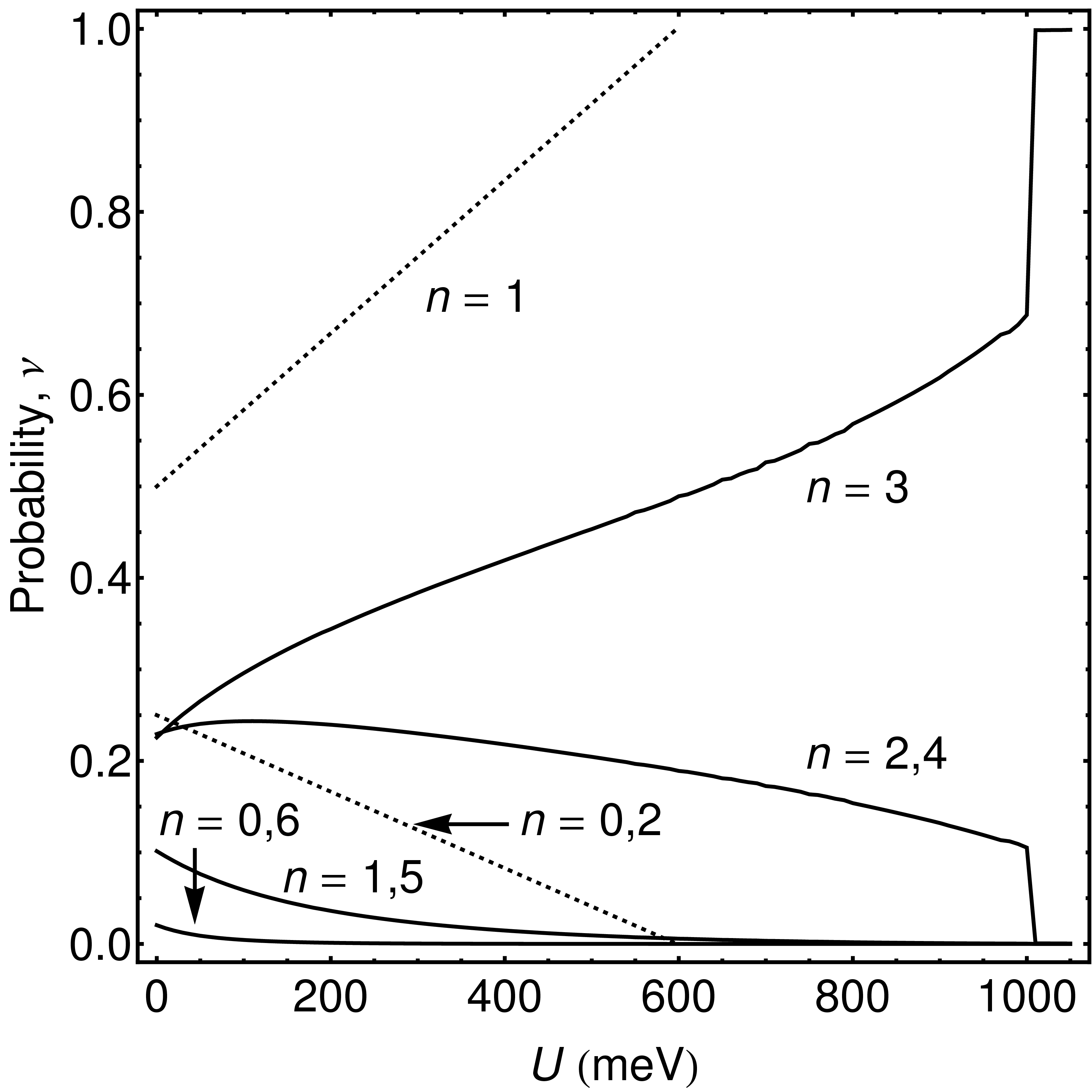}\\
(b) \\
\includegraphics[bb = 0 0 3320 2152, width=8.6cm]{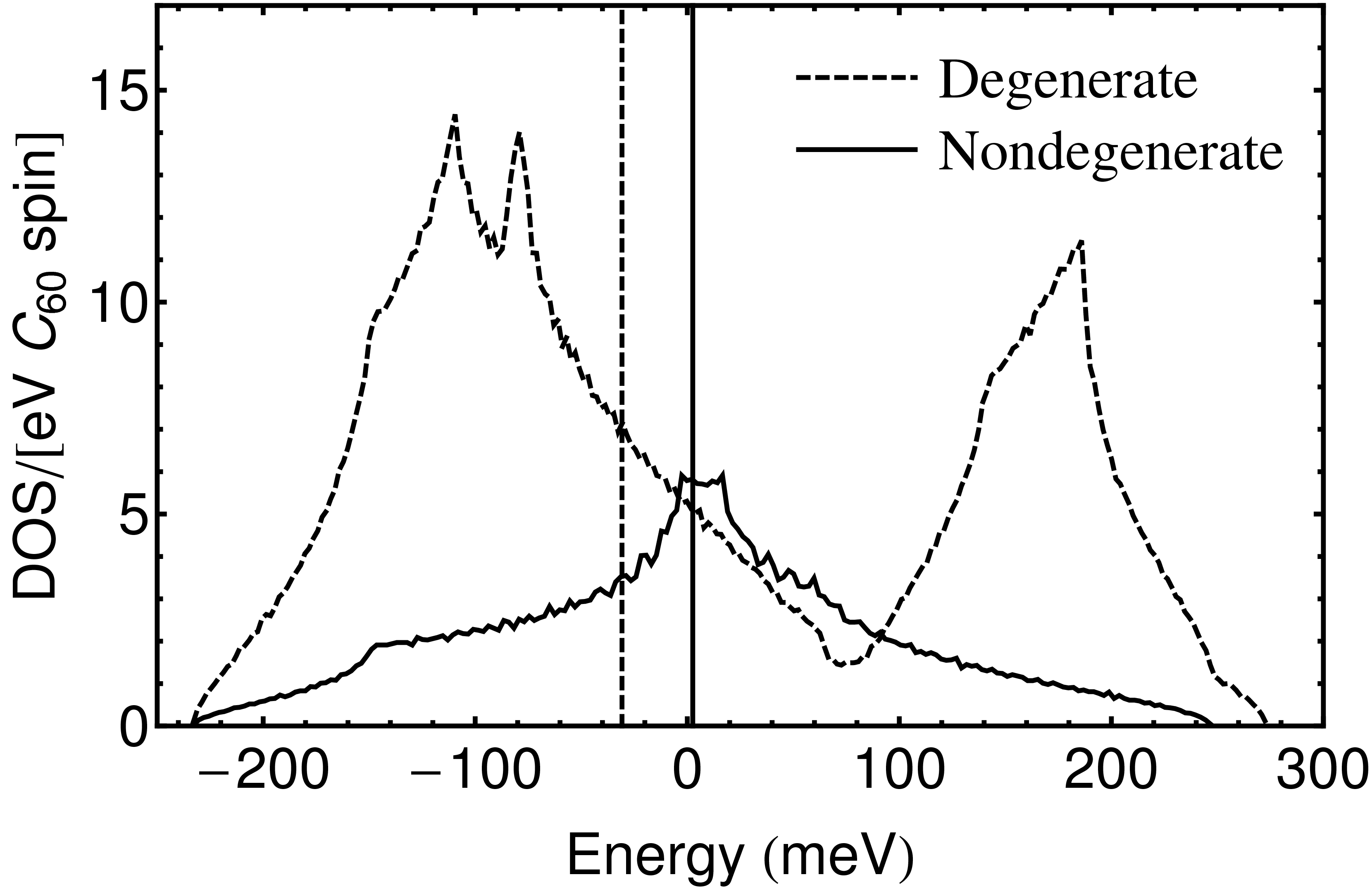}
\end{tabular}
\end{center}
\caption{
(a) The probabilities of the electron configurations appearing in Gutzwiller wave function of 
half-filled cubic system, $\nu$,\cite{Vollhardt1984a,Ogawa1975a} as functions of $U$.
The dashed lines and solid lines correspond to $\nu$'s for non-degenerate (one electron per site) 
and three-fold degenerate systems (three electrons per site), respectively.
(b) The DOS's for the non-degenerate (solid line) and degenerate (dashed line) systems.
Vertical lines show the position of the corresponding Fermi levels.
}
\label{Fig:nondeg}
\end{figure}
 
\section{Discussion and Conclusions}
\label{Sec:Discussion}
The vibronic interaction and the electron correlation in $A_3$C$_{60}$ are concomitantly treated 
%In order to concomitantly treat the JT interaction and the electron correlation,
by a new approach proposed here based on 
self-consistent Gutzwiller's ansatz with orbital-specific variational parameters.
%The orbital specific parameters enables different treatments 
%of the correlations in different subbands separated by the JT distortion.
The present Gutzwiller's calculations %include the on-site vibronic and bielectronic interactions 
with realistic vibronic constants, Hund's rule coupling 
and parameters of the LUMO band 
predict that both the static and the dynamical JT deformations arise in $A_3$C$_{60}$.
Since the electron correlation quenches the band energy, 
the localization of the electrons is enhanced, and consequently, the JT distortions on C$_{60}^{3-}$
sites is facilitated. 
It is shown that the dynamical JT instability appears 
for smaller on-site Coulomb repulsion, $U < 500$ meV than the static one (Fig. \ref{FigDJT}b).
Due to the existence of the dynamical JT distortion, 
the adiabatic LUMO band splits into three subbands (Fig. \ref{Fig:DOS}). 
%So far, the electron correlation in fullerides has been investigated based on the 
%degenerate band. \cite{Lu1994a, Gunnarsson1996a}
%Gunnarsson {\it et al.} considered that $N$-fold orbital degeneracy enhances the 
%number of paths of electron transfer and the band width by $\sqrt{N}$,
%leading to the increase of the critical $U$ for the metal-insulator transition, $U_c^{\rm MI}$.
%%The degenerate band has been considered to play a role to enhance the band energy, 
%%and thus the critical $U$ for the Mott-Hubbard transition.
%%Nonetheless, the JT effect and its role is lacking.
%Nonetheless, the dynamical JT deformation splits the LUMO band. 
%Therefore, the electron correlation should be taken into account based on the split band.
An indirect experimental evidence for the existence of dynamical JT effect in fullerides is
given by NMR spectroscopy of Cs$_3$C$_{60}$, showing that features attributed to 
dynamical Jahn-Teller effect in its insulating phase persist when this material is brought into 
metallic phase by applying an external pressure. \cite{arcon}

\subsection{Correlation in split bands}
\label{Sec:correlation}
The results of the present work do not support the established view that the electron correlation 
in fullerides takes place in a degenerate LUMO band.
As was shown by Gunnarsson {\it et al}., \cite{Gunnarsson2004a} 
Lu \cite{Lu1994a} and Han {\it et al}. \cite{Han1998a} the orbital degeneracy of the band/metal sites
increases the critical ratio $U/w$ for Mott-Hubbard metal-insulator transition where $w$ is 
the width of the band. 
Gunnarsson {\it et al}. has found that this ratio is $1.5-2.5$ for $A_3$C$_{60}$, which is 
significantly larger than the critical ratio $U/w \approx 1$ for Mott-Hubbard transition in lattices
with orbitally non-degenerate sites. \cite{Gunnarsson1996a, Gunnarsson2004a}
With the bandwidth $w \approx 0.5-0.6$ eV \cite{Gunnarsson2004a} (Fig. \ref{Fig:DOS}a)
and the estimated $U \approx 1.3-1.6$ eV \cite{Antropov1992a} 
it was natural to conclude that the orbital degeneracy of the LUMO band, 
leading to large critical values of $U/w$, is the reason for K$_3$C$_{60}$
and Rb$_3$C$_{60}$ to remain metals. \cite{Gunnarsson2004a}
This picture has become a basis for the interpretation of metal-insulator transition in fullerides, 
\cite{Gunnarsson2004a, Iwasa2003a} in particular, in Cs$_3$C$_{60}$. 
\cite{Ganin2008a, Takabayashi2009a, Ganin2010a}
Contrary to that, the JT-split correlated state derived here exhibits the Mott-Hubbard transition at a lower critical ratio $U/w$. 
Indeed, Fig. \ref{Fig:nu} shows that the probability $n$ for $n=2,4$ configurations goes to zero at 
$U >$ 700 meV, signaling the localization of electron on fullerene sites.
Thus we obtain a critical ratio $U/w = 1.4$ which is smaller than predicted for assumed perfectly 
degenerate LUMO band. \cite{Gunnarsson1996a, Gunnarsson2004a}
This, however, does not imply automatically an insulating state for K$_3$C$_{60}$ since the 
upper recent estimate for $U$ in this fulleride is 750 meV, \cite{nomura2012}
and the actual value can be significantly lower as discussed below (Sec. \ref{Sec:parameter}).
On the other hand, it would be incorrect to view the JT effect in the LUMO band
as simply leading to its enlarging (Fig. \ref{Fig:DOS}b) which increases the critical $U$
within (enlarged) single-band picture.
As a matter of fact, the electron correlation and 
the metal-insulator transition in fullerides develops mainly in the middle adiabatic subband.
%Since the subbands are separated from each other in the presence of the JTE, 
The role of the middle band in the Mott-Hubbard transition can be qualitatively reproduced by
single-band model.
The band energy of the single-band model, which includes only one of the $t_{1u}$ orbitals
(the corresponding DOS is shown in Fig. \ref{Fig:nondeg}a), is obtained as $E_{\rm t} = -74.7$ meV. 
Using the formula of the critical $U$ for non-degenerate band within Gutzwiller's approximation, 
\cite{Vollhardt1984a}
%we obtain $U_c^{\rm MI} = 8|E_{\rm t}| = 598$ meV,
%which is close to $U^{\rm MI}_c \approx 700$ meV for $A_3$C$_{60}$ obtained in the present work 
we obtain $U = 8|E_{\rm t}| = 598$ meV,
which is close to $U \approx 700$ meV for $A_3$C$_{60}$ obtained in the present work 
(Fig. \ref{Fig:nu}). 
The latter is larger than the estimate for the single-band model by about 100 meV
due to the remaining hybridization of the split bands.
From this analysis, one may conclude that the Mott-Hubbard transition mainly develops in the middle band.
%The same scenario may apply to low-symmetry fullerides such as intercalated (CH$_3$NH$_3$)K$_3$C$_{60}$. 
%In the fulleride, CH$_3$NH$_3$K$^+$ is hybridized with fullerene orbital, which 
%enhances the deformation of the C$_{60}^{3-}$. \cite{Potocnik2012a}
%Due to the larger JT deformation, the gap between the subbands is enlarged,
%hence the effect of the electron correlation becomes stronger in the middle band than $A_3$C$_{60}$.
%electron correlation becomes stronger in the middle band. 
%As a result, the low-symmetry fulleride is Mott-Hubbard insulator.

An important issue is the accuracy of the calculated ground state energy. 
We used here a six parameter Gutzwiller ansatz (\ref{Eq:pg}) in combination with Gutzwiller 
approximation for the calculation of total energy (\ref{Eq:Eg}). 
For comparison, Gunnarsson {\it et al}. \cite{Gunnarsson1996a}
used a single-parameter (conventional) Gutzwiller ansatz but calculated the total energy 
without approximation within variational Monte Carlo (VMC) approach.
Comparison with exact results obtained for small clusters of C$_{60}$ via exact diagonalization 
has shown that VMC reproduce the exact total energy with accuracy of 0.1 $\%$
(see Table 7.1 in Ref. \onlinecite{Gunnarsson2004a}), i.e., few meV of the total energy per one C$_{60}$
in fullerides.
The deviation from exact total energy will be certainly larger in the case of Gutzwiller approximation 
applied here, however, it will not cover the gain of the total energy due to JT splitting/orbital
disproportionation amounting many tens of meV (Fig. \ref{Fig:DJT}a).
Thus the main conclusion concerning dynamical JT instability in fullerides seems to be unaffected by this
approximation. This is further corroborated by the fact that the Gutzwiller wave function used in the present 
work is more flexible in variational sense than in the conventional Gutzwiller ansatz. 
Indeed, even in the case of degenerate LUMO band, conforming to cubic symmetry, the ansatz (\ref{Eq:pg})
involves two projection parameters. 
These are $A_{11}$ controlling the population of configurations 
$n_{\mathbf{m}\lambda \sigma} n_{\mathbf{m}\lambda -\sigma}$,
and $A_{12}$ controlling the population of configurations 
$n_{\mathbf{m}\lambda \sigma} n_{\mathbf{m}\lambda' \sigma'}$, $\lambda \ne \lambda'$.
That these are the only independent parameters allowed by the cubic symmetry can be understood if one 
generalizes the form (\ref{Eq:pg}) to arbitrary LUMO basis on fullerene sites. 
%(not coinciding with natural orbitals). 
%In this case $n_{\mathbf{m}\lambda \sigma}$ in Eq. (\ref{Eq:pg}) are replaced by elements of
%one-particle density matrix $\rho_{\mathbf{m}\alpha \beta \sigma}$, where $\alpha, \beta$ have 
%the sense of new directions of coordinate axes of fullerene sites (see Fig. \ref{Fig:xyz}).
Although, in general, $n_{\mathbf{m}\lambda \sigma}$ in Eq. (\ref{Eq:pg}) are replaced by elements of
one-particle density matrix, only the diagonal remains nonzero because of the cubic symmetry.
The terms in exponential of Eq. (\ref{Eq:pg}) then become of the form 
%$A_{\alpha \beta \gamma \delta} \rho_{\mathbf{m}\alpha \beta \sigma} \rho_{\mathbf{m}\gamma \delta \sigma'}$,
%in which the ``elasticity'' tensors $A_{\alpha \beta \gamma \delta}$ will be characterized by only 
%two independent parameters in the case of cubic symmetry. \cite{Landau1986a}
$A_{\alpha \alpha \beta \beta} n_{\mathbf{m}\alpha \sigma} n_{\mathbf{m}\beta \sigma'}$,
in which the ``elasticity'' tensors $A_{\alpha \alpha \beta \beta}$ will be characterized by only 
two independent parameters in the case of cubic symmetry. \cite{Landau1986a}
The second variational parameter in the Gutzwiller wave function changes drastically the description 
of Mott-Hubbard transition in the cubic band.
Thus, in a conventional single-parameter Gutzwiller ansatz within the Gutzwiller approximation
the critical $U/w = 4$ in the case of threefold orbital degeneracy of sites. \cite{Lu1994a}
This critical ratio is reduced to 2 in the case of Gutzwiller ansatz applied here (Fig. \ref{Fig:nondeg}b)
which is much closer to values obtained by Monte Carlo treatment. \cite{Gunnarsson1996a, Gunnarsson2004a}
It would be of interest to use in future the Gutzwiller ansatz for 
static and dynamic JT effect on sites proposed here as a trial functions 
in variational (VMC) and diffusion (projection) Monte Carlo (DMC) methods 
\cite{Gunnarsson2004a}
which give more accurate description of ground state energy.

\subsection{Parameters of the LUMO model}
\label{Sec:parameter}
Another important aspect concerns the values of relevant parameters for the model description of 
LUMO band in fullerides.
Given the large number of such parameters, their accurate knowledge is of primary importance
for realistic description of electronic properties of fullerides.
Recently, it was proven that the DFT calculated vibronic constants of C$_{60}^-$ with a hybrid B3LYP
functional compare well with those extracted from photoemission spectroscopy. \cite{Iwahara2010a}
Then the vibronic constants for the C$_{60}^{3-}$ anion and the exchange parameter
calculated within the same DFT functional should be reliable as well. 
Concerning the transfer Hamiltonian, the parameters of nearest-neighbor and next-nearest-neighbor
tight-binding model (Table \ref{Table:t}) reproduce well the dispersion of the LUMO bands calculated
within LDA (generalized gradient approximation, GGA) (Fig. \ref{Fig:band}). 
It was shown by GW calculation that the interband electronic interaction can 
enhance the LUMO ($t_{1u}$) bandwidth in fullerites by 30 \% \cite{Shirley1993a}
while the intraband interaction reduces the LUMO bandwidth in $A_3$C$_{60}$. \cite{Gunnarsson1997b}
However, in the latter case the GW approximation is ill-defined due to strong correlation effects
in the LUMO band. \cite{Aryasetiawan2004a, Gunnarsson2004a}

The parameter assessed with less certainty in the model Hamiltonian (\ref{Eq:H})-(\ref{Eq:Hbi})
is the intra-fullerene electron repulsion $U_\parallel$. 
Recent calculations of this parameter by constrained random-phase approximation (cRPA) 
\cite{Aryasetiawan2004a} with GGA band energies and wave functions give in the low-frequency limit 
$U_\parallel \alt $ 1 eV for the series of $A_3$C$_{60}$, the smallest being $U_\parallel = 820$ meV
for K$_3$C$_{60}$. \cite{nomura2012}
It is interesting to note that the estimated $U_\parallel$ in Ref. \onlinecite{nomura2012} 
gives for all fullerides values $\alt$ 1 eV, while former estimations made for fullerite 
(pure C$_{60}$ crystal) give larger values. \cite{Gunnarsson2004a, Antropov1992a, Lof1992a}
This is explained by the fact that in $A_3$C$_{60}$ fullerides 
the LUMO Wannier orbitals occupy larger volume due to hybridization with alkali atoms. \cite{nomura2012}
In these calculations a non-interacting polarization function was used that excluded polarization 
processes within the LUMO band.
Including the latter, i.e., considering the full screening in the non-interacting metallic regime, 
\cite{Aryasetiawan2004a} further reduces $U_\parallel$ by ca one order of magnitude in the low
frequency limit. \cite{nomura2012}
A similar strong effect of metallic screening (arising from the LUMO band) was predicted also for a 
simplified treatment of metallic polarization. \cite{Chakravarty1991a, Lammert1995a}
One should note that RPA is generally not expected to perform well in the limit of strong correlation, 
as well as the GW approximation mentioned above.
To check the screening capability of the correlated LUMO band, Koch {\it et al}. \cite{Koch1999a}
did VMC and DMC calculations of an induced charge arising in response to a test charge for a threefold 
degenerate LUMO model of $A_3$C$_{60}$.
They found that RPA performs surprisingly well till $U/w \alt$ 2, and even at the point of 
Mott-Hubbard transition (corresponding to $U/w \approx 2.5$ in their model), when the LUMO electrons
become localized, the screening charge is reduced by not more than 40 \% with respect to RPA
screening charge calculated for non-correlated LUMO band (Fig. 2 in Ref. \onlinecite{Koch1999a}). 
At the same extent is expected to be reduced the screening of the electron repulsion parameter
by the intra-LUMO band interaction, which means that this screening is significant in the entire metallic
phase of fullerides and should be taken into account for realistic assessment of $U_\parallel$. 
A rigorous model for the LUMO band, in which other degrees of freedom are excluded, requires
frequency-dependent electron repulsion parameters. \cite{Aryasetiawan2004a}
From it an effective static Hubbard model 
(involving a frequency-independent $U$) can be derived by fitting the 
self-energy in a low-frequency domain, not exceeding the width of uncorrelated LUMO band. 
\cite{Aryasetiawan2004a}
%The resulting static $U$ will basically correspond to intra LUMO band screening of low-frequency 
%interaction screened by other bands, obtained from cRPA interaction. 
%\cite{Aryasetiawan2004a, nomura2012}
%Clearly, at this stage the strength of correlation and JT splitting in the LUMO band will influence the 
%extent of intraband screening and, therefore, the value of $U$. 
%This means that, starting with partially screened interaction $W_r(\omega)$ 
%\cite{Aryasetiawan2004a}
%obtained in preliminary cRPA calculations, 
Note that the derivations of the static $U_\parallel$, $U_\perp$ in Eq. (\ref{Eq:Hbi})
should be done self-consistently with the derivation of the ground state following the iterative equations 
(\ref{Eq:step1})-(\ref{Eq:step2}). 

Experimentally $U$ can be assessed from Auger spectroscopy. \cite{Lof1992a, Bruhwiller1993a}
The estimates are 1.4 $\pm$ 0.2 eV for pure C$_{60}$ and band insulator K$_6$C$_{60}$,
\cite{Bruhwiller1993a} %, comment1} 
\footnote{
It was mentioned that Auger spectroscopy mostly probes near-surface layers, \cite{Bruhwiller1993a}
while bulk values of $U$ should be reduced by ca $0.2-0.3$ eV.
}
and 0.6 $\pm$ 0.3 eV for metallic K$_3$C$_{60}$. 
The much smaller value of $U$ in K$_3$C$_{60}$ reflects most probably the additional 
strong screening from half-filled LUMO band. 
\footnote{The authors of Ref. \onlinecite{Bruhwiller1993a} disregard the possibility of metallic screening
due to the LUMO band and interpret the Auger spectrum of K$_3$C$_{60}$ on the basis of difference between 
intra-LUMO ($U$) and core ($1s$)-LUMO ($U_{\rm core}$) electron repulsion parameters. 
However, they did not explain why $U$ and $U_{\rm core}$ should be different in K$_3$C$_{60}$
and have the same value in pure C$_{60}$ and K$_6$C$_{60}$.}
An instructive example of the sensitivity of $U$ on intra LUMO-band screening is offered by non-cubic 
fullerides NH$_3$K$_3$C$_{60}$ \cite{Rosseinsky1993a} and (CH$_3$NH$_2$)K$_3$C$_{60}$. \cite{Ganin2006a}
Contrary to the parent K$_3$C$_{60}$ fulleride, which is a metal, these compounds are antiferromagnetic 
Mott-Hubbard insulators. 
The main effect of spacers, NH$_3$ and CH$_3$NH$_2$, respectively, is the removal of degeneracy of three LUMO
orbitals on fullerene sites, which apparently reduces the orbital $U/w$ from its value in cubic K$_3$C$_{60}$
predicted for threefold degenerate LUMO band, \cite{Gunnarsson1996a, Gunnarsson2004a}
causing the Mott-Hubbard transition. 
Manini {\it et al}. \cite{Manini2002a} have checked this possibility via DMFT calculations
of a model twofold degenerate band and found that the calculated splitting of the $t_{1u}$ orbitals 
is indeed sufficient to induce the Mott-Hubbard transition. 
\footnote{
This splitting was identified with the splitting of the LUMO bands at the $\Gamma$-point
obtained via LDA calculations of NH$_3$K$_3$C$_{60}$ \cite{Manini2002a} and (CH$_3$NH$_2$)K$_3$C$_{60}$.
\cite{Potocnik2012a}
In both cases the obtained splitting is smaller than the JT splitting of the orbitals in K$_3$C$_{60}$, 
which means that LDA calculation do not grasp at all or seriously underestimate the JT effect in 
these fullerides, which was also the case in other similar calculations. \cite{Capone2000a}
}
However, the persistence of strong JT distortions in metallic fullerides, established in this work, 
calls for another interpretation. 
The crystal anisotropy induced by the spacers will enhance the splitting of the LUMO bands 
(Fig. \ref{Fig:DOS}b) thus reducing the intra-LUMO band screening of electron repulsion.
This results in the increase of $U_\parallel$ and $U_\perp$ in Eq. (\ref{Eq:Hbi}), which is the 
reason why the non-cubic fullerides are Mott-Hubbard insulators.

To conclude this part, several theoretical arguments and relevant experimental data argue in the 
favor of non-negligible intra-LUMO band screening of Hubbard $U$ parameter, which is thus expected
to be well below 1 eV. 
The latter is also a necessary condition for metallicity of fullerides in the presence of strong
JT distortions (Fig. \ref{Fig:nu}).

\subsection{Polaronic effects}
\label{Sec:polaron}
The simple form of dynamical vibronic wave function (\ref{Eq:G+DJT}) was derived under two simplifying 
assumptions.
First, the polaronic effect was neglected which seems to be justified for fullerides. 
%The polaronic effect is expected to be small in fullerides. 
Indeed, while the static JT energy of C$_{60}^{2-}$ and C$_{60}^{4-}$ is larger than 
in C$_{60}^{3-}$ by an amount $E_{\rm JT}^{(1)} = \hslash \omega g^2/2 = 50.2$ meV, 
the difference in the gain is compensated by the loss of stabilization energy from the dynamical contribution.
In the strong vibronic coupling limit, the dynamical JT contribution of C$_{60}^{3-}$ is $3\hslash \omega/2$
and that of C$_{60}^{2-}$ and C$_{60}^{4-}$ is $\hslash \omega$
because the trough is three-dimensional in the former case and two-dimensional in the latter.
Then $1/3$ of the dynamical contribution of C$_{60}^{3-}$ is lost 
if JT relaxation accompanies the electron transfer. 
Using the data from the numerical diagonalizations, % (\ref{Fig:DJT}), 
the loss of the dynamical JT contribution is estimated as $-E_{\rm DJT}/3 \approx 30$ meV.
Therefore, the binding energy of JT polaron (the energy gain arising from full JT relaxation)
%fluctuation of the numbers of electrons in C$_{60}^{3-}$ is accompanied by the gain 
is $\Delta E = -E_{\rm JT}^{(1)} - E_{\rm DJT}/3 \approx -20$ meV.
%Due to the gain, the probabilities of the configurations with two or four electrons increase,
%%which leads to the larger critical $U_c^{\rm MI}$.
%however, 
Compared with the total JT stabilization energy of $\approx 240$ meV the JT polaronic effect 
appears to be small. 
One should take into account that the JT polaronic effect 
%also gives rise to vibronic reduction\cite{Bersuker1989a} of the transfer parameters:
is accompanied by the Franck-Condon reduction of the band energy, which means that the JT polaron 
will only show up when the band energy is reduced by correlation effects
under 20 meV, i.e., close to Mott-Hubbard transition. 
On the other hand the stabilization energy of one electron after total symmetric fullerene distortions 
does not exceed 20 meV, i.e., is negligible either. \cite{Iwahara2010a}
In these estimations the relaxation due to displacements of alkali atoms has not been included, 
which is unimportant for $A_3$C$_{60}$ but can be significant in insulating $A_4$C$_{60}$ and 
A$_6$C$_{60}$. \cite{Wehrli2004a}
As for the second assumption of weak hybridization of the bands belonging to different
$t_{1u}$ orbitals (Fig. \ref{Fig:twosite}), 
it seems to be only justified in the strongly correlated limit.
When it is not the case, the $\Omega$-dependence of the coefficients $U_{\mathbf{m}li}$ 
in Eq. (\ref{Eq:a+}) cannot be neglected and ultimately the rotations of JT deformation on 
different fullerene sites (Eq. (\ref{Eq:chirot})) cannot be separated. 
This means that in metallic Cs$_3$C$_{60}$ the rotation of JT deformations occurs independently 
on different fullerene sites, while in K$_3$C$_{60}$ these rotations are more probable to be correlated. 
In the latter case the wave function (\ref{Eq:G+DJT}) does not represent a close solution and should
be rather considered as a variational function which nevertheless will correspond to lower total energy than the static JT solution (\ref{Eq:gwf}).

\subsection{Summary}
The main achievements of this work can be summarized as follows:
\begin{enumerate}
\item We have developed an approach for the investigation of correlated JT metals
based on self-consistent Gutzwiller approximation.
\item The concomitant treatment of JT effect and electron correlation in metallic fullerides 
$A_3$C$_{60}$ proves the existence of dynamical JT instability in their ground state. 
The JT distortions arise due to strong reduction of the band energy by electron correlation 
effects and achieve an amplitude close to the value in a free C$_{60}^{3-}$ ion. 
\item The JT instability induces strong overall enlargement of the uncorrelated LUMO band and its splitting 
in three components corresponding to individual adiabatic orbitals on fullerene sites. 
The results call for reconsideration of the role played by orbital degeneracy in the physics 
of metallic fullerides.
\item JT distortions together with electron correlation induce disproportionation of 
electron density between subbands corresponding to different adiabatic orbitals on fullerene sites.
Besides the JT splitting there is also a bielectronic contribution to the separation of these
subbands which vanishes in the limit of strong correlation. 
Importantly, the orbital disproportionation does not exist as a pure electronic low-symmetry instability 
in the absence of JT effect on fullerene sites ($g=0$), in which case the correlated LUMO band will have
a perfect cubic symmetry for any $U$.
\end{enumerate}
Finally, we note that a similar analysis can be applied to other correlated metals with JT active sites.

%The JTE and the kinetic energy is treated concomitantly on the basis of the 
%self-consistent method within Gutzwiller's ansatz.
%The static JT distortion is enhanced by the on-site Coulomb coupling because both of the 
%couplings reduce the kinetic energy.
%In $A_3$C$_{60}$, 
%the change of the kinetic energy accompanied by the rotation of the JT deformation 
%is smaller than the lowest vibronic excitation, thus the deformation is freely rotating.
%The dynamical JT stabilization enhances the deformation and the orbital disproportionation.
%The energy band splits into three subbands due to the JTE.
%The main contribution to the kinetic energy is from the middle subband
%unlike the case of the degenerate band.
%The evidence of the dynamical JTE in metallic fullerides will be found in 
%various spectra such as photoelectron spectra and optical conductivity.

%%%%%%%%%%%%%%%%%%%%%%%%%%%%%%%%%%%%%%%%%%%%%%%%%%%%%%%%%%%%%
%
% Acknowledgement
%
%%%%%%%%%%%%%%%%%%%%%%%%%%%%%%%%%%%%%%%%%%%%%%%%%%%%%%%%%%%%%
\section*{Acknowledgment}
N. I. would like to acknowledge the financial support from the Flemish Science Foundation (FWO)
and the GOA grant from KU Leuven.
We would like to thank Denis Ar\v{c}on for useful discussions.

%%%%%%%%%%%%%%%%%%%%%%%%%%%%%%%%%%%%%%%%%%%%%%%%%%%%%%%%%%%%%
%
% Appendix
%
%%%%%%%%%%%%%%%%%%%%%%%%%%%%%%%%%%%%%%%%%%%%%%%%%%%%%%%%%%%%%

\appendix
\section{Tight-binding parametrization of the LUMO band structure of K$_3$C$_{60}$}
\label{Sec:Ht}
We assume that all C$_{60}$'s in fcc K$_3$C$_{60}$ lattice are equally orientated 
in a similar fashion shown in Figure \ref{Fig:A1} (Fig. \ref{Fig:fccK3C60}). 
Using the unit vectors of fcc lattice, 
\begin{eqnarray}
 \mathbf{a}_1 &=& \frac{a}{2} \left(\mathbf{e}_y + \mathbf{e}_z\right), \qquad
 \mathbf{a}_2 = \frac{a}{2} \left(\mathbf{e}_z + \mathbf{e}_x\right), 
\nonumber\\
 \mathbf{a}_3 &=& \frac{a}{2} \left(\mathbf{e}_x + \mathbf{e}_y\right),
\end{eqnarray}
the displacements $\Delta \mathbf{m}$ of the nearest neighbor sites from site $\mathbf{m}$ 
are written as
\begin{eqnarray}
 \Delta \mathbf{m} &=& 
 \left(
  \mathbf{a}_3, \mathbf{a}_1 - \mathbf{a}_2, -\mathbf{a}_3, -(\mathbf{a}_1 - \mathbf{a}_2), 
 \right.
\nonumber\\
 &&
 \left.
  \mathbf{a}_1, \mathbf{a}_2 - \mathbf{a}_3, -\mathbf{a}_1, -(\mathbf{a}_2 - \mathbf{a}_3), 
 \right.
\nonumber\\
 &&
 \left.
  \mathbf{a}_2, \mathbf{a}_3 - \mathbf{a}_1, -\mathbf{a}_2, -(\mathbf{a}_3 - \mathbf{a}_1)
 \right).
\end{eqnarray}
The next nearest neighbors are displaced by vectors
\begin{eqnarray}
 \Delta \mathbf{m} &=& 
 a \left(
  \mathbf{e}_x, -\mathbf{e}_x, \mathbf{e}_y, -\mathbf{e}_y, \mathbf{e}_z, -\mathbf{e}_z
 \right).
\end{eqnarray}
Here, $a$ is the lattice constant of a simple cubic lattice and
$\mathbf{e}_x, \mathbf{e}_y, \mathbf{e}_z$ correspondingly are unit vectors 
directed along tetragonal $x, y, z$ axes (Fig. \ref{Fig:fccK3C60}).
The tight-binding Hamiltonian has the form:
\begin{eqnarray}
 \hat{H}_{\rm t} &=& \sum_\mathbf{m}\sum_{\lambda \sigma} 
 \epsilon \hat{n}_{\mathbf{m}\lambda \sigma} 
 + \sum_{\mathbf{m}} \sum_\sigma 
 \left(\hat{H}_{\mathbf{m}\sigma}^{\rm nn} + \hat{H}_{\mathbf{m}\sigma}^{\rm nnn} \right), 
\end{eqnarray}
where the nearest-neighbor part is 
%\begin{widetext}
\begin{eqnarray}
 \hat{H}_{\mathbf{m} \sigma}^{\rm nn} &=& 
  t_1
  \left[ \sum_{i=1}^4 \hat{c}_{\mathbf{m}+\Delta \mathbf{m}_i x \sigma}^\dagger \hat{c}_{\mathbf{m}x\sigma} 
   + \sum_{i=5}^8     \hat{c}_{\mathbf{m}+\Delta \mathbf{m}_i y \sigma}^\dagger \hat{c}_{\mathbf{m}y\sigma} 
  \right.
\nonumber\\
  &+&
  \left.
     \sum_{i=9}^{12}  \hat{c}_{\mathbf{m}+\Delta \mathbf{m}_i z \sigma}^\dagger \hat{c}_{\mathbf{m}z\sigma} 
  \right]
\nonumber\\
 &+& t_3
  \left[ \sum_{i=1}^4 \hat{c}_{\mathbf{m}+\Delta \mathbf{m}_i y \sigma}^\dagger \hat{c}_{\mathbf{m}y\sigma} 
   + \sum_{i=5}^8     \hat{c}_{\mathbf{m}+\Delta \mathbf{m}_i z \sigma}^\dagger \hat{c}_{\mathbf{m}z\sigma} 
  \right.
\nonumber\\
  &+&
  \left.
     \sum_{i=9}^{12}  \hat{c}_{\mathbf{m}+\Delta \mathbf{m}_i x \sigma}^\dagger \hat{c}_{\mathbf{m}x\sigma} 
 \right]
\nonumber\\
 &+& t_4
  \left[ \sum_{i=1}^4 \hat{c}_{\mathbf{m}+\Delta \mathbf{m}_i z \sigma}^\dagger \hat{c}_{\mathbf{m}z\sigma} 
   + \sum_{i=5}^8     \hat{c}_{\mathbf{m}+\Delta \mathbf{m}_i x \sigma}^\dagger \hat{c}_{\mathbf{m}x\sigma} 
  \right.
\nonumber\\
  &+&
  \left.
     \sum_{i=9}^{12}  \hat{c}_{\mathbf{m}+\Delta \mathbf{m}_i y \sigma}^\dagger \hat{c}_{\mathbf{m}y\sigma} 
 \right]
\nonumber\\
 &-& t_2
  \left[ 
     \sum_{i=1}^4 (-1)^{i} 
     \left(\hat{c}_{\mathbf{m}+\Delta \mathbf{m}_i x \sigma}^\dagger \hat{c}_{\mathbf{m}y\sigma} 
     \hat{c}_{\mathbf{m}+\Delta \mathbf{m}_i y \sigma}^\dagger \hat{c}_{\mathbf{m}x\sigma} \right)
\right.
 \nonumber\\
  &+&
\left.
     \sum_{i=5}^8 (-1)^{i}
     \left(\hat{c}_{\mathbf{m}+\Delta \mathbf{m}_i y \sigma}^\dagger \hat{c}_{\mathbf{m}z\sigma} 
   + \hat{c}_{\mathbf{m}+\Delta \mathbf{m}_i z \sigma}^\dagger \hat{c}_{\mathbf{m}y\sigma} \right)
\right.
 \nonumber\\
  &+&
\left.
     \sum_{i=9}^{12} (-1)^{i}
     \left(\hat{c}_{\mathbf{m}+\Delta \mathbf{m}_i z \sigma}^\dagger \hat{c}_{\mathbf{m}x\sigma} 
   + \hat{c}_{\mathbf{m}+\Delta \mathbf{m}_i x \sigma}^\dagger \hat{c}_{\mathbf{m}z\sigma} \right)
 \right],
\nonumber\\
\label{Eq:Htnn}
\end{eqnarray}
and the next-nearest-neighbor part is
\begin{eqnarray}
 \hat{H}_{\mathbf{m}\sigma}^{\rm nnn} &=& 
  t_5\left(
  \hat{c}_{\mathbf{m}+a\mathbf{e}_x x \sigma} \hat{c}_{\mathbf{m}x\sigma} 
  + \hat{c}_{\mathbf{m}-a\mathbf{e}_x x \sigma} \hat{c}_{\mathbf{m}x\sigma}
\right.
 \nonumber\\
  &+&
\left.
    \hat{c}_{\mathbf{m}+a\mathbf{e}_y y \sigma} \hat{c}_{\mathbf{m}y\sigma} 
  + \hat{c}_{\mathbf{m}-a\mathbf{e}_y y \sigma} \hat{c}_{\mathbf{m}y\sigma}
\right.
 \nonumber\\
  &+&
\left.
    \hat{c}_{\mathbf{m}+a\mathbf{e}_z z \sigma} \hat{c}_{\mathbf{m}z\sigma}
  + \hat{c}_{\mathbf{m}-a\mathbf{e}_z z \sigma} \hat{c}_{\mathbf{m}z\sigma}
  \right)
 \nonumber\\ 
 &+&
  t_6\left( \hat{c}_{\mathbf{m}+a\mathbf{e}_x y \sigma} \hat{c}_{\mathbf{m}y\sigma}
  + \hat{c}_{\mathbf{m}-a\mathbf{e}_x y \sigma} \hat{c}_{\mathbf{m}y\sigma}
\right.
 \nonumber\\
  &+&
\left.
    \hat{c}_{\mathbf{m}+a\mathbf{e}_y z \sigma} \hat{c}_{\mathbf{m}z\sigma} 
  + \hat{c}_{\mathbf{m}-a\mathbf{e}_y z \sigma} \hat{c}_{\mathbf{m}z\sigma}
\right.
 \nonumber\\
  &+&
\left.
    \hat{c}_{\mathbf{m}+a\mathbf{e}_z x \sigma} \hat{c}_{\mathbf{m}x\sigma} 
  + \hat{c}_{\mathbf{m}-a\mathbf{e}_z x \sigma} \hat{c}_{\mathbf{m}x\sigma}
  \right)
\nonumber\\
 &+&
  t_7\left( \hat{c}_{\mathbf{m}+a\mathbf{e}_x z \sigma} \hat{c}_{\mathbf{m}z\sigma} 
 + \hat{c}_{\mathbf{m}-a\mathbf{e}_x z \sigma} \hat{c}_{\mathbf{m}z\sigma}
\right.
 \nonumber\\
  &+&
\left.
    \hat{c}_{\mathbf{m}+a\mathbf{e}_y x \sigma} \hat{c}_{\mathbf{m}x\sigma} 
  + \hat{c}_{\mathbf{m}-a\mathbf{e}_y x \sigma} \hat{c}_{\mathbf{m}x\sigma}
\right.
 \nonumber\\
  &+&
\left.
    \hat{c}_{\mathbf{m}+a\mathbf{e}_z y \sigma} \hat{c}_{\mathbf{m}y\sigma} 
  + \hat{c}_{\mathbf{m}-a\mathbf{e}_z y \sigma} \hat{c}_{\mathbf{m}y\sigma}
  \right),
\label{Eq:Htnnn}
\end{eqnarray}
%\end{widetext}
In Eq. (\ref{Eq:Htnn}), 
$\Delta \mathbf{m}_i$ indicates $i$th nearest neighbor. 
%and $\epsilon$ is the LUMO level.
%Fitting the LUMO band obtained by density functional theory (DFT) calculation to the model (Fig. S1), we derive LUMO levels $\epsilon$ 
%and transfer parameters (Table S1).
%For the calculations described in the main text, we put $\epsilon = 0$.

The DFT calculation of the band structure of K$_3$C$_{60}$ 
was performed using {\sc Quantum ESPRESSO 3.0} package
with the pseudopotentials C.pbe-mt gipaw.UPF and K.pbe-mt fhi.UPF. \cite{QE-2009}
The lattice constant of K$_3$C$_{60}$ was taken from Ref. \onlinecite{Stephens1991a} 
and the structure of C$_{60}$ of Ref. \onlinecite{Iwahara2010a} was used.

\begin{figure}[tb]
\begin{center}
 \includegraphics[bb = 0 0 3320 2240, width=8.6cm]{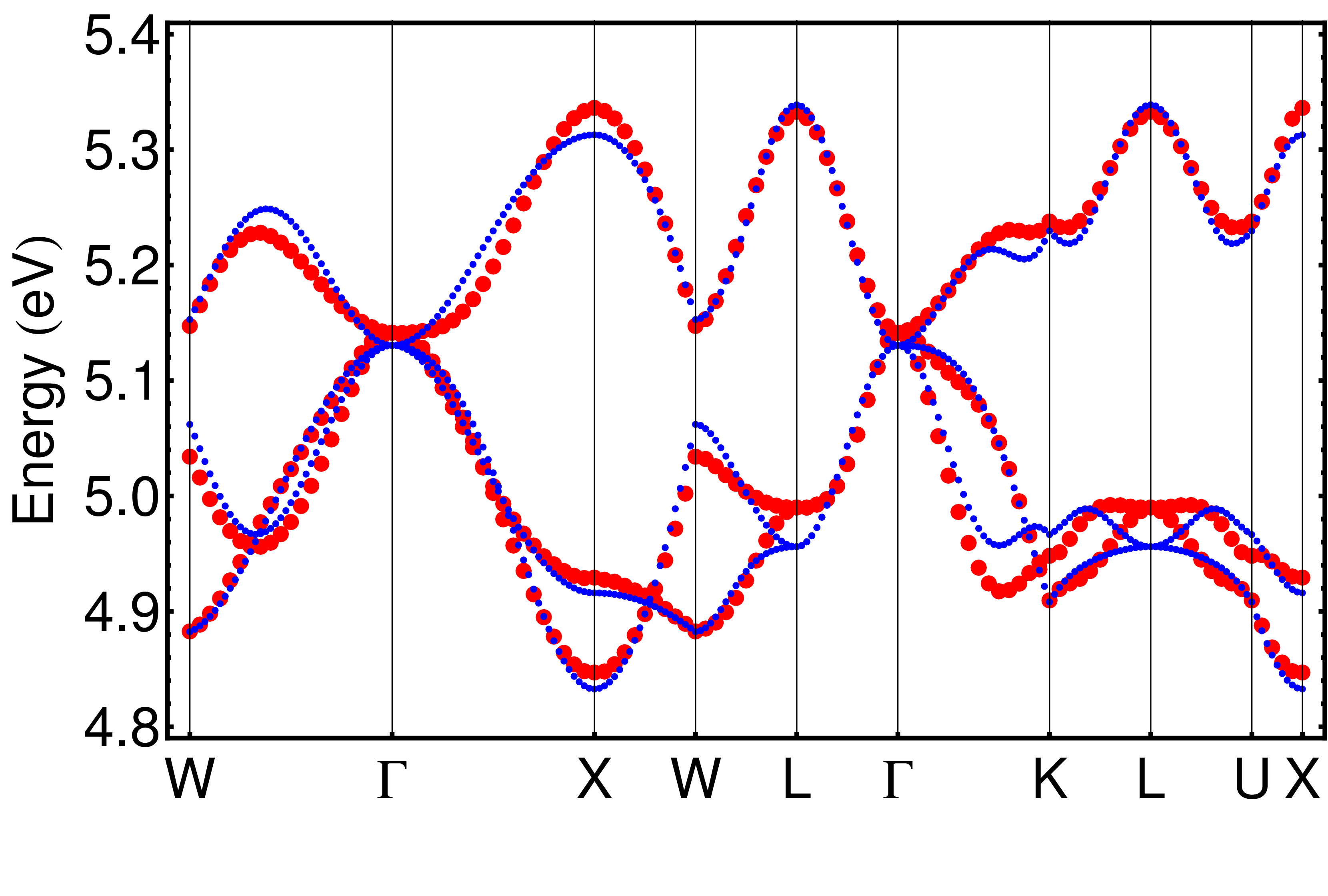}
\end{center}
\caption{(color online) Band structure of fcc K$_3$C$_{60}$ in eV. 
The red and blue points are obtained from DFT calculations and model Hamiltonian, respectively.
The symmetric points $(k_x, k_y, k_z)$ are as follows:
$\Gamma = (0,0,0)$, $X = (2\pi/a,0,0)$, $W = (2\pi/a, \pi/a, 0)$, 
$K = (3\pi/2a, 3\pi/2a, 0)$, $L = (\pi/a, \pi/a, \pi/a)$, $U = (2\pi/a, \pi/2a, \pi/2a)$.
}
\label{Fig:band}
\end{figure}

The band structures from the DFT calculation (red) and the fitted tight-binding 
Hamiltonian (blue) are shown in Fig. \ref{Fig:band}.
The transfer parameters derived from the DFT calculation are tabulated in Table \ref{Table:t}.
The present values are close to the recent estimates with optimized structure. \cite{nomura2012}
%This implies that the totally symmetric expansion of C$_{60}$ structure in K$_3$C$_{60}$ is not large. 

\begin{table}[tb]
\caption{LUMO level $\epsilon$ (eV), transfer parameters, and band width $w$ (meV) for K$_3$C$_{60}$.}
%\caption{Transfer parameters, and band width $w$ for K$_3$C$_{60}$ (in meV).}
\label{Table:t}
\begin{ruledtabular}
\begin{tabular}{ccccccccc}
 $\epsilon$ & $t_1$ & $t_2$ & $t_3$ & $t_4$ & $t_5$ & $t_6$ & $t_7$ & $w$ \\
% $t_1$ & $t_2$ & $t_3$ & $t_4$ & $t_5$ & $t_6$ & $t_7$ & $w$ \\
\hline
 5.066 & 43.3 & -31.9 & -6.2 & -16.6 & -9.6 & -2.0 & 2.7 & 505.9\\
% 43.3 & -31.9 & -6.2 & -16.6 & -9.6 & -2.0 & 2.7 & 505.9\\
\end{tabular}
\end{ruledtabular}
\end{table}

\section{Gutzwiller reduction factors for the LUMO bands and bielectronic energy in $A_3$C$_{60}$}
\label{Sec:q}
To derive the form of Eq. (\ref{Eq:Eg}), we apply the Gutzwiller's approximation 
extending the one for the nondegenerate band. \cite{Vollhardt1984a, Ogawa1975a}
Within the Gutzwiller's approximation, physical quantities are described in terms of the 
probability $\nu_\Gamma$ that one-site electron configuration $\Gamma$ appears in the 
Gutzwiller's wave function $|\Psi_{\rm G}\rangle$. \cite{Vollhardt1984a, Ogawa1975a}

The occupation number $n_x$ in spin orbital $x\sigma$ 
is described by the probabilities $\nu_\Gamma$ as follows:
\begin{eqnarray}
 n_x &=& \frac{1}{2}
         [2\nu_{x}
        + 2\nu_{x\bar{x}} + 4\nu_{xy} + 4\nu_{zx}
\nonumber\\
        &+& 4\nu_{xy\bar{x}} + 2\nu_{zx\bar{z}} + 4\nu_{zx\bar{x}} + 2\nu_{xy\bar{y}} + 8\nu_{xyz}
\nonumber\\
        &+& 2\nu_{xy\bar{x}\bar{y}} + 2\nu_{zx\bar{z}\bar{x}} + 8\nu_{xyz\bar{x}} + 4\nu_{xyz\bar{y}} 
         + 4\nu_{xyz\bar{z}} 
\nonumber\\
        &+& 4\nu_{xyz\bar{x}\bar{y}} + 2\nu_{xyz\bar{y}\bar{z}} + 4\nu_{xyz\bar{z}\bar{x}}
        + 2 \nu_{xyz\bar{x}\bar{y}\bar{z}}
        ],
\label{Eq:nG}
\end{eqnarray}
where, $1/2$ is due to the spin degrees of freedom, 
and $\lambda$ and $\bar{\lambda}$ $(\lambda = x,y,z)$ indicate spin orbitals 
$(\lambda \uparrow)$ and $(\lambda \downarrow)$, respectively.
Since we consider the metallic phase, $\nu_\Gamma$ does not depend on the spin part of $\Gamma$.
For example, $\nu_x = \nu_{\bar{x}}$.
$n_y$ and $n_z$ are obtained by cyclic permutation of the indices $(x,y,z)$ in Eq. (\ref{Eq:nG}).

The Gutzwiller's reduction factors $q_{xx}$ appearing in Eq. (\ref{Eq:Et}) is given by 
\begin{eqnarray}
 q_{xx} &=& \frac{1}{n_{x}(1-n_{x})} 
\left(
  \sqrt{\nu_0 \nu_x} 
+ \sqrt{\nu_{\bar{x}} \nu_{x\bar{x}}} 
+ 2\sqrt{\nu_y \nu_{xy}} 
\right.
\nonumber\\
&+& 2\sqrt{\nu_z \nu_{zx}} 
+ 2\sqrt{\nu_{y\bar{x}} \nu_{xy\bar{x}}}
+ \sqrt{\nu_{z\bar{z}} \nu_{zx\bar{z}}}
+ 2\sqrt{\nu_{z\bar{x}} \nu_{zx\bar{x}}}
\nonumber\\
&+&
  \sqrt{\nu_{y\bar{y}} \nu_{xy\bar{y}}}
+ 4\sqrt{\nu_{yz} \nu_{xyz}}
+ \sqrt{\nu_{y\bar{x}\bar{y}} \nu_{xy\bar{x}\bar{y}}}
\nonumber\\
&+&
  \sqrt{\nu_{z\bar{z}\bar{x}} \nu_{zx\bar{z}\bar{x}}}
+ 4\sqrt{\nu_{yz\bar{x}} \nu_{xyz\bar{x}}}
+ 2\sqrt{\nu_{yz\bar{y}} \nu_{xyz\bar{y}}}
\nonumber\\
&+& 
  2\sqrt{\nu_{yz\bar{z}} \nu_{xyz\bar{z}}}
+ 2\sqrt{\nu_{yz\bar{x}\bar{y}} \nu_{xyz\bar{x}\bar{y}}}
+ \sqrt{\nu_{yz\bar{y}\bar{z}} \nu_{xyz\bar{y}\bar{z}}}
\nonumber\\
&+& 
\left.
  2\sqrt{\nu_{yz\bar{z}\bar{x}} \nu_{xyz\bar{z}\bar{x}}}
+ \sqrt{\nu_{yz\bar{x}\bar{y}\bar{z}} \nu_{xyz\bar{x}\bar{y}\bar{z}}}
 \right)^2,
\label{Eq:qxx}
\end{eqnarray}
where $\Gamma = 0$ means the configuration with no electron.
$q_{yy}$ and $q_{zz}$ are obtained by 
cyclic permutation of the indices $(x,y,z)$ in Eq. (\ref{Eq:qxx}).
For $q_{\lambda \lambda'} (\lambda \ne \lambda')$, following relation holds:
\begin{eqnarray}
 q_{\lambda \lambda'} = \sqrt{q_{\lambda \lambda} q_{\lambda' \lambda'}}.
\end{eqnarray}

The bielectronic energy is 
\begin{eqnarray}
 E_{\rm bi} &=&
  U_\parallel (\nu_{x\bar{x}} + \nu_{y\bar{y}} + \nu_{z\bar{z}})
\nonumber\\
  &+& (U_\perp - J_{\rm H}/2) (4\nu_{xy} + 4\nu_{yz} + 4\nu_{zx})
\nonumber\\
  &+& (3U_\perp + J_{\rm H}) (2\nu_{xy\bar{x}} + 2\nu_{yz\bar{y}} + 2\nu_{zx\bar{z}}
\nonumber\\
  &+& 2\nu_{zx\bar{x}} + 2\nu_{xy\bar{y}} + 2\nu_{yz\bar{z}})
\nonumber\\
  &+& (3U_\perp - 3J_{\rm H}/2) 8\nu_{xyz}
\nonumber\\
  &+& (6U_\perp + 2J_{\rm H}) (\nu_{xy\bar{x}\bar{y}}
  + \nu_{yz\bar{y}\bar{z}} + \nu_{zx\bar{z}\bar{x}})
\nonumber\\
  &+& (6U_\perp - J_{\rm H}/2) (4\nu_{xyz\bar{x}} + 4\nu_{xyz\bar{y}} + 4\nu_{xyz\bar{z}})
\nonumber\\
  &+& (10U_\perp) (2\nu_{xyz\bar{x}\bar{y}} + 2\nu_{xyz\bar{y}\bar{z}} + 2\nu_{xyz\bar{z}\bar{x}})
\nonumber\\
  &+& (15U_\perp) \nu_{xyz\bar{x}\bar{y}\bar{z}}.
%  (Uperp + 2.0D0 * JH) * (nu(5) + nu(6) + nu(7))                        
%   + (Uperp - 0.5D0 * JH) * 4.0D0 * (nu(8) + nu(9) + nu(10))            
%   + (3.0D0 * Uperp + JH)                                               
%   * 2.0D0 * (nu(11) + nu(12) + nu(13) + nu(14) + nu(15) + nu(16))      
%   + (3.0D0 * Uperp - 1.5D0 * JH) * 8.0D0 * nu(17)                      
%   + (6.0D0 * Uperp + 2.0D0 * JH) * (nu(18) + nu(19) + nu(20))          
%   + (6.0D0 * Uperp - 0.5D0 * JH) * 4.0D0 * (nu(21) + nu(22) + nu(23))  
%   + (10.0D0 * Uperp) * 2.0D0 * (nu(24) + nu(25) + nu(26))              
%   + (15.0D0 * Uperp) * nu(27)
\end{eqnarray}

The Coulomb contribution (\ref{Eq:epsilonbi}) to the subband energy level is given by 
\begin{eqnarray}
 \epsilon_{{\rm bi},x} &=& \frac{U}{n_x} \left[
  \left(\nu_{x\bar{x}} + 2\nu_{xy} + 2\nu_{zx}\right)
 \right.
\nonumber\\
 &+&
  2\left(2\nu_{xy\bar{x}} + 2\nu_{zx\bar{x}} + \nu_{xy\bar{y}} + \nu_{zx\bar{z}} + 4\nu_{xyz} \right)
\nonumber\\
 &+&
  3\left(\nu_{xy\bar{x}\bar{y}} + \nu_{zx\bar{z}\bar{x}} + 4\nu_{xyz\bar{x}} + 2\nu_{xyz\bar{y}} + 2\nu_{xyz\bar{z}}\right)
\nonumber\\
 &+&
 4\left(2\nu_{xyz\bar{x}\bar{y}} + 2\nu_{xyz\bar{z}\bar{x}} + \nu_{xyz\bar{y}\bar{z}}\right)
\nonumber\\
 &+&
 \left.
 5\nu_{xyz\bar{x}\bar{y}\bar{z}}
\right].
\label{Eq:epsilonbix}
\end{eqnarray}
Here, we choose the ordered JT distortion (\ref{Eq:q}).
$\epsilon_{{\rm bi}, y}$ and $\epsilon_{{\rm bi}, z}$ are obtained by 
cyclic permutation of the indices $(x,y,z)$ in Eq. (\ref{Eq:epsilonbix}).
$\Delta \epsilon_{\rm bi}$ in Eq. (\ref{Eq:DeltaEdisp}) is obtained as 
$\Delta \epsilon_{\rm bi} = \epsilon_{{\rm bi}, z} - \epsilon_{{\rm bi}, x}$.

\bibliographystyle{apsrev4-1}
\bibliography{a3c60jt}

\end{document}